\documentclass[journal]{IEEEtran}

\usepackage{algorithm,amsbsy,amsmath,amssymb,epsfig,bbm,mathrsfs,multirow,amsthm}
\usepackage{array, multirow,graphicx}
\usepackage{graphicx}
\usepackage{graphics}
\usepackage{subcaption}
\usepackage{xcolor}
\usepackage{epstopdf}
\usepackage{color}
\usepackage{soul}
\usepackage{bbm}
\usepackage{supertabular}
\usepackage{longtable}
\usepackage{cite}
\usepackage{algpseudocode}
\usepackage{mathtools}
\usepackage{pdfpages}
\usepackage{booktabs}
\usepackage{enumitem}
\usepackage{makecell}

\usepackage{pdfrender}
\newcommand*{\boldcheckmark}{%
  \textpdfrender{
    TextRenderingMode=FillStroke,
    LineWidth=.5pt, 
  }{\checkmark}%
}

\usepackage[compact]{titlesec}  
\titlespacing{\section}{0pt}{2pt}{2pt}
\titlespacing{\subsection}{0pt}{2pt}{2pt}
\titlespacing{\subsubsection}{0pt}{2pt}{2pt}

\newcommand{\beq}{\begin{equation}}
\newcommand{\eeq}{\end{equation}}
\newcommand{\bitm}{\begin{itemize}}
\newcommand{\ba}{\begin{array}}
\newcommand{\ea}{\end{array}}
\newcommand{\eitm}{\end{itemize}}
\newcommand{\beqn}{\begin{eqnarray}}
\newcommand{\eeqn}{\end{eqnarray}}
\newcommand{\beqno}{\begin{eqnarray*}}
\newcommand{\eeqno}{\end{eqnarray*}}
\newcommand{\bma}{\begin{displaymath}}
\newcommand{\ema}{\end{displaymath}}
\newcommand{\bnu}{\begin{enumerate}}
\newcommand{\enu}{\end{enumerate}}
\newcommand{\bce}{\begin{center}}
\newcommand{\ece}{\end{center}}
\newcommand{\btb}{\begin{tabular}}
\newcommand{\etb}{\end{tabular}}

\begin{document}
\title{Graph Neural Networks for Next-Generation-IoT: Recent Advances and Open Challenges}

\author{Nguyen~Xuan~Tung, Le~Tung~Giang, Bui~Duc~Son, Seon~Geun-Jeong,  Trinh~Van~Chien, Won~Joo~Hwang, \textit{Senior Member}, \textit{IEEE}, Lajos~Hanzo, \textit{Life Fellow, IEEE} 
\thanks{Nguyen Xuan Tung is with Faculty of Interdisciplinary Digital Technology, PHENIKAA University, Yen Nghia, Ha Dong, Hanoi 12116, Viet Nam (e-mail: tung.nguyenxuan@phenikaa-uni.edu.vn)
\newline
Le Tung Giang and Seon Geun Jeong are with the Department of Information Convergence Engineering, Pusan National University, Busan 46241, Republic of Korea (e-mail: giang.lt2399144@pusan.ac.kr; wjdtjsrms11@pusan.ac.kr). 
\newline
Bui Duc Son is with the School of Electrical and Electronic Engineering, Hanoi University of Science and Technology, Hanoi 100000, Vietnam. (e-mails:son.bd200524@sis.hust.edu.vn).
\newline
Trinh Van Chien is with the School of Information and Communications Technology (SoICT), Hanoi University of Science and Technology (HUST), Vietnam (e-mail: chientv@soict.hust.edu.vn). 
\newline
Won-Joo Hwang is with the School of Computer Science and Engineering, Center for Artificial Intelligence Research, Pusan National University, Busan 46241, South Korea (e-mail: wjhwang@pusan.ac.kr).
\newline
Lajos Hanzo is with the Department of Electronics and Computer Science, University of Southampton, Southampton SO17 1BJ, U.K. (e-mail: lh@ecs.soton.ac.uk)}
}

\maketitle
\begin{abstract}
\textcolor{black}{Graph Neural Networks (GNNs) have emerged as a powerful framework for modeling complex interconnected systems, hence making them particularly well-suited to address the growing challenges of next-generation Internet of Things (NG-IoT) networks. Despite increasing interest in this area, existing studies remain fragmented, and there is a lack of comprehensive guidance on how GNNs can be systematically applied to NG-IoT systems. As NG-IoT systems evolve toward 6G, they incorporate diverse technologies such as massive MIMO, reconfigurable intelligent surfaces (RIS), terahertz (THz) communication, satellite systems, mobile edge computing (MEC), and ultra-reliable low-latency communication (URLLC). These advances promise unprecedented connectivity, sensing, and automation but also introduce significant complexity, requiring new approaches for scalable learning, dynamic optimization, and secure, decentralized decision-making. This survey provides a comprehensive and forward-looking exploration of how GNNs can empower NG-IoT environments structured as ten open research questions that span the relevant theoretical foundations, practical deployments, and emerging integration pathways. We commence by exploring the fundamental paradigms of GNNs and articulating the motivation for their use in NG-IoT networks. Besides, to further justify their suitability, we intrinsically connect GNNs for the first time with the family of low-density parity-check (LDPC) codes, modeling the NG-IoT as dynamic constrainted graphs where GNNs harness belief propagation for convergence and interpretability through density evolution and EXIT charts. We highlight the distinct roles of node-, edge-, and graph-level tasks in tackling key challenges and demonstrate the GNNs' ability to overcome the limitations of traditional optimization methods. Following this, we examine the application of GNNs across core NG-enabling technologies and their integration with distributed frameworks to support privacy-preservation and distributed intelligence. We then delve into the challenges posed by adversarial attacks, offering insights into defense mechanisms to secure GNN-based NG-IoT networks. Lastly, we examine how GNNs can be integrated with emerging technologies like integrated sensing and communication (ISAC), satellite-air-ground-sea integrated networks (SAGSIN), and quantum computing. Our findings highlight the transformative potential of GNNs in improving efficiency, scalability, and security within NG-IoT systems, paving the way for future advances. Finally, we summarize the key lessons learned throughout the paper and outline promising future research directions, along with a set of design guidelines aimed at facilitating the development of efficient, scalable, and secure GNN models tailored for NG-IoT applications.}
\end{abstract}


\section*{Glossary}

\tablefirsthead{}
\tablehead{}
\tabletail{}
\tablelasttail{}
\begin{supertabular}{p{15mm}p{65mm}}
6G & Sixth Generation \\ 
AIoT & Artificial Intelligence of Things \\ 
APs & Access Points \\ 
CDF & Cumulative Distribution Function \\ 
CLOPS & Circuit Layer Operations Per Second \\ 
D2D & Device-to-Device \\ 
DDPG & Deep Deterministic Policy Gradient \\ 
DNN & Deep Neural Networks \\ 
DP & Differential Privacy \\
DRL & Deep Reinforcement Learning \\ 
EXIT & Extrinsic Information Transfer \\
FL & Federated Learning \\ 
GAEs & Graph Autoencoders \\ 
GATs & Graph Attention Networks \\ 
GCNs & Graph Convolutional Networks \\ 
\end{supertabular}
\begin{supertabular}{p{15mm}p{65mm}}
GraphSAGE & Graph Sample And Aggregation \\ 
GRLO & Reinforcement Learning-Based Offloading \\ 
GRU & Gated Recursive Unit \\ 
GWCN & Graph-Weighted Convolution Network \\ 
HeGNNs & Heterogeneous Graph Neural Networks \\ 
HoGNNs & Homogeneous Graph Neural Networks \\ 
HQGNN & Hybrid Quantum Graph Neural Network \\ 
IIoT & Industrial Internet of Things \\ 
IoT & Internet of Things \\ 
IoV & Internet of Vehicles \\ 
ISAC & Integrated Sensing And Communication \\
LDPC & Low-Density Parity-Check \\
LEO & Low Earth Orbit \\ 
LLM & Large Language Model \\ 
MEC & Mobile Edge Computing \\ 
MIMO & Multiple Input Multiple Output \\ 
\end{supertabular}
\begin{supertabular}{p{15mm}p{65mm}}
NG & Next Generation \\ 
NISQ & Noisy Intermediate-Scale Quantum \\ 
QGNNs & Quantum Graph Neural Networks \\ 
QML & Quantum Machine Learning \\ 
RIS & Reconfigurable Intelligent Surfaces \\ 
RL & Reinforcement Learning \\ 
SAGSINs & Satellite-Air-Ground-Sea Integrated Networks \\ 
SFCs & Service Function Chains \\ 
THz & Terahertz \\ 
UAVs & Unmanned Aerial Vehicles \\ 
URLLC & Ultra-Reliable Low Latency Communications \\ 
VQC & Variational Quantum Circuits \\ 
WMMSE & Weighted Minimum Mean Squared Error \\
\end{supertabular}
\vspace{10pt}
    
    \section{Introduction}
    \label{sec:introduction}
     
    {\color{black}The Internet of Things (IoT) has revolutionized the way we interact with our environment, supporting a vast network of interconnected devices that communicate and exchange data seamlessly \cite{IoT_survey_Lin_Mar_2017, Jahanbakht9328873}. As we move toward the next-generation IoT (NG-IoT) paradigm, driven by the emergence of 6G, the integration of technologies such as massive MIMO, reconfigurable intelligent surfaces (RIS), satellites, terahertz (THz) communications, mobile edge computing (MEC), and ultra-reliable low-latency communication (URLLC) opens up unprecedented possibilities for intelligent, low-latency, and context-aware connectivity \cite{6G_IoT_survey_NguyenDC_Aug_2021}. At the same time, recent advances like integrated sensing and communication (ISAC), satellite-air-ground-sea integrated networks (SAGSINs), and quantum-enhanced computing will significantly expand the system scale, heterogeneity, and complexity of NG-IoT environments. Despite their potential, the rapid development of NG-IoT networks involves higher complexity and dynamism, which introduce critical challenges in terms of scalability, robustness, and real-time decision-making. As network topologies become increasingly heterogeneous and time-varying, achieving efficient resource allocation, topology-aware learning, and resilient inference under dynamic wireless environments has become essential.}

    \begin{table*}[h]
    \centering
    \caption{Overview of Developments from Graph Theory to Graph Neural Networks (GNNs)}
    \label{Table:Graph_Developments}
    \begin{tabular}{|p{0.05\textwidth}|p{0.1\textwidth}|p{0.25\textwidth}|p{0.35\textwidth}|p{0.12\textwidth}|}
    \hline
    \textbf{Year} & \textbf{Category} & \textbf{Significance} & \textbf{Key Application areas} & \textbf{Reference} \\
    \hline
    1956 & Classical Graph Theory & Provided foundational algorithms for pathfinding in graphs & Shortest Path Algorithms (Routing, Logistics) & Dijkstra, E. W. \cite{Dijkstra1959}  \\
    \hline
    1977 & Centrality Measures & Introduced measures to quantify node importance in graphs & Network Analysis (Node Importance in Social and Communication Networks) & Linton C. Freeman \cite{Linton1977} \\
    \hline
    1996 & Spectral Graph Theory & Leveraged eigenvalues and eigenvectors for graph partitioning and embedding & Network Partitioning (Clustering, Community Detection) & Fiedler, M. \cite{Fiedler1973}, Cvetkovic, D.M. \textit{et al.} \cite{cvetkovic1980spectra}  \\
    \hline
    2005 & Graph Neural Networks & Proposed using neural networks for general graphs, extending deep learning to graph-structured data & Machine Learning (Node Classification, Link Prediction, Graph-Level Regression, Network Communication) & Gori, M. \textit{et al.} \cite{Gori2005} \\
    \hline
    2014 & Random Walks & Enabled learning of node embeddings by random walks, pioneering unsupervised learning on graphs & Extends convolutional neural networks to graph data for feature learning (Natural Language Processing, Recommender Systems) & Perozzi, B. \textit{et al.} \cite{Perozzi2014} \\
    \hline
    2017 & Graph Convolutions & Applied convolution operations to graph data, revolutionizing graph learning with scalability and efficiency & Extends convolutional neural networks to graph data for feature learning (Social Networks, Citation Networks) & Thomas N. Kipf and Max Welling \cite{Thomas2017} \\
    \hline
    2018 & Graph Attention Networks & Introduced attention mechanisms to graph learning for more adaptive representation learning & Introduced attention mechanisms for graph learning (Recommender Systems, Social Networks, Computational Biology) & Velickovic, P. \textit{et al.} \cite{Velickovic2018} \\
    \hline
    2019 & Quantum GNNs & Integrated quantum computing to improve large-scale graph processing and optimization & Integrated quantum computing for large-scale graph processing and optimization (Quantum Computing, Cryptography, Large-Scale Networks) & Liao, Y. \textit{et al.} \cite{Guillaume2019} \\
    \hline
    {\color{black}2023} & {\color{black}GraphGPT} & {\color{black}Combined large language models(LLMs) with graph structural learning via instruction tuning, enabling cross-domain generalization} & {\color{black}Generalized graph representation across datasets with zero-shot and supervised learning using text-graph alignment (Zero-Shot Learning, Instruction-Tuned LLMs, Cross-Modal Graph Learning)} & {\color{black}Tang, J. \textit{et al.} \cite{Tang_GraphGPT}} \\
    \hline
    {\color{black}2024} & {\color{black}Retrieval-Augmented GNNs} & {\color{black}Unified retrieval + GNN framework to inject non-parametric knowledge into graph learning, improving generalization for unseen entities or relations} & {\color{black}Knowledge Graphs, Open-World QA, Recommender Systems} & {\color{black}Han, H. \textit{et al.} \cite{han2025retrievalaugmentedgenerationgraphsgraphrag}}\\
    \hline
    \end{tabular}
    \end{table*}

    \begin{table*}[h]
    \centering
    \caption{Comparison of Optimization Approaches, Traditional Deep Learning, and GNNs in NG-IoT Networks}
    \label{tab:comparison}
    \begin{tabular}{|m{2.8cm}|m{4.5cm}|m{4.5cm}|m{4.5cm}|}
    \hline
    \textbf{Criterion}                       & \textbf{Optimization Approaches}               & \textbf{Traditional Deep Learning}           & \textbf{Graph Neural Networks (GNNs)}         \\ \hline
    \textbf{Scalability}                     & Low scalability, often limited by problem size and complexity. & Limited scalability, the model works on trained network size. & High scalability, able to handle variable network sizes. \\ 
    \hline
    \textbf{Computational complexity}        & High, especially for large-scale problems. & High, requires large training datasets. & Moderate to low, efficient at handling graph-structured data with fewer training samples. \\ 
    \hline
    \textbf{Practical implementation}        & Moderate, difficult for large network size. & Moderate requires high memory to save models for different network sizes. & Practical and flexible. \\ 
    \hline
    \textbf{Integration with emerging technologies} & Limited requires specific modifications for different technologies. & Moderate can be adapted but with significant effort and computational cost. & High easily integrates. \\ 
    \hline
    \end{tabular}
    \end{table*}

    {\color{black}Traditional approaches, such as optimization-based methods and conventional deep learning, have been at the forefront of efforts to enhance 6G-IoT technologies \cite{Liwen10539120, Mahmood9861650, Vaezi9711564, Ferrag10255264, Salh9389782, Wang8957702}. However, these methods are increasingly inadequate in addressing these rapidly evolving multifaceted demands. While these approaches have yielded notable improvements, these are often attained at the cost of eroded scalability and excessive complexity. For instance, MEC and massive MIMO require sophisticated resource allocation and interference management, which are computationally intensive and, hence, challenging to implement in real-time scenarios. Similarly, the deployment of RIS and THz communications necessitates agile environmental adaptation and robust signal processing techniques, which traditional optimization methods struggle to handle efficiently.

    Graph Neural Networks (GNNs) have emerged as a promising technique of circumventing these limitations, offering a novel approach to modeling the complex relationships and dependencies inherent in NG-IoT networks. Unlike traditional optimization or deep learning methods that operate on structured, Euclidean data, GNNs are designed to learn over irregular, non-Euclidean domains such as graphs. This allows GNNs to naturally represent the complex, multi-hop relationships among entities in wireless networks, such as IoT devices, edge servers, UAVs, and base stations, where the data structure is inherently graph-like \cite{GNN_Survey_Dong_Apr_2023, RA_Wireless_Communications_Theory_to_Practice_Yifei_May_2023}. To better understand the evolution of GNNs and their capabilities, Table~\ref{Table:Graph_Developments} highlights their evolution from early graph theory to advanced GNN models. 
    
    Furthermore, GNNs can dynamically adapt to evolving topologies without retraining, handle varying-sized networks, and support decentralized inference, all of which are critical for NG-IoT deployments. Their message-passing architecture captures both local and global contexts, enabling robust predictions even under partial observability or uncertain channel conditions. By contrast, traditional optimization techniques often suffer from high complexity and limited generalization capability, while deep neural networks lack the inductive relational biases required for graph-structured environments. To illustrate this contrast, Table~\ref{tab:comparison} provides a comparative analysis of optimization-based approaches, conventional deep learning, and GNNs in terms of scalability, computational complexity, integration flexibility, and real-world applicability.}
    
    {\color{black}Existing surveys have significantly contributed to the understanding of GNNs and their application in wireless communications and in the IoT. For example, Lee \textit{et al.} \cite{GNN_Survey_Lee_Oct_2022} discussed the potential of GNNs in wireless communications, focusing on how graphical models are constructed and on their application in wireless networks. As a further advance, Ivanov \textit{et al.} \cite{GNN_Survey_Ivanov_Aug_2022} provided insights into resource allocation using GNNs for integrated space and terrestrial networks, while Tam \textit{et al.} \cite{GNN_Survey_Tam_Sep_2022} provided a review of GNN applications in areas such as network management, offloading strategies, routing optimization, virtual network function orchestration, and resource allocation. Additionally, Suarez \textit{et al.} \cite{GNN_Survey_Suarez_Aug_2022} focused on specific use cases of GNNs in communication networks. The most recent survey was offered by Sabarish \textit{et al.} \cite{moorthy2024surveygraphneuralnetwork}, who have explored GNN applications in IoT networks, highlighting their benefits in spectrum awareness, data fusion, and network intrusion detection. While these surveys have made considerable progress, this treatise provides an up-to-date critical appraisal of the relevant follow-up advances. Besides, these surveys tend to focus on isolated use cases or specific layers of the network stack, lacking a holistic exploration of how GNNs can address the growing complexity of next-generation IoT (NG-IoT) systems. Most prior studies provided technical summaries, but fall short of offering a practical framework or decision-making guide for selecting appropriate GNN models for different NG-IoT tasks.} 
    
    {\color{black}Moreover, critical issues such as deployment feasibility, system cost, energy efficiency, and communication overhead, which are essential for real-world GNN deployment, have not been discussed. The role of decentralized structures and federated learning, which are increasingly important in privacy-sensitive distributed IoT networks, also remains underexplored. Additionally, wireless networks, due to their open and distributed nature, are particularly vulnerable to threats such as eavesdropping, jamming, and adversarial attacks \cite{Zou7467419, Hoang10428002, Zhang9149584}. Yet, most surveys inadequately tackle GNN robustness under such threats, leaving a critical knowledge gap in the context of secure and resilient NG-IoT operations.} 
    
    {\color{black}Furthermore, prior review papers tend to focus on specific aspects of GNN applications, missing a holistic perspective that integrates multiple technologies and their interactions. The future potential and emerging applications of GNNs are frequently overlooked, limiting the scope to existing technologies. Importantly, the question of why GNNs are needed in NG-IoT networks has often been overlooked in previous studies. In contrast, this paper aims to comprehensively address this gap by presenting an integrated survey that explores both the theoretical and practical dimensions of GNN deployment in next-generation IoT systems. We formulate ten open questions that collectively examine the core challenges and opportunities of applying GNNs in NG-IoT environments. These include foundational discussions, including the justification for GNNs based on the structural characteristics of NG-IoT networks and a novel theoretical bridge to low-density parity-check (LDPC) codes, as well as strategies for model selection, convergence analysis, and interpretability. Moreover, we investigate real-world deployment challenges, including scalability, cost-efficiency, security, and application diversity, along with the role of GNNs in enabling distributed and federated learning across heterogeneous network infrastructures. We also examine how GNNs can be integrated with future-proof technologies such as integrated sensing and communication, space-air-ground-sea integrated networks, and quantum computing. Each question is explored in depth, providing an overview of the existing body of research, open challenges, and future directions. Through this lens, we aim to equip researchers and practitioners with a holistic understanding of how GNNs can empower the next wave of intelligent IoT systems. A summary of these open questions is provided in Table~\ref{Table: summary_open_questions}. The main contributions of this paper are summarized as follows:}
    \begin{figure}[t]
        \includegraphics[trim=0cm 0cm 0cm 0.7cm, clip=true, width=\columnwidth]{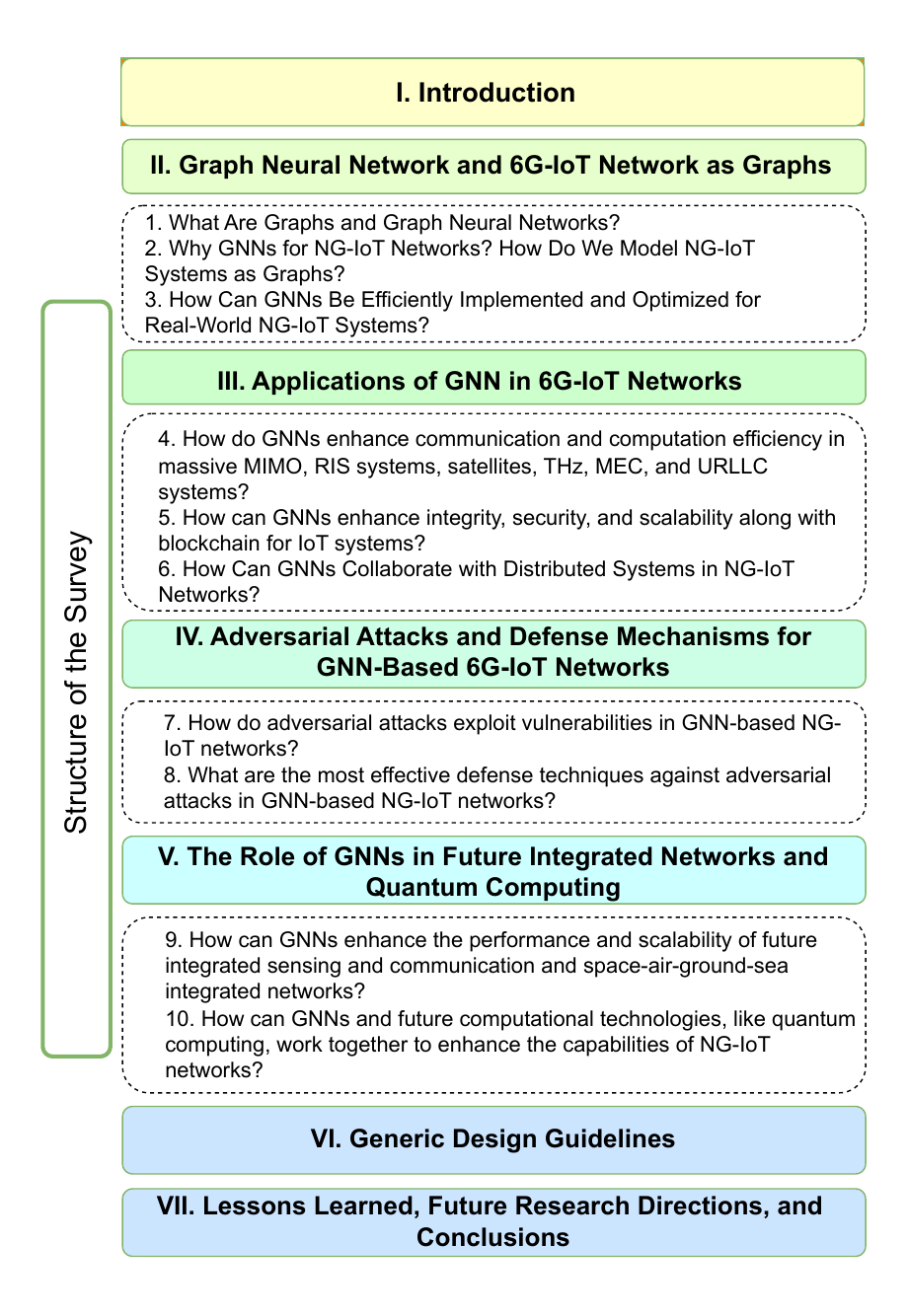}
        \caption{{Structure of this paper.}}
        \label{fig: Table of contents}
    \end{figure}

    \begin{table*} \scriptsize%
        \centering
        \def\arraystretch{1.25}
        \caption{Summary of Open Questions on GNN-Based NG-IoT Networks.}
        \label{Table: summary_open_questions}
        \begin{tabular}{|p{0.15\textwidth}|p{0.25\textwidth}|p{0.23\textwidth}|p{0.25\textwidth}|}
        \hline
        \textbf{Open Questions} & \textbf{Background} & \textbf{State-of-the-art} & \textbf{Challenges and Future Directions} \\
        \hline
        1) What Are Graphs and Graph Neural Networks? & Graph definition and Graph Neural Networks (GNNs) paradism & \begin{itemize}[leftmargin=*, nosep]
            \item Graph data, GNN Paradigm, and GNN variants \cite{azizian2021expressivepowerinvariantequivariant, ZHOU202057, Wu2023}  
            \item The explainability and scalability of GNNs \cite{Yuan2023, III_B_GNN_Comprehensive_Survey_Zonghan_Mar_2021} 
            \end{itemize}
            & \begin{itemize}[leftmargin=*, nosep]
            \item Model depth and over-smoothing
            \item Scalability of GNN models
            \item Security
            \item Privacy-specific to GNN
        \end{itemize} \\
        \hline
        2) Why GNNs for NG-IoT Networks? How Do We Model NG-IoT Systems as Graphs? & NG-IoT networks feature dynamic topology, heterogeneity, and non-Euclidean structures, making them well-suited for graph modeling. GNNs enable effective learning at node, edge, and graph levels to address these complexities. & \begin{itemize}[leftmargin=*, nosep]
            \item Node-level task \cite{III_B_Node_Abode_D_2023, III_B_Node_Li_B_2024, III_B_Node_Giang_L_2024, III_B_Node_Lee_M_2021, III_B_Node_Chen_T_2022, III_B_Node_Wang_Z_2023, III_B_Node_Lyu_S_2024, III_B_Node_Zhao_N_2022, III_B_Node_Xiao_J_2023, III_B_Node_Huoh_T_2023}
            \item Edge-level task \cite{III_B_Edge_Liu_X_2021, III_B_Edge_Sun_Z_2023, RA_Vertex_vs_Edge_Peng_Jan_2024, III_B_Edge_Wang_Y_2023, III_B_Edge_Liu_S_2023}
            \item Graph-level task \cite{III_B_Graph_Yang_Y_2023, III_B_Graph_Wu_Z_2022, III_B_Graph_Wang_G_2023, III_B_Graph_Asheralieva_A_2023}
        \end{itemize} & \begin{itemize}[leftmargin=*, nosep]
            \item High computational and memory requirements.
            \item Divide large graphs into subgraphs for local learning and then aggregate
            \item Adapting DE and EXIT chart for analyze GNN and model design
        \end{itemize} \\
        \hline
        3) How Can GNNs Be Efficiently Implemented and Optimized for Real-World NG-IoT Systems? & Deploying GNN in real NG-IoT environments is affected by multiple aspects, including model deployment, multi-modal data, socio-economic dimentions of GNN including cost consideration and sustainability. & \begin{itemize}[leftmargin=*, nosep]
            \item Model Optimization for Edge and Cloud Deployment \cite{Zhang3686539, Wang3614955, xie2024lightweight, wu2024edgefreestructureawareprototypeguidedknowledge}
            \item Multi-modal data processing in NG-IoT networks \cite{LI2024119815, Zhang10860124}
            \item Cost consideration \cite{Shuvo9985008, Duc3341145, Murshed3469029, strubell-etal-2019-energy}
        \end{itemize} & \begin{itemize}[leftmargin=*, nosep]
            \item Resource constraints at edge devices
            \item Need for real-time, scalable inference
            \item Multimodal data integration
            \item Deployment cost and energy efficiency
        \end{itemize} \\
        \hline
        4) How do GNNs enhance communication and computation efficiency in massive MIMO, RIS systems, satellites, THz, MEC, and URLLC systems? & Massive MIMO, RIS Systems, Satellite, THz Communication, MEC, and URLLC are essential for enhancing NG-IoT system performance, coverage, and efficiency & 
        \begin{itemize}[leftmargin=*, nosep]
            \item Massive MIMO \cite{RA_Wireless_Communications_Theory_to_Practice_Yifei_May_2023, cfmMIMO_IoT_Li_Jun_2023, Tung10750215, cf_mMIMO_Salaun_Dec_2022, RA_Vertex_vs_Edge_Peng_Jan_2024, cf_mMIMO_Ranasinghe_Dec_2021,mMIMO_IoT_Chien_Jan_2024,Coordinated_Sum_Rate_Deep_Unrolling_Schynol_Apr_2023}
            \item RIS Systems \cite{RIS_Singh_Dec_2023, RIS_Zhang_May_2022, RIS_Satellite_IoT_Cao_Feb_2022, Coordinated_Sum_Rate_Deep_Unrolling_Schynol_Apr_2023}
            \item Satellite \cite{SAGS_GNN_IoT_Huang_Nov_2023, SAGS_GNN_IoT_Chen_Mar_2023, SAGS_GNN_IoT_He_Jan_2024, SAGS_GNN_Wang_Apr_2023HotICN, SAGS_GNN_IoT_Asheralieva_Jul_2023, SAGS_GNN_IoT_Tekbyk_Jul_2021, SAGS_GNN_IoT_Chen_Mar_2023}
            \item THz Communication \cite{THz_GNN_IoT_Zhang_Sep_2023, THz_GNN_IoT_Mar_2024, THz_GNN_Li_Oct_2023}
            \item MEC \cite{Edge_general_Sun_Oct_2021, Edge_general_Wang_Oct_2023, Edge(UAV)_general_Li_Jun_2022, Edge_Transpo_Xu_Nov_2023, Edge_Transpo_Zhou_Jan_2023, Edge_Transpo_Liu_Apr_2022, Edge_health_Fei_Jan_2024, Edge_smarthome_Sun_Apr_2023, Edge_industrial_Tang_Oct_2023}
            \item URLLC \cite{URLLC_IoT_GNN_Liu_Oct_2023, URLLC_IoT_GNN_Jiaqi_Feb_2024, URLLC_IoT_GNN_Liu_Dec_2021, III_B_Graph_Gu_Y_2024}
        \end{itemize} 
        & 
        \begin{itemize}[leftmargin=*, nosep]
            \item Heterogeneity of network's entities
            \item Dynamic resource allocation with diverse constraints
            \item Scalability
        \end{itemize} \\
        \hline
        5) How can GNNs enhance integrity, security, and scalability along with blockchain for IoT systems? & Blockchain is a promising technology for ensuring data integrity and security in IoT systems. GNNs can help enhance blockchain-based IoT systems by improving node classification, enhancing security, and supporting scalable solutions & 
        \begin{itemize}[leftmargin=*, nosep]
            \item User privacy \cite{GTxChainIoT_security_Cai_Jul_2023}
            \item Application distribution among IoT networks \cite{Blockchain_IoT_GNN_Kim_Nov_2020}
            \item Malicious node detection \cite{Blockchain_IoT_GNN_Ziyu_Aug_2020}
        \end{itemize}
        & 
        \begin{itemize}[leftmargin=*, nosep]
            \item Computational and communication burden
            \item Ensuring data integrity and security in decentralized environments
            \item Lightweight GNN models design
            \item Privacy-preserving mechanisms
        \end{itemize} \\
        \hline
        6) How Can GNNs Collaborate with Distributed Systems in NG-IoT Networks? & Distributed learning and GNNs together enable scalable, privacy-aware intelligence across NG-IoT edge devices. 
        &
        \begin{itemize}[leftmargin=*, nosep]
            \item Distributed architecture for GNN in NG-IoT networks \cite{jianping2024federated, 10026810, 10433561, 10734080, khanna2025grl, 10533257, 10855737, 9919905, Edge_general_Zeng_Jul_2023}
            \item GNN for the distributed systems in NG-IoT networks \cite{10364739, III_B_Node_Wang_Z_2023, 10437172, III_B_Graph_Gu_Y_2024}
        \end{itemize}
        & 
        \begin{itemize}[leftmargin=*, nosep]
            \item Unstable training and poor convergence.
            \item Lack scalability in large or dense networks.
            \item Synchronization and consistency
            \item Evolving graphs demand online updates due to varying environment.
        \end{itemize} \\
        \hline
        7) How Do Adversarial Attacks Exploit Vulnerabilities in GNN-Based NG-IoT Networks? & GNN models are vulnerable to adversarial attacks, where small changes to input data can degrade performance. In NG-IoT networks, these attacks threaten data integrity, disrupt services, and pose security risks in various applications. 
        &
        \begin{itemize}[leftmargin=*, nosep]
            \item Adversarial  Homogeneous  Graph Neural Network \cite{ma2020towards, sharma2023task, dai2018adversarial, chang2020restricted, wang2020evasion}
            \item Adversarial Heterogeneous Graph Neural Network \cite{sun2020adversarial, zhao2024hgattack}
        \end{itemize}
        & 
        \begin{itemize}[leftmargin=*, nosep]
            \item Adversarial HoGNN defense
            \item Trade-off between deconstructing performance and computational complexity
            \item Explore joint attack methods involving multiple algorithms.
        \end{itemize} \\
        \hline
        8) What are the most effective defense techniques against adversarial attacks in GNN-based NG-IoT networks? & Defense strategies are crucial to safeguarding GNNs from adversarial threats to ensure robust deployment of GNNs in critical NG-IoT applications, including smart cities, healthcare, and autonomous systems & 
        \begin{itemize}[leftmargin=*, nosep]
            \item Adversarial Homogeneous GNN defense \cite{zhu2019robust, zhang2020gnnguard}
            \item Adversarial Heterogeneous GNN defense \cite{zhang2022robust, 10124876}
        \end{itemize} 
        & 
        \begin{itemize}[leftmargin=*, nosep]
            \item Adversarial training
            \item Defensive distillation: Distill knowledge from a complex model to a simple one
            \item Hybrid defense approaches to improve robustness.
        \end{itemize} \\
        \hline
        9) How can GNNs enhance the performance and scalability of future integrated sensing and communication and space-air-ground-sea integrated networks? & Future integrated networks like SAGSINs and ISAC are crucial for ensuring seamless connectivity across multiple domains. GNNs can enhance the performance and scalability of these networks by optimizing communication and sensing processes and handling complex cross-domain interactions. & 
        \begin{itemize}[leftmargin=*, nosep]
            \item Integrated Communications and Sensing \cite{OpenIssue_JSAC_Lee_Oct_2022, THz_GNN_Li_Oct_2023}
            \item Space-air-ground-sea integrated networks \cite{Edge_Transpo_Liu_Apr_2022}
        \end{itemize} 
        & 
        \begin{itemize}[leftmargin=*, nosep]
            \item Scalability and heterogeneity in ISAC and SAGSINs
            \item Privacy and security issues in integrated communication and sensing
            \item Lack of standardized protocols for seamless integration of GNNs
        \end{itemize} \\
        \hline
        10) How can GNNs and future computational technologies, like quantum computing, work together to enhance the capabilities of NG-IoT networks? & Quantum computing has the potential to address the computational limitations of GNNs, providing enhanced capabilities for processing complex graph data in NG-IoT networks & 
        \begin{itemize}[leftmargin=*, nosep]
            \item Quantum computing and quantum circuits \cite{10274707, JEONG2024608, verdon2019quantum}
            \item Variational quantum circuit for GNN \cite{10499715Zheng_QuantumGNN_VQC}
            \item Hybrid quantum graph neural network \cite{liao2024graph}
        \end{itemize}
        & \begin{itemize}[leftmargin=*, nosep]
            \item Noisy intermediate-scale quantum (NISQ) devices
            \item Standard optimal variational quantum circuits for implementing QGNNs guaranteeing efficient circuit design and scalability
            \item Quantum cryptography with GNN
        \end{itemize} \\
        \hline
        \end{tabular}
    \end{table*}

    \begin{table*} \scriptsize%
    \centering
    \def\arraystretch{1.25}
    \caption{Related surveys on the applications of GNN in NG-IoT networks versus our study.}
    \label{Table: Survey comparisons}
    \begin{tabular}    {|p{0.18\textwidth}|p{0.23\textwidth}|p{0.03\textwidth}|p{0.03\textwidth}|p{0.03\textwidth}|p{0.03\textwidth}|p{0.03\textwidth}|p{0.03\textwidth}|p{0.03\textwidth}|p{0.03\textwidth}|b{0.08\textwidth}|}
    
    
    \hline
     & &  \makecell[l]{\cite{GNN_Survey_He_Nov_2021}} & \makecell[l]{\cite{GNN_Survey_Suarez_Aug_2022}} & \makecell[l]{\cite{GNN_Survey_Tam_Sep_2022}} & \makecell[l]{\cite{GNN_Survey_Ivanov_Aug_2022} } & \makecell[l]{\cite{GNN_Survey_Lee_Oct_2022}} & \makecell[l]{\cite{GNN_Survey_Yuxi_Mar_2023}} & \makecell[l]{\cite{GNN_Survey_Dong_Apr_2023}} &  \makecell[l]{\cite{moorthy2024surveygraphneuralnetwork}} & \textbf{Our research} \\
    \hline
    \makecell[{{p{0.15\textwidth}}}]{Categorizing NG-IoT network problems by graph problems} &  & \makecell{$\checkmark$} &  & \makecell{$\checkmark$} &  \makecell{$\checkmark$} & \makecell{$\checkmark$} &  &  & \makecell{$\checkmark$} & \makecell{$\boldcheckmark$} \\
    \hline
    \makecell[{{p{0.15\textwidth}}}]{Efficient deploy GNN in real NG-IoT environments} & Model deployment, multi-modal data, socio-economic dimentions &  &  &  &  &  &  &  &  & \makecell{$\boldcheckmark$} \\
    \hline
    \multirow{7}{*}{GNN for NG technologies}  & Massive MIMO & \makecell{$\checkmark$} &  &  & \makecell{$\checkmark$} & \makecell{$\checkmark$} &  &  & & \makecell{$\boldcheckmark$}  \\
    \cline{2-11}
       & RIS &  &  &  &  &  &  &  & & \makecell{$\boldcheckmark$} \\
    \cline{2-11}
                     & Satellite Communication & \makecell{$\checkmark$} & \makecell{$\checkmark$}  &  & \makecell{$\checkmark$} &  &  & \makecell{$\checkmark$} & & \makecell{$\boldcheckmark$} \\
    \cline{2-11}
                     & THz &  &  &  &  &  &  &  & \makecell{$\checkmark$} & \makecell{$\boldcheckmark$} \\
    \cline{2-11}
                     & URLLC &  &  & \makecell{$\checkmark$} &  &  &  &  & & \makecell{$\boldcheckmark$} \\
    \cline{2-11}
                    & Edge computing &  &  & \makecell{$\checkmark$} &  &  & \makecell{$\checkmark$} & \makecell{$\checkmark$} & \makecell{$\checkmark$} & \makecell{$\boldcheckmark$}  \\
    \cline{2-11}
                     & Blockchain &  &  & \makecell{$\checkmark$} &  &  & \makecell{$\checkmark$} &  & & \makecell{$\boldcheckmark$} \\
    \cline{2-11}
                     & Distributed systems  &  &  &  &  &  &  &  & & \makecell{$\boldcheckmark$} \\
    \hline
    Adversarial attack and defense on GNN &  &  &  &  &  &  &  &  & & \makecell{$\boldcheckmark$} \\
    \hline
    \multirow{3}{*}{\makecell[{{p{0.18\textwidth}}}]{GNN with future NG-IoT networks and technologies}} & Integrated communication and sensing &  &  &  &  &  &  &  & & \makecell{$\boldcheckmark$} \\
    \cline{2-11}
                     & Space-air-groud-sea integrated networks &  &   &  & \makecell{$\checkmark$} &  &  &  & & \makecell{$\boldcheckmark$} \\
    \cline{2-11}
                     & Quantum GNN &  &  &  &  &  &  &  & & \makecell{$\boldcheckmark$} \\
    \hline
    \end{tabular}
    \end{table*}
    
    {\color{black}\begin{itemize}
        \item We design an overview of graph neural networks in NG IoT systems, discussing their definitions, paradigms, and key use cases. To justify the use of GNNs, we analyze the unique characteristics of NG-IoT networks and bridge them to the factor graph with the low-density parity-check (LDPC) code to show how GNNs serve as learnable generalizations of belief propagation. This connection supports exploring the GNN convergence with density evolution and training interpretability via EXIT charts. We explain the roles and benefits of node-, edge-, and graph-level tasks in wireless network problems, offering practical examples to guide their application in various NG-IoT scenarios. We compare the performance of different GNN models, including our proposed hybrid quantum GNN harnessed for power allocation in cell-free massive MIMO, demonstrating the efficiency of GNNs and the potential of quantum GNNs for future research.
        \item We explore key factors influencing the efficient deployment of GNNs in real-world NG-IoT environments, including on-device implementation, multimodal data integration, cost considerations, and sustainability.
        \item We present a comprehensive review of GNN applications, carefully categorized by the core technologies driving NG advances. These include massive MIMO schemes, RIS, Satellite, THz communications, MEC, URLLC, and blockchain. We also discuss the integration of GNNs with distributed systems, highlighting how they mutually enhance scalability, privacy, and collaborative intelligence in NG-IoT networks. 
        \item To fully harness the advantages of GNNs, we conduct an in-depth examination of adversarial attacks and defense techniques in GNNs, providing essential insights into the security challenges and solutions for deploying GNNs in NG-IoT networks.
        \item We explore the potential of GNNs in shaping NG-IoT networks, focusing on identifying adversarial attacks and proposing defense techniques. We also examine the use of GNNs in future integrated networks, including ISAC and SAGSINs. Additionally, we discuss the role of quantum computing in NG systems, highlighting how the combination of quantum and GNN can enhance GNN capabilities and the challenges of implementing them.
        \item Table~\ref{Table: Survey comparisons} offers a detailed comparison between our work and other state-of-the-art surveys in the field, emphasizing the unique contributions and advances presented in this paper.
    \end{itemize}}
      
    \emph{Paper Organization:} {\color{black}The remainder of the paper is organized as follows. Section~\ref{Section: Graph Neural Network and 6G-IoT Network as Graphs} provides an overview of GNNs, including their definitions and fundamental paradigms. We highlight the essential role of GNNs in NG-IoT networks and describe how to represent NG-IoT environments as graph structures. Various graph task types, including node-, edge-, and graph-level, are introduced, followed by a simulation study on power allocation in cell-free massive MIMO systems to compare the performance of different GNN models. In Section~\ref{GNN4_6G}, we survey GNN applications across key NG technologies, including massive MIMO, RIS, satellite communications, THz, MEC, URLLC, and blockchain, along with their integration with distributed systems. Section~\ref{Section: Adversarial Graph Neutral Network} delves into adversarial GNNs, examining potential attack methods and defense strategies in GNN-based NG networks, with a focus on both homogeneous and heterogeneous graph settings. In Section~\ref{Challenges}, we discuss the integration of GNNs with emerging NG technologies, including ISAC, SAGSINs, and quantum graph neural networks. Section~\ref{Sec: Generic Design Guidelines} will provide a set of generic design guidelines. Finally, Section~\ref{Conclusion} concludes the paper. For convenience, we treat each of the eight open questions in the form of an identical structure: \textit{1)} Background; \textit{2)} State-of-the-art; \textit{3)} Challenges and future directions.}

\section{Foundations of GNNs for NG-IoT}
    \label{Section: Graph Neural Network and 6G-IoT Network as Graphs}
    {\color{black}In this section, we address Open Questions 1, 2, and 3. We commence by introducing the fundamental concept of graphs, distinguishing between key types such as directed vs. undirected and homogeneous vs. heterogeneous graphs. We then provide an overview of Graph Neural Networks (GNNs), including their working principles and prominent architectures, such as Graph Convolutional Networks (GCNs) \cite{Thomas2017}, Graph Attention Networks (GATs) \cite{Velickovic2018}, Graph Autoencoders (GAEs) \cite{pan2018adversarially}, and Graph Spatial-Temporal Networks (GSTNs) \cite{song2020spatial}. Each of these models is designed to address specific challenges associated with learning from graph-structured data. Following this, we explore Open Question 2 by examining why GNNs are particularly suitable for NG-IoT networks. We analyze the unique characteristics of NG-IoT environments and illustrate how these systems can be effectively modeled as graphs. We then bridge NG-IoT traits to graph-based modeling through the factor graphs inspired by LDPC decoding. GNNs, as learnable generalizations of belief propagation, enable convergence analysis via density evolution (DE) and offer training insights via the extrinsic information transfer (EXIT) charts, reinforcing their suitability for dynamic, scalable, and structured NG-IoT environments. The advantages of node-, edge-, and graph-level tasks are discussed in the context of optimizing key wireless functions, such as power allocation, user association, and network deployment, with practical examples provided. Subsequently, we present simulation results comparing the performance of different GNN architectures, including a proposed hybrid quantum GNN model, which demonstrates promising potential for enhancing wireless communication systems. Finally, in addressing Open Question 3, we investigate the challenges of deploying GNNs in real-world NG-IoT environments. This includes considerations for model implementation, multi-modal data processing, and socio-economic dimensions of GNN. Therein, we explore various types of costs, including computational, energy, and infrastructure costs, that must be considered to ensure cost-effective and scalable GNN adoption in NG-IoT environments. Therein, we analyze key cost considerations, including deployment, operational, and infrastructure expenses, as well as the sustainability challenges associated with deploying GNNs in NG-IoT environments.}

    \subsection*{Open Question 1: What Are Graphs and Graph Neural Networks?}
    \subsubsection{Background:}
    \textbf{Graph Definition:}
    \begin{figure}[t]
        \includegraphics[trim=0cm 0cm 0cm 0cm, clip=true, width=0.5\textwidth]{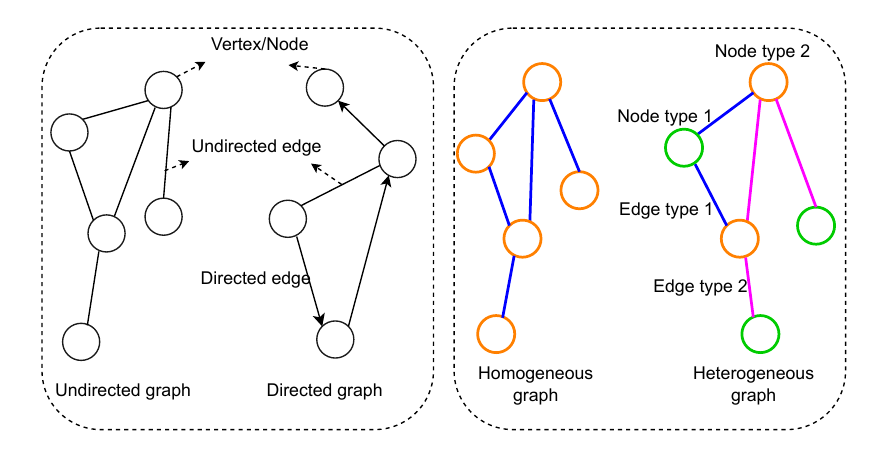}
        \caption{The left figure separates the family of graphs into undirected and directed types. The right figure classifies the graph into homogeneous and heterogeneous types.} 
        \label{fig: GraphDataTypes}
    \end{figure}
    \begin{figure}[t]
        \includegraphics[trim=0cm 0.5cm 0cm 0.3cm, clip=true, width=0.5\textwidth]{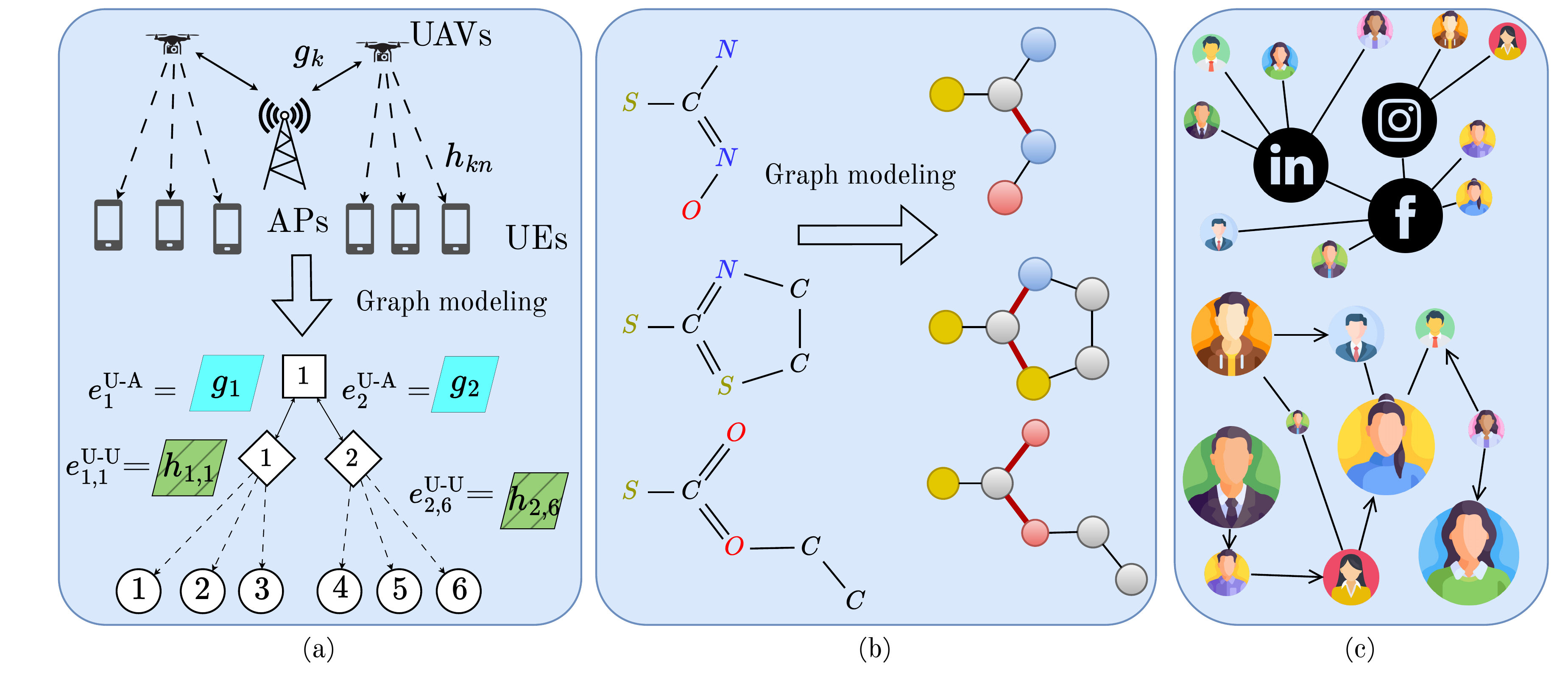}
        \caption{\textcolor{black}{Examples of graphical data representations from various fields. (a) Communication system. (b) Molecules. (c) Social network.}}
        \label{fig: WirelessGraphExample}
    \end{figure}
    A graph is a mathematical representation consisting of a set of vertices (also called nodes) and edges (also called links) that connect pairs of vertices. In general, a graph is represented by a tuple $\mathcal{G} = (\mathcal{V},\mathcal{E})$. The set of $|\mathcal{V}|$ vertices is denoted by $\mathcal{V} = \{1,...,v,...|\mathcal{V}|\}$, while the set of edges is defined based on the graph types: \textit{Undirected graph}: $\mathcal{E} = \{(i,j) | i,j \in \mathcal{V}, i\neq j\}$, where $(i,j)$ represents an undirected connection between vertices $i$ and $j$; \textit{Directed graph}: $\mathcal{E} = \{(i,j) | i,j \in \mathcal{V}, i\neq j\}$, where $(i,j)$ is the edge directed from vertex $i$ to vertex $j$. Graphs can also be categorized into two types based on the nature of their nodes and edges: \textit{Homogeneous graph}: Consists of a single type of node and a single type of edge; \textit{Heterogeneous graph}: These graphs include multiple types of nodes and edges. The illustrations of different types of graphs are represented in Fig.~\ref{fig: GraphDataTypes}. This diversity better represents complex real-world networks, such as wireless communication systems, where different devices and connections exist. The \textit{adjacency matrix} of the graph is represented as $\mathbf{A} \in \{0;1\}^{|\mathcal{V}| \times |\mathcal{V}|}$, where $\mathbf{A}_{ij}=1$ if ${(i,j)}\in \mathcal{E}$. For an undirected graph, $\mathbf{A}$ is symmetric, while for a directed graph, it may not be. The \textit{node feature matrix} $\mathbf{X}$ is a $|\mathcal{V}| \times F_n$ matrix, where each row corresponds to a vector of features for a node. Similarly, the \textit{edge feature matrix} $\mathbf{E}$ is a $|\mathcal{E}| \times F_e$ matrix, where each row corresponds to features of an edge. \textcolor{black}{By storing information in the nodes and edges, the graph can capture the complexity of real-world networks, as illustrated in Fig.~\ref{fig: WirelessGraphExample}. In Fig.~\ref{fig: WirelessGraphExample}(a), a hierarchical wireless communication network is represented as a graph, where each system entity, including user equipment (UEs), unmanned aerial vehicles (UAVs), and access points (APs), is modeled as a node. Communication links between entities are depicted as edges, with different edge types capturing distinct connection types. For example, the edge between an AP and the $i$-th UAV is denoted as $e^{\textrm{U-A}}{1} = [g_i]$, while the edge between the $i$-th UAV and the $j$-th UE is denoted as $e^{\textrm{U-U}}{ij} = [h_{ij}]$, representing their respective channel gains.
    Fig.~\ref{fig: WirelessGraphExample}(b) illustrates another graph-based example: a chemical structure. Here, atoms such as Carbon (C), Oxygen (O), Nitrogen (N), and Sulfur (S) are represented as nodes, while chemical bonds (single, double, or aromatic) are represented as undirected edges. Fig.~\ref{fig: WirelessGraphExample}(c) shows graph modeling in social networks, where individuals are nodes and their interactions, such as friendships, follows, or message exchanges, form the edges. These edges can be directed or undirected, depending on the nature of the relationship. For instance, a “follow” on platforms like Instagram or Twitter is a directed edge, while a mutual friendship results in an undirected edge. Such graph representations are widely used in applications like recommendation systems, influence propagation, and community detection.}
    
    \begin{figure*}[t]
        \centering
        \includegraphics[trim=0.5cm 0cm 0.6cm 0cm, clip=true, width=7.0in]
        {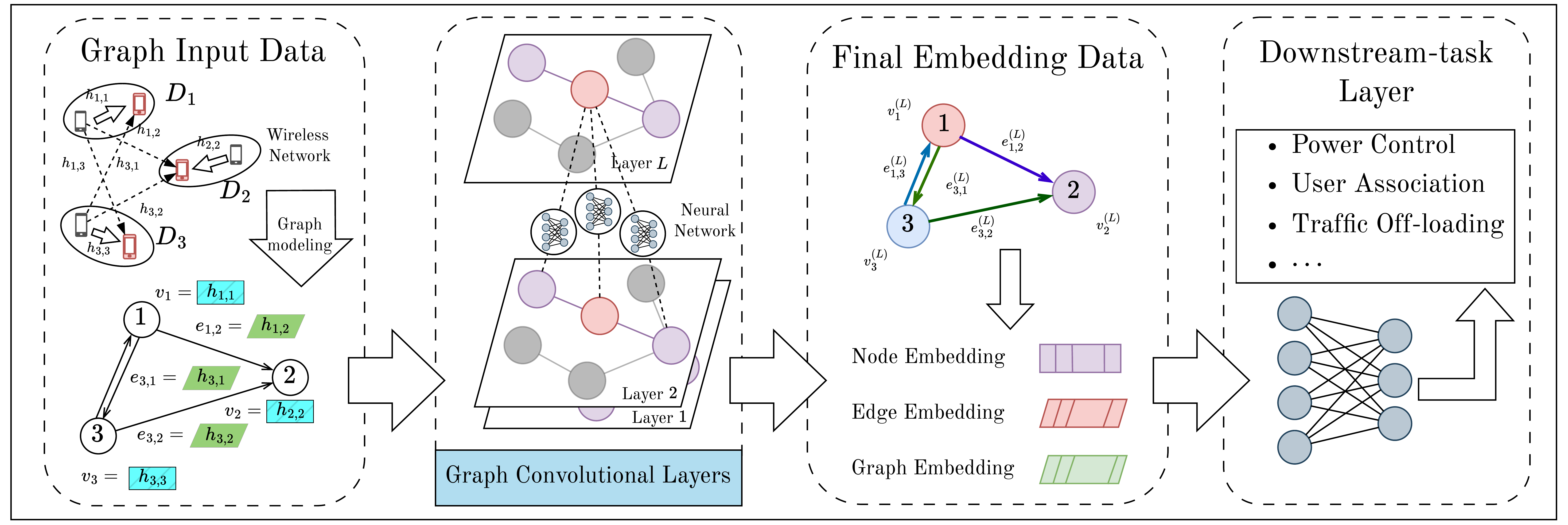}
        \caption{{\color{black}A general pipeline for solving wireless network problems using GNNs. The pipeline illustrates how graph-structured data, such as users, base stations, and links, is processed via graph convolutional layers to generate embeddings that facilitate downstream tasks like user association, resource allocation, and interference management.}}
        \label{fig: General_GNN_pipeline}
        \vspace{-15pt}
    \end{figure*}
    {\color{black}\textbf{Graph Neural Network Paradigms:} 
    \begin{itemize}
        \item \textit{What is GNNs?}: GNN is a specialized neural network designed for processing and analyzing graph data. GNNs are effective at extracting deep-level topological information, unveiling critical and intricate data characteristics, and enabling efficient data processing \cite{Gori2005}. The core idea of GNNs is to learn a mapping function to generate node, edge, or graph representations, known as embedding vectors, based on initial graph information.
        
        \item \textit{How GNNs work?}: {\color{black}As the GNN processes the graph, each node's embedding vector is iteratively updated by aggregating its own features with information gleaned from neighboring nodes and edges, as shown in the graph convolutional layer block of Fig.~\ref{fig: General_GNN_pipeline}. In the context of wireless networks, this framework allows the modeling of complex relationships of users as nodes and their links (e.g., interference, connectivity, or cooperation) as edges. GNNs can learn meaningful representations that capture spatial-, temporal-, and resource-related dependencies, thereby enabling efficient solutions for tasks such as user association, frequency and power allocation, relay selection, and mobility prediction. The final node or edge embeddings are then input to the downstream layer in support of task-specific decisions.} In contrast, MLP treats each node independently without leveraging structural relationships, aggregating information only from fully connected layers, while CNN applies local filters to structured grid-based data, capturing spatial dependencies within a fixed neighborhood but lacking adaptability to irregular graph structures. Ultimately, each node and edge obtains an embedding that captures the broader network context. The GNN model then processes these node and edge representations to produce node-level, edge-level, or graph-level features for tasks such as node clustering, link prediction, or graph classification. Fig.~\ref{fig: General_GNN_pipeline} illustrates the general pipeline of a GNN model, comprising input data, graph convolutional layers, and a downstream task layer.
        \item \textit{Types of GNNs and their functionality}: GNNs can be categorized based on architecture and function, including Graph Convolutional Networks (GCNs) \cite{Thomas2017}, Graph Attention Networks (GATs) \cite{Velickovic2018}, Graph Autoencoders (GAEs) \cite{pan2018adversarially}, and Graph Spatial-Temporal Networks (GSTNs) \cite{song2020spatial}. GCNs aggregate node features via convolution operations, making them scalable for large graphs. GATs introduce attention mechanisms to assign dynamic weights to edges, enhancing the model’s ability to capture complex relationships. GAEs, designed for unsupervised learning, use an encoder-decoder framework for graph reconstruction and anomaly detection \cite{atkinson2021anomaly}. GSTNs integrate spatial and temporal information, making them ideal for time-evolving data like traffic prediction \cite{fang2021spatial, guo2019attention} and dynamic network analysis \cite{lan2022dstagnn}.

        GNNs also integrate with Reinforcement Learning (RL) for decision-making in graph-structured problems \cite{shan2021reinforcement, Munikoti10161704}. In such hybrid models, GNNs extract graph features, which RL algorithms leverage for sequential decision-making. This synergy is particularly effective in applications such as network routing \cite{ALMASAN2022184}, robotics \cite{yao2022graph, lu2021mgrl}, and multi-agent systems \cite{chen2021graph}.
    \end{itemize}}

    {\color{black}\textbf{Privacy-specific to GNNs:}
    The relational learning paradigm of GNNs introduces distinct privacy challenges that are fundamentally different from those in traditional deep learning architectures. In contrast to conventional neural networks that process isolated data points, GNNs inherently expose sensitive information through their graph-structured computations, hence resulting in vulnerabilities. Structural privacy risks arise from the inherent nature of message passing in GNNs, where iterative aggregation may inadvertently expose critical topological features, such as bridge links in social networks, proprietary dependencies in industrial IoT, or cluster memberships in critical infrastructure, potentially revealing sensitive relationships and operational structures \cite{Zhang10693287, DecentralizedPrivacyCriticalConnections2023}. In these scenarios, protecting such high-impact connections is far more important than safeguarding non-essential links, whose disclosure typically has negligible operational impact. Attribute inference further aggravates these risks. The inference of embeddings and/or gradients can be exploited by adversaries for reconstructing sensitive node or edge attributes, posing significant privacy threats. Indeed, recent studies show that node attributes can be reconstructed from gradient updates \cite{sinha2024gradientinversionattackgraph} or final embeddings, while edge properties, such as weights and connection types, may be deduced by observing multiple GNN layers \cite{wang2022grouppropertyinferenceattacks, PrivacyPreservingGraphEmbedding2023}.
    
    To mitigate these risks, advanced privacy-preserving techniques have been proposed. For example, Yuan \textit{et al.} \cite{PersonalizedDPGNN2023} introduced locally private GNN training frameworks by allocating privacy budgets based on node degrees and aggregation perturbation methods to enforce personalized differential privacy (DP) during message passing, reducing leakage while maintaining utility. Furthermore, personalized DP mechanisms allow heterogeneous privacy budgets, enabling fine-grained protection across different nodes and edges. As a further advance, Li \textit{et al.} \cite{PrivacyPreservingGraphEmbedding2023} conceived node-level privacy-preserving embeddings for protecting sensitive attributes, while Sun \textit{et al.} \cite{DecentralizedDPSubgraph2023} developed decentralized DP approaches for the estimation of subgraph statistics, ensuring the protection of critical graph connections without sacrificing accuracy. Incorporating these state-of-the-art DP solutions represents a promising direction for enhancing privacy in GNN-based NG-IoT networks.}
    

    \subsubsection{State-of-the-art:}
    {\color{black}The development of GNNs has resulted in numerous advances across domains such as social networks, wireless networks \cite{Fan9139346}, biology \cite{Li9585532}, and recommendation systems \cite{Wu3535101}. Early contributions, such as \cite{azizian2021expressivepowerinvariantequivariant}, explored the robustness of GNNs in processing graph data independently of node permutations, ensuring that GNNs can operate efficiently across a wide variety of graph structures. A comprehensive taxonomy of GNNs was provided by Zhou \textit{et al.} \cite{ZHOU202057}, where the authors categorized GNN models into paradigms like GCNs, GATs, and GAEs. Further building on this, Wu \textit{et al.} \cite{Wu2023} focused their attention on the application of GNNs in recommendation systems, demonstrating how GNNs enhance recommendation accuracy by leveraging user-item interaction graphs.

    The explainability of GNNs has become a crucial area of research, Yuan \textit{et al.} \cite{Yuan2023} emphasized the importance of making GNNs interpretable, especially in sensitive areas like healthcare and finance. They introduced explainability methods such as GNNExplainer and PGExplainer, which help identify key subgraphs and node features, hence improving the transparency of GNN models. Moreover, Keyulu \textit{et al.} \cite{xu2019powerfulgraphneuralnetworks} examined the limitations of shallow GNNs and proposed deeper architectures for capturing more complex graph structures and node dependencies, contributing to the scalability of GNNs in large datasets. Similarly, Wu \textit{et al.} \cite{III_B_GNN_Comprehensive_Survey_Zonghan_Mar_2021} augmented the understanding of GNNs, especially for dynamic and spatio-temporal graphs, focusing on traffic prediction and network analysis. 

    Despite these advances, challenges such as scaling GNNs to larger datasets, improving interpretability, and enhancing their application in real-time systems remain. Future research is expected to focus on overcoming these obstacles for enabling GNNs to efficiently handle larger datasets and operate in more complex, real-time scenarios.}

    \subsubsection{Challenges and future directions:}
    {\color{black}\begin{itemize}
        \item \textbf{Model Depth and Oversmoothing}: A major challenge in GNNs is the oversmoothing issue, where deeper layers render node representations to become indistinguishable. As highlighted by Keyulu \textit{et al.} \cite{xu2019powerfulgraphneuralnetworks} and Xu \textit{et al.} \cite{Wu2020}, adding too many layers may result in all nodes having similar embeddings, hence reducing the model’s ability to differentiate between them. Solutions like residual connections and multi-scale GNNs are needed to retain expressiveness in deeper models. A possible solution is to incorporate multi-scale attention mechanisms that could allow GNNs to learn both local and global features effectively.
        \item \textbf{Scalability}: GNNs struggle with scalability, particularly when applied to large graphs. The computational cost escalates for larger networks. Zhou \textit{et al.} \cite{ZHOU202057} emphasized the need for techniques like graph sampling, mini-batch training, and graph sparsification to make GNNs more efficient for large-scale applications. To address this, research into advanced graph partitioning algorithms, combined with edge and node-compression techniques, can further reduce the computational burden of large networks. 
        \item \textbf{Security}: Ensuring the robustness of GNNs, particularly in adversarial environments, remains an open problem, where small perturbations to the graph structure or node features can lead to incorrect predictions. Recent research has focused on developing adversarial training methods and robust GNN architectures that can guard against topology attacks. However, the existing defenses are still limited, and further advances are necessary in this area to construct more secure and reliable GNN models \cite{xu2019topologyattackdefensegraph}. Strengthening GNN security requires more robust adversarial training to defend against both structure- and feature-based attacks. Future work should explore quantum-safe encryption for secure GNN communication, especially in critical fields like cybersecurity. Additionally, explainable AI methods can enhance transparency and improve the detection of adversarial threats.
        \item {\color{black}\textbf{Privacy}: While adversarial security has received growing attention, privacy risks specific to GNNs remain underexplored. The message-passing mechanisms of GNNs can inadvertently expose sensitive topological or contextual information, such as bridge links or critical community structures, even in the absence of attacks \cite{Zhang10693287}. Generic anonymization or link-hiding methods may protect non-essential edges but fail to safeguard high-impact relationships. Addressing this challenge requires research into context-aware privacy-preserving methods that specifically target critical connections and sensitive attributes in graph data.}
    \end{itemize}}
    
    \subsection*{Open Question 2: Why GNNs for NG-IoT Networks? How Do We Model NG-IoT Systems as Graphs?}
    \setcounter{subsubsection}{0}
    \begin{figure*}[t]
        \includegraphics[width=\linewidth]
        {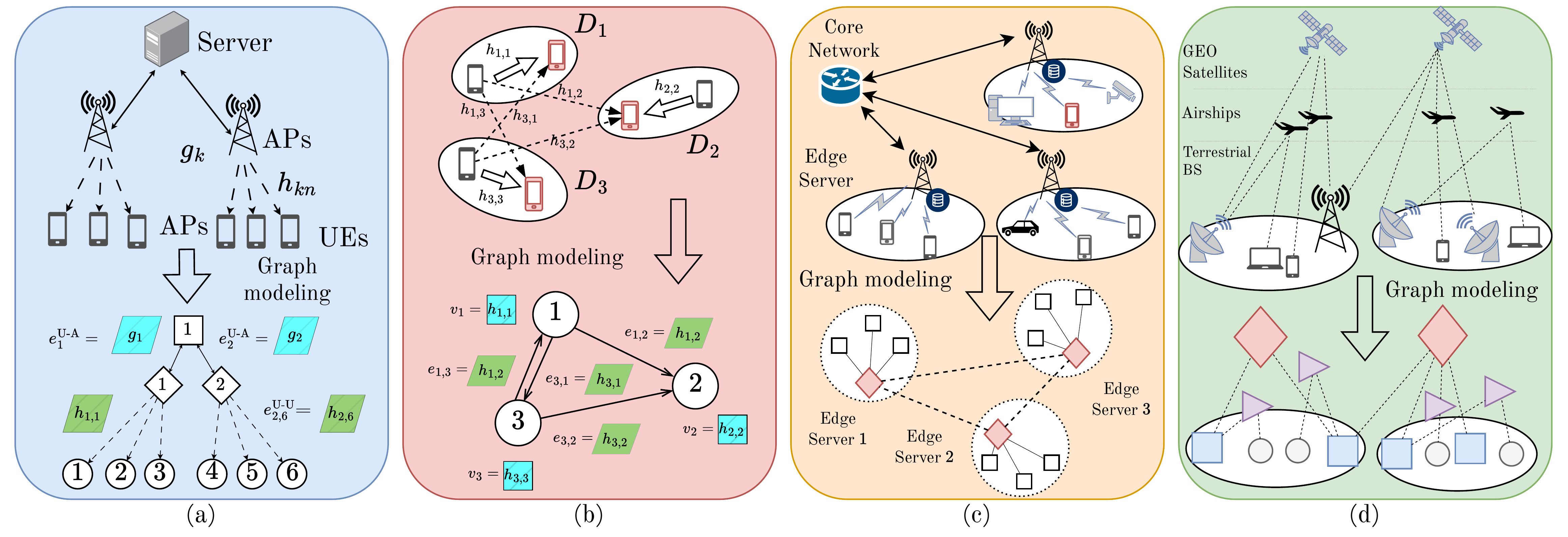}
        \caption{\textcolor{black}{Graph representation of some NG-IoT networks: (a) Centralized cellular network, (b) Decentralized D2D network, (c) MEC-enable network, (d) Integrated satellite and terrestrial network.}}
        \label{fig: GNN_Survey-Q2}
        \vspace{-15pt}
    \end{figure*}
    \subsubsection{Background:}
    {\color{black}\textbf{Why GNNs for NG-IoT networks?}: 
    \begin{itemize}
        \item[\textit{i)}] \textit{Real-World Challenges of NG-IoT Networks:} NG-IoT environments are marked by high complexity and scale, which traditional optimization and deep learning approaches struggle to handle their key characteristics include: 
        \begin{itemize}
            \item {Dynamic topology}: Devices frequently move, join, or leave the network, leading to constantly changing connectivity and traffic patterns.
            \item {Heterogeneous Architecture:} NG-IoT integrates diverse components such as IoT sensors, mobile devices, MEC servers, RIS, base stations, UAVs, and satellites into a unified but multi-layered network.
            \item {Non-Euclidean Data:} Communication signals and traffic flow originate from complex, irregular domains that do not conform to grid-like structures (e.g., sensor graphs or channel states).
            \item {Massive Scale and Connectivity:} NG-IoT supports ultra-dense device deployments, resulting in graphs having large numbers of nodes and edges.
            \item {Interdependent Operations:} Tasks such as routing, resource allocation, and scheduling rely heavily on node-to-node dependencies and on the contextual awareness of the entire system.
        \end{itemize}
        To address the complexities of NG-IoT networks, modeling and optimization solutions must satisfy critical requirements. Models must be capable of capturing both local and global dependencies across heterogeneous network elements to reflect the interactions in the system. Solutions must handle the dynamically evolving topologies of NG-IoT networks. Given the massive scale of IoT deployments and limited computational capability at the edge, scalability and efficiency are non-trivial. The solutions should remain tractable while maintaining performance across large, distributed graphs. Moreover, context-aware processing is needed to integrate real-time information, such as traffic load, latency, and device states, into the decision-making process. Finally, the models must exhibit strong generalization to unseen scenarios, as NG-IoT systems regularly face novel configurations with new devices and unpredictable traffic behaviors.
        
        \item[\textit{ii)}] \textit{NG-IoT as a Dynamic Factor Graph:} The aforementioned characteristics inherently suggest that NG-IoT systems should be modeled as dynamic graphs, where learning must be distributed, adaptive, and structure-aware. To formalize this, we draw inspiration from factor graphs, a structured inference framework used in error correction coding, such as low-density parity-check (LDPC) codes \cite{Bonello6083556, Babar7336474, Jinghu1495850, Hajjar2014}. In LDPC decoding, a bipartite factor graph is constructed with variable nodes representing individual bits of a codeword and factor nodes enforcing parity-check constraints. The decoding process involves iterative message passing among these nodes. Each variable node updates its belief (e.g., log-likelihood ratio) based on incoming messages from neighboring factor nodes, while each factor node enforces local constraints by aggregating the messages it receives and sending back more reliable feedback. This results in a back-and-forth iterative refinement of estimates that converges under well-defined conditions, hence enabling accurate signal recovery even under hostile impairments. This message-passing and constraint-checking mechanism mirrors the interactions in NG-IoT systems, where we have:
        \begin{itemize}
            \item[$\circ$] Variable nodes: Network entities (e.g., devices, satellites, APs) represent states (e.g., power, queue length, channel quality);
            \item[$\circ$] Factor nodes: Network constraints like maximum transmission power or network-wide interference budgets;
            \item[$\circ$] Edges: Logical or physical relationships, for example, data dependencies, interference coupling, or task offloading chains. 
        \end{itemize}
        However, in contrast to LDPC codes, the topology here is time-varying, with new devices joining and existing ones disconnecting dynamically. This dynamic graph structure naturally aligns with the types of graphs processed by GNNs. Moreover, edges may represent heterogeneous relationships, such as line-of-sight channels, virtual task links, or shared computational dependencies, reinforcing the need for multi-relational and adaptive modeling.
        \item[\textit{iii)}] \textit{GNNs as generalized LDPC decoding:}
        The LDPC codes and modern GNNs share a common foundation, message-passing on graphs, but take fundamentally different approaches. The LDPC's fixed message passing:
        \begin{itemize}
            \item[$\circ$] Variable-to-Factor: Each device combines messages from all connected constraints using predefined rules.
            \item[$\circ$] Factor-to-Variable: Constraints compute responses using exact, hand-crafted formulas.
        \end{itemize}
        Meanwhile, GNNs' learned message-passing as
        \begin{itemize}
            \item[$\circ$] Neural message functions: Devices and constraints learn optimal communication strategies through training.
            \item[$\circ$] Dynamically weights inputs based on their importance and network conditions.
        \end{itemize}
        \begin{figure}[t]
            \includegraphics[trim=1.0cm 0.1cm 0.6cm 0cm, clip=true, width=0.5\textwidth]{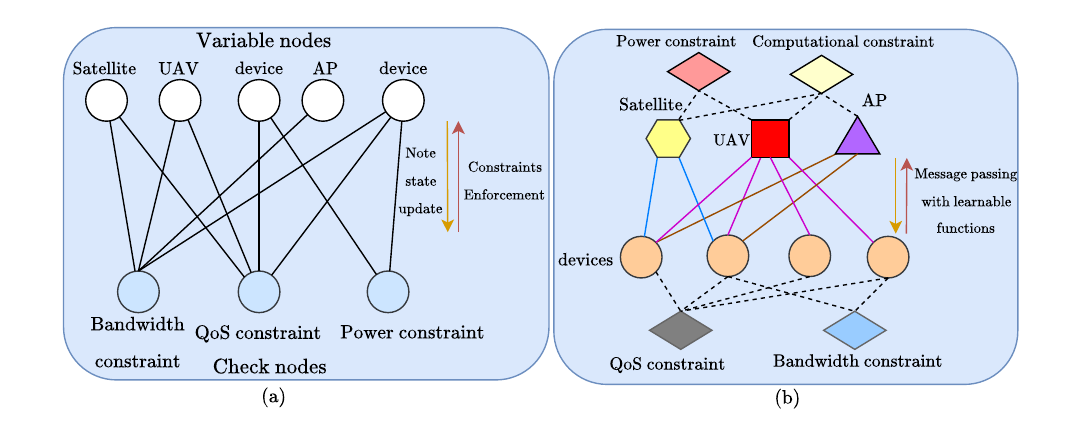}
            \caption{\textcolor{black}{Bridging LDPC decoding and GNN-based learning in NG-IoT networks. (a) LDPC-style factor graph representation of NG-IoT systems with fixed message-passing between device states and network constraints. (b) GNN-based generalization using learnable message passing over a dynamic, heterogeneous graph with implicit constraint handling.}}
            \label{fig: GNN_Survey-LDPC}
        \end{figure}
        While the belief propagation operates on static factor graphs using fixed update rules such as the sum-product algorithm, GNNs extend this paradigm by introducing learnable message-passing mechanisms tailored for dynamic and heterogeneous graphs like those found in NG-IoT networks. The update process is similar in spirit, but improved flexibility and adaptiveness. Fig.~\ref{fig: GNN_Survey-LDPC}(a) illustrates a traditional LDPC-inspired factor graph, where variable nodes (e.g., devices, satellites, UAVs, APs) interact with check nodes (e.g., QoS, power, and bandwidth constraints) through fixed message-passing rules. Fig.~\ref{fig: GNN_Survey-LDPC}(b) presents a GNN-based generalization, where diverse node types exchange learnable messages over a dynamic, multi-relational graph. Constraint enforcement becomes soft as well as adaptive, and message updates are learned rather than predefined. This transition highlights how GNNs extend the structure and harness the convergence principles of LDPC decoding to meet the scalability, heterogeneity, and flexibility demands of next-generation IoT systems. Table~\ref{tab:GNN_LDPC_Duality} presents a structured comparison between traditional LDPC decoding and GNN-based processing in NG-IoT networks. This duality highlights how GNNs generalize the belief propagation framework used in LDPC codes by replacing fixed update rules with learned message functions, enabling dynamic, data-driven processing over heterogeneous and evolving network topologies. The table outlines key analogies across graph types, node and edge semantics, message-passing mechanisms, convergence tools, and application domains, thus bridging well-established coding theory with modern GNN-based learning for IoT.
        \begin{table*}[t]
            \centering
            \caption{GNN-LDPC Duality in NG-IoT Systems}
            \label{tab:GNN_LDPC_Duality}
            \begin{tabular}{|p{0.15\textwidth}|p{0.35\textwidth}|p{0.42\textwidth}|}
            \hline
            \textbf{Aspect} & \textbf{LDPC Decoding} & \textbf{GNNs in NG-IoT Networks} \\
            \hline
            \textbf{Graph Type} & Static bipartite factor graph (variables $\leftrightarrow$ parity checks) & Dynamic multigraph (devices, APs, servers, constraints, etc.) \\
            \hline
            \textbf{Variable Nodes} & Bits (e.g., codeword elements $x_i$) & Device states (e.g., power, location, queue length, CSI) \\
            \hline
            \textbf{check nodes} & Parity-check constraints & Physical/logical constraints (e.g., interference limits, delay, energy budgets) \\
            \hline
            \textbf{Edges} & Binary connections between variables and checks & Typed edges for wireless links, task dependencies, or shared resource constraints \\
            \hline
            \textbf{Message Type} & Log-likelihood ratios (LLRs) & Learnable embeddings  \\
            \hline
            \textbf{Message Passing} & Sum-product algorithm (fixed) & Trainable functions  \\
            \hline
            \textbf{Learning} & No learning (hand-crafted rules) & Data-driven learning via backpropagation \\
            \hline
            \textbf{Convergence Tool} & Density evolution (analytical) & Neural density evolution \cite{cordonnier2025convergencemessagepassinggraph}; convergence under Lipschitz conditions \\
            \hline
            \textbf{Information Flow} & EXIT charts for decoding visualization & EXIT-inspired metrics to diagnose layer-wise training dynamics \\
            \hline
            \textbf{Applications} & Error correction (bit decoding) & Power control, scheduling, routing, anomaly detection in NG-IoT \\
            \hline
            \end{tabular}
        \end{table*}

        \item[\textit{iv)}] \textit{Convergence analysis via density evolution and EXIT charts for learning analysis:} 
        \begin{itemize}
            \item[$\circ$] \textit{Density evolution (DE) \cite{Bonello6083556}}: LDPC's DE tracks message distributions across iterations to predict decoding thresholds. For GNNs, Matthieu \textit{et al.} \cite{cordonnier2025convergencemessagepassinggraph} has shown that, under mild assumptions (e.g., Lipschitz continuity), GNN message-passing admits a state-space evolution analogous to DE. This shows that GNN updates can be viewed as stochastic operators acting over a structured graph domain, mirroring DE in LDPC systems. Therefore, the iterative message-passing process converges to a fixed-point embedding. For instance, DE can help evaluate the stability and performance of GNN-based algorithms in resource allocation or routing under dynamic conditions. 
            \item[$\circ$] \textit{EXIT (Extrinsic Information Transfer) charts}: EXIT charts \cite{Hajjar2014} constitute another core concept in LDPC decoding, used to visualize how mutual information between variable and check nodes evolves during iterations. In the GNN context, such visualization is not yet standard, but analogous metrics have been explored. For instance, mutual information tracking between hidden representations across layers has been proposed in \cite{oono2021graphneuralnetworksexponentially}, where Kenta \textit{et al.} showed how deeper GNNs suffer from representation collapse. By adapting EXIT-style visual tools to plot mutual information layer by layer, practitioners can better diagnose issues like over-smoothing or underfitting, and design more robust and interpretable GNNs for NG-IoT applications. Furthermore, they can also provide tangible design guidelines, concerning the best converging activation order of handling the nodes' inputs. They can also be used for mitigating the signaling overhead by identifying the nodes in the immediate neighborhood that are capable of expediting convergence, while neglecting those providing marginal improvements. 
            
            \item[$\circ$] \textit{Bridging insight}: Although DE and EXIT charts originated from coding theory, they provide a solid analytical foundation for understanding the GNN behavior. GNNs can be viewed as learnable generalizations of belief propagation, where instead of fixed message-update rules, the aggregation and update functions are parameterized by neural networks. This flexibility introduces challenges in theoretical analysis but also allows tools from LDPC research to be reinterpreted to provide convergence guarantees and learning diagnostics for GNN-based solutions. To elaborate a little further from an LDPC perspective code having short cycles in their factor-graph invariably perform poorly. Hence further research is required for eliminating short cycles in the GNNs. By the same token, LDPC codes having rates between 1/3 and 2/3 tend to have the best performance. Hence the factor-graph-based design of GNNs has a high promise for future research.
        \end{itemize}

    \end{itemize}
    Given the aforementioned characteristics and analytical findings, GNNs offer a naturally compatible and theoretically grounded framework for NG-IoT networks:
    \begin{itemize}
        \item \textit{Structure-aware modeling:} GNNs operate directly on graph-structured data, allowing them to represent complex relationships among heterogeneous network entities (e.g., devices, APs, UAVs, satellites).
        
        \item \textit{Dynamic message passing:} Unlike static models, GNNs use learned message-passing mechanisms that adapt to evolving network topologies and conditions, enabling real-time, context-aware, and distributed decision-making.
        
        \item \textit{Theoretical grounding:} By generalizing belief propagation from LDPC codes, GNNs can leverage insights from density evolution and EXIT charts to analyze convergence behavior and guide the design of stable, efficient architectures.
        
        \item \textit{Scalability and generalization:} GNNs naturally generalize across varying graph sizes and structures, making them robust to unseen network conditions. They remain scalable through techniques such as graph sampling, partitioning, and hierarchical modeling.
        
        \item \textit{Training diagnostics and interpretability:} EXIT charts provides mutual information tracking that helps identify and mitigate issues like over-smoothing and representation collapse in deep GNNs, improving learning efficiency and interpretability.

    \end{itemize}}


    {\color{black}\textbf{Graph representation of NG-IoT networks}:
    
    To address NG-IoT network challenges effectively, they must first be formulated as graph-based problems. To do so, this involves key steps: 
    \begin{enumerate}
        \item \textit{Identifying graph components}: Define key elements for the NG-IoT network and map them to graph components.
        \begin{itemize}
            \item Nodes: User equipment, base stations, access points, UAVs, satellites, MEC servers, RIS, etc.
            \item Edges: Communication links, interference channel, data dependencies.
            \item Graph types: Directed vs. undirected, homogenous vs. heterogeneous graphs.
        \end{itemize}
        \item \textit{Mapping network data to graph information}: Associating network data with the corresponding graph elements.
        \begin{itemize}
            \item Node features: Device type, mobility, energy consumption, transmission power, individual data. 
            \item Edge features: Link quality, latency, interference level, channel gain.
            \item Global network characteristics such as connectivity density or overall traffic load.
        \end{itemize}
        \item \textit{Graph construction based on network characteristics}
        \begin{itemize}
            \item Centralized Networks (e.g., Cellular Networks)
            \begin{itemize}
                \item Description: Rely on centralized infrastructure like base stations for communication.
                \item Key Features: Wide coverage, high reliability, scalable.
                \item Graph Representation:
                \begin{itemize}
                    \item[i)] Vertices: Base stations, IoT devices, and core network elements.
                    \item[ii)] Edges: Links between devices and base stations and base stations to the core.
                \end{itemize}
                \item Example: Fig.~\ref{fig: GNN_Survey-Q2}(a) depicts a typical cellular system where UEs connect to APs, and APs forward data to a central server. This forms a heterogeneous graph with nodes representing UEs, APs, and the server, and edges modeling wireless and wired communication links. Such structure supports centralized resource control and coordination.
            \end{itemize}
        
            \item Decentralized Networks (e.g., D2D, Mesh/Ad-hoc Networks)
            \begin{itemize}
                \item Description: Enable direct device communication without centralized infrastructure.
                \item Key Features: Low latency, energy-efficient, resilient.
                \item Graph Representation:
                \begin{itemize}
                    \item[i)] Vertices: IoT devices.
                    \item[ii)] Edges: Direct, dynamic links between devices.
                \end{itemize}
                \item Example: Fig.~\ref{fig: GNN_Survey-Q2}(b) shows devices communicating directly without central infrastructure. Each device is a node, and edges represent dynamic, peer-to-peer links. This homogeneous graph structure captures local interactions in energy-efficient and latency-sensitive applications.
            \end{itemize}
        
            \item Edge-Enabled Networks (e.g., MEC-enabled Networks)
            \begin{itemize}
                \item Description: Use edge computing for localized data processing.
                \item Key Features: Low latency, real-time analytics, AI/ML integration.
                \item Graph Representation:
                \begin{itemize}
                    \item[i)] Vertices: IoT devices, edge servers, cloud data centers.
                    \item[ii)] Edges: Links between devices and edge servers, and edge servers to the cloud.
                \end{itemize}
                \item Example: Fig.~\ref{fig: GNN_Survey-Q2}(c) showcases a network architecture enhanced with mobile edge computing, where edge servers serve as intermediaries between end devices and the cloud. The graph includes nodes for IoT devices, edge servers, and cloud infrastructure. Edges connect devices to their respective edge servers and edge servers to the centralized cloud. 
            \end{itemize}
        
            \item Hybrid Networks (e.g., Integrated Terrestrial-Satellite Networks)
            \begin{itemize}
                \item Description: Combine terrestrial and satellite communication for global coverage.
                \item Key Features: Ubiquitous connectivity, high reliability.
                \item Graph Representation:
                \begin{itemize}
                    \item[i)] Vertices: IoT devices, base stations, satellites, ground stations.
                    \item[ii)] Edges: Links between terrestrial and satellite components.
                \end{itemize}
                \item Example: Fig.~\ref{fig: GNN_Survey-Q2}(d) illustrates an integrated terrestrial infrastructure (e.g., base stations and IoT devices) with aerial and satellite elements (e.g., UAVs, GEO satellites). The graph representation includes heterogeneous nodes, such as, IoT terminals, terrestrial base stations, UAVs, and satellites, with edges capturing both ground and space communication links. 
            \end{itemize}
        \end{itemize}
        \item \textit{Problem formulation using graph structures (Incorporating code, edge, and graph-level tasks)}
        \begin{itemize}
            \item {Node level:}
            Node-level tasks are particularly advantageous in scenarios where the primary focus is on optimizing and configuring individual nodes and their features within a wireless network. These tasks involve associating each variable with a node entity in the graph, making them well-suited for problems that require attention to the configuration and performance of specific network components, such as user equipment, access points, and base stations \cite{III_B_6G_Survey_Mostafa_Jul_2020, III_B_6G_Survey_Wei_Feb_2021, III_B_Node_Wang_Z_2023}. Common node-level tasks include node classification, node clustering, and node regression, where each node’s unique characteristics play a vital role in the network’s overall functionality. 
            
            \item {Edge level:} Formulating wireless network problems as edge-level tasks within GNNs allows models to focus on capturing the interactions within connected nodes, which is essential for wireless communication that relies on device-to-device connections. Particularly in wireless networks, the quality of communication links is affected by factors such as distance, interference, and environmental conditions. By representing these links as edges in a graph, GNNs can capture the intricate relationships and dependencies among these variables, enabling more accurate predictions of link quality and improved resource management strategies. This approach is well-suited for tasks like link prediction, interference management, and resource scheduling, where the performance of individual links is the primary concern.
    
            \item {Graph level:} Graph-level tasks involve obtaining a global representation of the entire graph, which can capture comprehensive information that node-level and edge-level tasks might miss. This global perspective is vital for tasks that demand a holistic understanding of the network's structure and behavior. By summarizing the representations of all nodes and edges, graph-level embedding vectors enable the model to leverage global hidden features beyond the scope of node-level and edge-level approaches. 
            
        \end{itemize}
    \end{enumerate}}


    \subsubsection{State-of-the-art:}\label{subsubsection: Ques2-existing literatures-simulation result}
    {\color{black}Table \ref{tab:III_B} provides a summary of existing applications of GNNs on various IoT network scenarios with three types of tasks in the graph, highlighting their effectiveness in solving different types of wireless network challenges.}
    
    \begin{table*}[t]
    \def\arraystretch{1.5}
    \centering
    \caption{Existing application of GNNs on various wireless networks 
    \label{tab:III_B}categorized by types of tasks}
    \begin{tabular}{| p{0.08\textwidth}| p{0.2\textwidth}| p{0.2\textwidth} | 
    p{0.2\textwidth} |p{0.2\textwidth} |}
    \hline
    \textbf{Task Level} 	& \textbf{Reference}  & \textbf{Network Architecture} & \textbf{Considered Problem}   & \textbf{Graph Data}	      \\ \hline
    \multirow{11}{*}{Node-level} 
				& D. Abode \textit{et al.} \cite{III_B_Node_Abode_D_2023} (2023)   & Industrial Wireless Subnetworks (IWS) & Power control	&Undirected + Homogeneous  \\ \cline{2-5}
				& B. Li \textit{et al.} \cite{III_B_Node_Li_B_2024} (2024)   &Cellfree-massive MIMO & Power control & Directed + Heterogeneous     \\ \cline{2-5}
				& L. Giang \textit{et al.} \cite{III_B_Node_Giang_L_2024} (2024)   &Wireless Sensor Networks & Power control & Undirected + Heterogeneous     \\ \cline{2-5}
				& M. Lee \textit{et al.} \cite{III_B_Node_Lee_M_2021} (2021)   &D2D & Link scheduling	& Directed + Homogeneous    \\ \cline{2-5}
                    & T. Chen \textit{et al.} \cite{III_B_Node_Chen_T_2022} (2022)   &D2D & Link scheduling	&Directed + Homogeneous  \\ \cline{2-5} 
				& Z. Wang \textit{et al.} \cite{III_B_Node_Wang_Z_2023} (2023)   &RIS & Power allocation + RIS phase-shift	&Undirected + Heterogeneous   \\ \cline{2-5}  
				& S. Lyu \textit{et al.} \cite{III_B_Node_Lyu_S_2024} (2024)   &RIS & Beamforming + RIS phase-shift	&Undirected + Homogeneous   \\ \cline{2-5}
				& N. Zhao \textit{et al.} \cite{III_B_Node_Zhao_N_2022} (2022)   &Cellular Network & Network traffic prediction	&Undirected + Homogeneous   \\ \cline{2-5}
				& J. Xiao \textit{et al.} \cite{III_B_Node_Xiao_J_2023} (2023)   & Security Network & Anomaly detection	&Undirected + Homogeneous  \\ \cline{2-5}
				& T. Huoh \textit{et al.} \cite{III_B_Node_Huoh_T_2023} (2023)   &Encrypted Network & Network traffic classification 	&Directed + Homogeneous   \\ \cline{1-5}
	\multirow{6}{*}{Edge-level} 
                & X. Liu \textit{et al.} \cite{III_B_Edge_Liu_X_2021} (2022)   & Massive URLLC & User association 	& Undirected + Heterogeneous (Bipartite graph) \\  \cline{2-5}
				& Z. Sun \textit{et al.} \cite{III_B_Edge_Sun_Z_2023} (2023) & Multi-access Edge Computing (MEC) & Computation offloading & Directed + Heterogeneous   \\ \cline{2-5}
				& Y. Peng \textit{et al.} \cite{RA_Vertex_vs_Edge_Peng_Jan_2024} (2024) & D2D/ MIMO & Power allocation 	& Undirected + Heterogeneous   \\ \cline{2-5}
				& Y. Wang \textit{et al.} \cite{III_B_Edge_Wang_Y_2023} (2023)   &Downlink cellular network & Power allocation 	&Undirected + Heterogeneous   \\ \cline{2-5}
                & S. Liu \textit{et al.} \cite{III_B_Edge_Liu_S_2023} (2023)   & MISO & Precoding design	&Undirected + Heterogeneous\\ \hline
                    
	\multirow{5}{*}{\parbox{0.1\textwidth}{Sub-Graph/ Graph-level}} 
                    & Y. Yang \textit{et al.} \cite{III_B_Graph_Yang_Y_2023} (2023) & Wireless Communication & Network deployment & Undirected + Homogeneous \\ \cline{2-5}
                    & Z. Wu \textit{et al.} \cite{III_B_Graph_Wu_Z_2022} (2022) & Indoor Localization System & Indoor localization & Undirected + Homogeneous \\ \cline{2-5}
                    & G. Wang \textit{et al.} \cite{III_B_Graph_Wang_G_2023} (2023) & MEC & Task offloading & Undirected + Homogeneous \\ \cline{2-5}
                    & A. Asheralieva \textit{et al.} \cite{III_B_Graph_Asheralieva_A_2023} (2023) & MEC & Malicious edge server detection & Directed + Homogeneous \\ \hline
    \end{tabular}
    \end{table*}
     
    \begin{itemize}
        {\color{black}\item \textbf{Node level:}
        Node-level GNN applications are widely used in wireless networks, especially in power allocation \cite{III_B_Node_Abode_D_2023, III_B_Node_Li_B_2024, III_B_Node_Giang_L_2024, III_B_Node_Lyu_S_2024}. For instance, in \cite{RA_Wireless_Communications_Theory_to_Practice_Yifei_May_2023}, the uplink power allocation of a cell-free massive MIMO IoT System can be formulated as a node-level task of an undirected graph, where users and access points (APs) are represented as nodes. This approach allows the model to generalize effectively across different network configurations, ensuring scalability. However, in downlink power allocation, the problem becomes more complex because each AP must allocate power vectors that are dependent on the number of served users, making the task more sensitive to changes in network size. To overcome this challenge, Shen \textit{et al.} \cite{RA_Scalable_Radio_Resource_Management_Yifei_Jan_2021} proposes to represent communication links as nodes, as illustrated in Fig. \ref{fig: LinkAsNode_Example}. This representation shifts the focus to the links, allowing the system to maintain scalability even as the network size varies. This method ensures that the power allocation process remains efficient and adaptable in dynamic network environments, where the number of connected devices may fluctuate. Furthermore, by representing each communication link as a node, the node-level tasks lend themselves to solving link scheduling problems, such as AP-IoT device association or frequency assignment problems. In this way, decisions regarding transceiver pair scheduling can be made independently based on the node’s features. This method enables efficient link scheduling, as demonstrated in studies like \cite{III_B_Node_Lee_M_2021, III_B_Node_Chen_T_2022}. 
        \begin{figure}[t]
            \resizebox{\columnwidth}{!}{\includegraphics[trim=0cm 0cm 0cm 0cm, clip=true, width = \columnwidth]{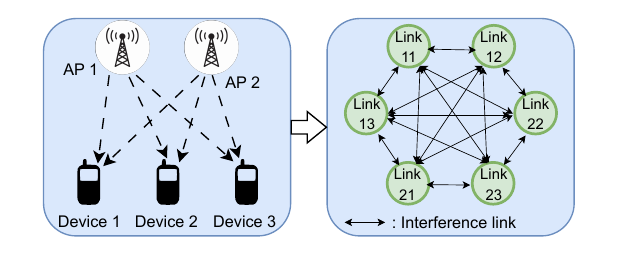}}
            \caption{An example of a wireless network graph, where each communication link is considered as a node.}
            \label{fig: LinkAsNode_Example}
            \vspace{-15pt}
        \end{figure}
        
        Node-level tasks also effectively predict user behavior by analyzing node attributes that capture activity and communication patterns. In \cite{III_B_Node_Zhao_N_2022}, a cellular traffic network is modeled as an undirected graph with mobile traffic data as node attributes. This approach enables GNNs to forecast traffic fluctuations accurately, enhancing network responsiveness and efficiency. Moreover, node-level tasks are beneficial in classification problems within wireless networks. In \cite{III_B_Node_Xiao_J_2023}, a GNN was applied to a binary classification problem to detect anomalies and insider threats. In this context, nodes represent activity log positions, while edges represent the contextual relationships between activities. This node-centric approach allowed the GNN to identify abnormal patterns effectively. Another example, Huoh \textit{et al.} \cite{III_B_Node_Huoh_T_2023} addressed encrypted network traffic classification by mapping each packet to a node and using packet relationships and meta-features as graph inputs. The authors represented the classification problem by both a node-level task as well as an edge-level task and revealed that the classification accuracy was higher when using node-level tasks than edge-level tasks. This indicates the effectiveness of node-level representations in certain wireless network applications, where capturing and leveraging node-specific information is crucial for achieving superior performance.
        
        \item \textbf{Edge level:} Edge-level GNNs have proven effective in addressing various wireless network challenges, especially in optimizing user association, which is crucial for enhancing system performance in wireless networks \cite{Liu7378276}. Liu \textit{et al.} \cite{III_B_Edge_Liu_X_2021} utilized the edge-GNN concept to predict the user-BS association. Similarly, in \cite{III_B_Edge_Sun_Z_2023}, the edge-level concept was used to represent the task offloading problem between wireless devices and mobile edge computing. In terms of resource allocation, the edge-level task formulation enhances interference management by directly modeling interactions between interfering links. In dense wireless networks, where devices compete for resources, this approach allows GNNs to predict and mitigate interference more accurately. By focusing on the edges, GNNs can optimize scheduling and power control, hence reducing interference and improving network performance \cite{RA_Vertex_vs_Edge_Peng_Jan_2024, III_B_Edge_Wang_Y_2023}. Peng \textit{et al.} \cite{RA_Vertex_vs_Edge_Peng_Jan_2024} revealed that edge-GNN could perform just as well as the node-GNNs in power allocation, with the added benefit of reduced training time. Another advantage of edge-level tasks in GNNs is their superior scalability in large and dynamic wireless networks. By focusing on the edges, GNNs can efficiently manage changes in network topology, including the addition or removal of links, without being constrained by the number of nodes. By contrast, node-level tasks may struggle to attain scalability, particularly in scenarios like downlink power allocation. Here, the output dimension at an AP node is typically fixed based on the number of users, limiting flexibility as the network grows. By associating the power allocation variable with the edge, as demonstrated in \cite{III_B_Edge_Wang_Y_2023, III_B_Edge_Liu_S_2023}, edge-level GNNs can more effectively adapt to varying network sizes, ensuring scalability.
        
        \item \textbf{Graph level:} Graph-level tasks have proven to be highly effective in optimizing large-scale wireless networks by providing a holistic understanding of the entire network structure. For instance, a novel GNN-based approach is proposed by Yang \textit{et al.} \cite{III_B_Graph_Yang_Y_2023} optimizing the deployment of network nodes to enhance the overall network throughput, treating the entire network as a unified entity. The authors model the network throughput as the maximum flow of the network and employ a GNN for learning the relationship between node deployment and network flow. Their simulations demonstrate that addressing wireless policies at a graph level significantly outperforms simpler node-level regression tasks, underscoring the importance of a global understanding of the network. Another example is found in \cite{III_B_Graph_Wu_Z_2022}, where a GNN-based federated learning framework is proposed for indoor fingerprint localization. The problem is modeled at two levels: the client level and the server level. For clients, each received signal strength sample is treated as a graph, and a GNN is utilized for predicting locations through a graph-level regression task. 
        
        Edge computing substantially benefits from graph-level tasks due to the need for a holistic view of the interconnected network, especially in task offloading. Wang \textit{et al.} \cite{III_B_Graph_Wang_G_2023} minimized the average offloading delay by using a Branch \& Bound (B\&B) algorithm, representing the process as an enumeration tree of edges and nodes. Briefly, the GNN processes the input state and action pair to derive the MEC system's reward through the final graph-level embedding, which is then used for optimizing the offloading strategy via the B\&B method. Similarly, Asheralieva \textit{et al.} \cite{III_B_Graph_Asheralieva_A_2023} model MEC networks relying on multiple edge servers as a directed multigraph, where the GNN produces a graph-level embedding used for managing security and efficiency. This approach illustrates how graph-level tasks can effectively handle complex problems that require considering the entire network, leading to accurate and robust solutions.}
        
        \item \textbf{Simulations:}
        
        \textit{Model architecture}: {\color{black}We consider the uplink power allocation for a cell-free massive MIMO system, where $M$ APs jointly serve $K$ users simultaneously. The goal is to allocate uplink transmission power for maximizing the minimum user rate across the system. The uplink data rate for each user is adopted from \cite{CellFree_vs_SmallCell_hien2017}. The global solution may indeed be obtained, but it is computationally expensive, especially for large-scale networks experiencing dynamic user and channel conditions. This makes real-time optimization challenging. Using GNN provides an efficient design alternative. 
        
        As suggested by Peng \textit{et al.} \cite{RA_Vertex_vs_Edge_Peng_Jan_2024}, the cell-free massive MIMO system can be represented as a heterogeneous graph. Therein, the associated max-min fairness problem can be formulated either as an edge-level task or a node-level task and then solved using the edge and node convolution-based methods, respectively. To evaluate the effectiveness of different GNN architectures for this task, we harnessed both models from \cite{RA_Vertex_vs_Edge_Peng_Jan_2024} plus proposed some new ones. Below is a summary of the models considered:
        \begin{itemize}
            \item \textbf{Node-GNN}: Updates node features, representing users and APs, to allocate power based on node-level characteristics.
            \item \textbf{Hybrid Quantum GNN (Our proposal)}: The HQGNN uses the GNN to preprocess and generate node embeddings, which are then handled by a deep quantum neural network (DQN) using quantum circuits for power allocation.
            \item \textbf{Edge-GNN}: Updates edge features, representing the connections between users and APs, to allocate power based on edge-level characteristics.
            \item \textbf{Graph-GNN}: Utilizes global graph pooling to aggregate features across all nodes' embeddings and then uses the graph embedding vector to predict the power allocation.
        \end{itemize}}

        \begin{figure}[t]
            \includegraphics[trim=3.8cm 8.5cm 1.5cm 8.5cm, clip=true, width=4.1in]{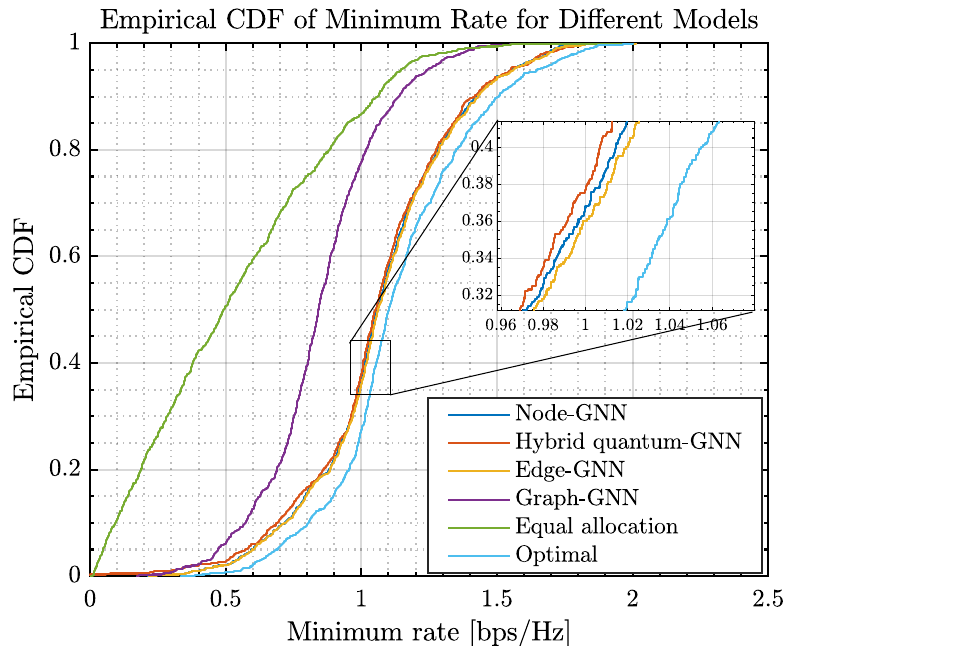}
            \caption{An example of a wireless network graph where each communication link is a node.}
            \label{fig: Max-min-fairness}
        \end{figure}
        \textit{Simulations and results}: {\color{black}In this simulation, we consider a cell-free massive MIMO system, where $6$ users are simultaneously served by $30$ access points. The APs and users are uniformly distributed in a square area of size $D \times D$ km$^2$. The path loss model follows a three-slope model based on \cite{Tang944859}, which accounts for different propagation conditions depending on the AP-UE distance. Only large-scale fading coefficients are used as inputs of the GNNs. We model the system by a bipartite graph associated with two types of nodes, including cellular users and access points. Fig.~\ref{fig: Max-min-fairness} presents the performance of various GNN architectures, illustrated using the cumulative distribution function (CDF) derived from $10,000$ realizations of a cell-free massive MIMO system.} 

        The results show that an edge-GNN and node-GNN constitute a pair of models capable of approximating the optimal solution. Briefly, these two models achieve a minimum rate above $1$bps/Hz in about $65$\% of the instances. The hybrid quantum GNN, which integrates quantum circuits into the Node-GNN architecture, also performs competitively. Quantitatively, this model tends to achieve $1$bps/Hz in about $64$\% of the instances. On average, the Edge-GNN, Node-GNN, and HQGNN models provide data throughputs of $1.0783$, $1.0733$, and $1.0640$ bps/Hz, respectively. By contrast, the optimal solution achieves an average throughput of $1.129$ bps/Hz, slightly outperforming all GNN models, but at the cost of significantly higher computational complexity, particularly as the network size escalates. The graph-GNN performs worst of all GNN models, with less than $40$\% of cases reaching a minimum rate of $1.0$ bps/Hz. This poor performance is attributed to its global aggregation of information, which fails to capture the crucial local node and edge interactions necessary for optimizing power allocation in the network. Overall, the results underscore the superiority of node-level GNN models, particularly of those utilizing attention mechanisms for solving the power control problem of cell-free massive MIMO systems. Furthermore, the hybrid quantum GNN offers an intriguing avenue for integrating quantum computing into GNN architectures.
    \end{itemize}
    
    \subsubsection{Challenges and future directions}
    {\color{black}Again, a significant challenge in modeling GNNs for wireless networks is ensuring scalability, particularly when dealing with networks that have inherent constraints, such as limited node resources or dynamic topology changes. As the graph size grows, the computational cost and memory usage increase, making it challenging to maintain the target performance in real-time. To address this challenge, a promising future direction is to partition the large graph into multiple subgraphs, allowing GNNs to learn locally within each subgraph. The information learned from each subgraph may then be aggregated, ensuring that the model is able to process graphs of varying sizes without losing generalization capabilities. This approach can mitigate scalability issues and maintain performance across different network scales. Moreover, integrating tools from coding theory, such as DE and EXIT charts, offers new opportunities to analyze and improve GNNs in NG-IoT networks. While DE provides convergence guarantees in LDPC decoding by tracking message distributions, its application to GNNs is still developing, particularly in the context of dynamic and heterogeneous graph structures. Similarly, using EXIT charts to monitor mutual information across GNN layers could help identify training bottlenecks like over-smoothing, though more formal metrics and empirical studies are needed. Future research could focus on adapting these tools to analyze GNN convergence, guide model design, and support scalable learning across distributed and partitioned network environments, thereby enabling more interpretable and robust GNN deployment.}

    \subsection*{Open Question 3: How Can GNNs Be Efficiently Implemented and Optimized for Real-World NG-IoT Systems?}
    \setcounter{subsubsection}{0}
    \subsubsection{Background:}
    {\color{black}NG-IoT systems require real-time decision-making, scalability, and energy efficiency. To efficiently deploy GNNs in NG-IoT systems while meeting these requirements, several key aspects must be considered: 
    
    \textbf{Model Optimization for Edge and Cloud Deployment:} Deploying GNN in NG-IoT systems requires careful consideration of computational optimization, energy usage, scalability, and resource-constrainted IoT devices, edge nodes, and cloud platforms:
    \begin{itemize}
        \item \textit{Embedded GNNs for lightweight devices:} Embedded GNNs are designed to operate on specialized hardware platforms with limited computational resources, such as AI chips, Field-Programmable Gate Arrays (FPGAs), and Edge Tensor Processing Units (TPUs). Unlike traditional DL models that rely on large-scale GPU or cloud-based processing, embedded GNNs enable on-device learning and inference, reducing the reliance on cloud services and minimizing latency. 
        \item \textit{Model compression techniques for efficient deployment:} To address computational constraints, several model compression techniques have been introduced to optimize GNN performance without sacrificing accuracy: 1) \textbf{Quantization} reduces numerical precision in model parameters, storing weights in lower-bit representations such as $8$-bit integers instead of $32$-bit floating points. This technique significantly reduces memory footprint and speeds up inference, making GNNs feasible for real-time applications in NG-IoT systems \cite{Wang3614955}; \textbf{Pruning} removes unnecessary connections and redundant parameters while preserving the graph structure. By selectively eliminating edges or nodes with minimal contribution to the model’s performance, pruning reduces computational overhead and enhances efficiency. This approach is particularly useful for energy-efficient deployment on edge devices, as demonstrated in recent work on sparse GNNs \cite{xie2024lightweight}; \textbf{Knowledge distillation} enables a smaller, compact GNN (student model) to learn from a larger, complex GNN (teacher model). This process allows for transferring knowledge from high-capacity models to lightweight versions while retaining predictive performance.
        \item \textit{Hardware acceleration for GNN inference:} Beyond software-level optimizations, specialized AI hardware has been developed to enhance the execution speed and energy efficiency of GNNs. Hardware accelerators such as Graphics Processing Units (GPUs), Tensor Processing Units (TPUs), Application-Specific Integrated Circuits (ASICs), and neuromorphic chips have been explored for GNN inference, offering improvements in parallelism and computation efficiency. GPUs and TPUs, in particular, have demonstrated substantial speedups in GNN message-passing operations, making them ideal for large-scale graph-based learning in NG-IoT networks \cite{Kiningham_1970}. 
    \end{itemize}

    \begin{figure}[t]
        \centering
        \includegraphics[width=\columnwidth]{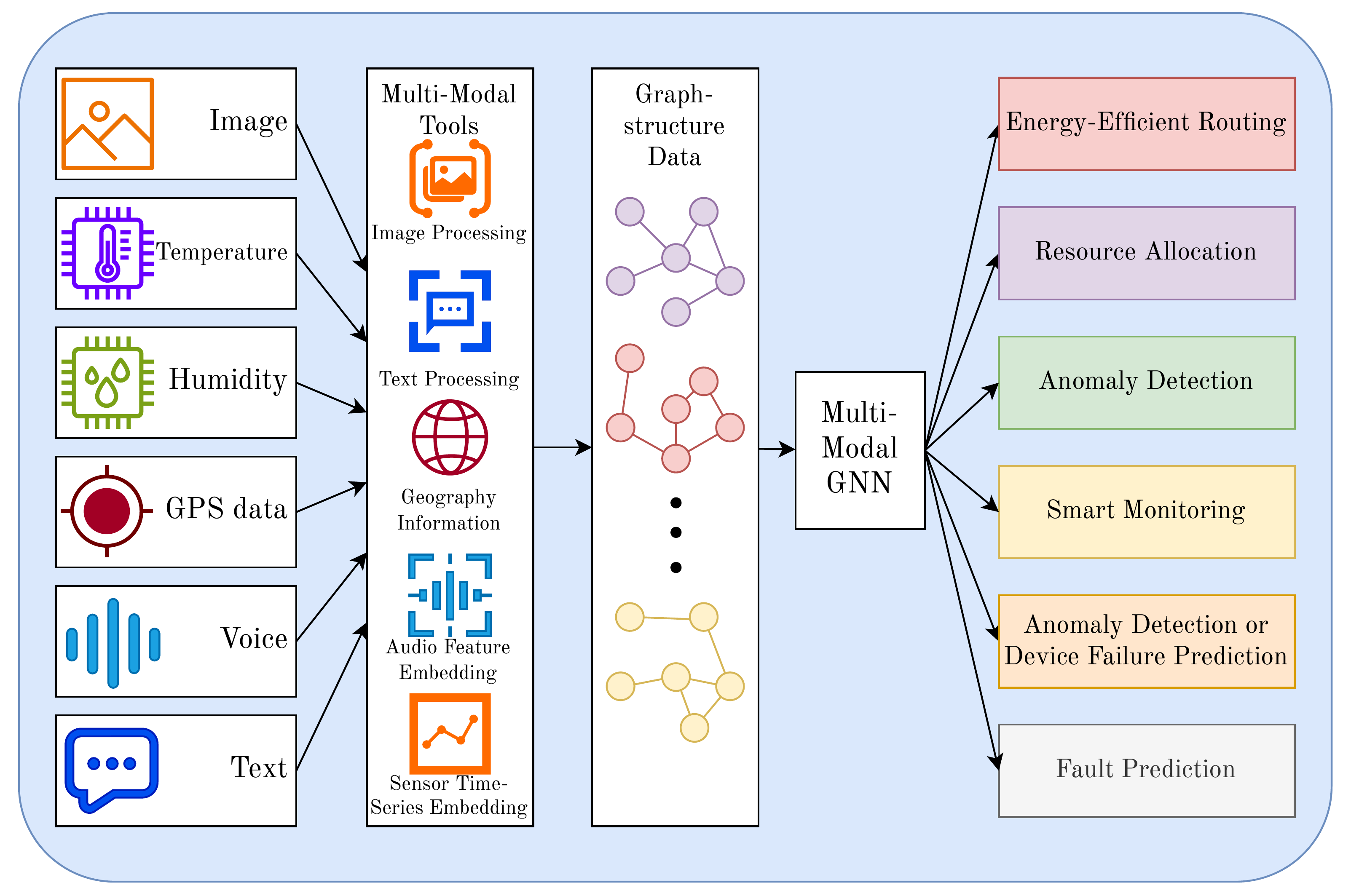}
        \caption{\textcolor{black}{Multi-Modal GNN Framework for NG-IoT Applications.}}
        \label{fig:II_3_MultiModalGNN}
    \end{figure}

    \textbf{Multi-modal data processing in NG-IoT networks:} Data processing plays a vital role in deploying GNNs in NG-IoT networks, which often involve multimodal data sources, such as sensor readings, network logs, and user interactions, images, or text, as illustrated in Fig.~\ref{fig:II_3_MultiModalGNN}. The integration of diverse data modalities presents challenges related to data synchronization and representation \cite{Wang3408317}. Traditional unimodal processing methods often fail to fully utilize the correlation and complementary information between different modalities. In contrast, GNNs can naturally model the relationships between heterogeneous data sources, capturing their interdependencies in complex IoT environments \cite{LI2024119815, Zhang10860124}. Fig.~\ref{fig:II_3_MultiModalGNN} illustrates how heterogeneous data sources, such as images, temperature, humidity, GPS, voice, and text, are processed through dedicated multi-modal tools (e.g., image processing, text analysis, and sensor embeddings) to generate graph-structured data. A multi-modal GNN then learns from these fused graph representations to support various NG-IoT tasks, including energy-efficient routing, resource allocation, anomaly detection, smart monitoring, device failure prediction, and fault detection.
    
    However, ensuring the scalability and efficiency of multimodal GNNs in real-world NG-IoT systems remains an open challenge, requiring further research into data pre-processing, representation learning, and computational optimization techniques.
    
    \textbf{Security, Privacy, and Scalability Challenges:} As NG-IoT applications involve large-scale, interconnected devices handling sensitive data, ensuring the privacy of IoT users and the security of IoT applications is critical. Indeed, IoT networks are inherently vulnerable to adversarial attacks and inference manipulation, hence necessitating robust security mechanisms. To address these concerns, encryption-based training, differential privacy techniques, and secure multi-party computation can enhance security for GNN deployment. Given the importance of these challenges, we provide an in-depth discussion on adversarial attacks and defense mechanisms for GNN-based NG-IoT networks in Section \ref{Section: Adversarial Graph Neutral Network}, analyzing different attack strategies and mitigation techniques.

    \textbf{Socio-Economic Dimensions of NG-IoT Applications:} The deployment of GNNs in NG-IoT networks introduces several socio-economic considerations, including cost, sustainability, and accessibility.
    \begin{itemize}
        \item \textit{Cost consideration:}  Deploying GNNs in NG-IoT networks involves significant costs, including training, deployment, and scalability. The training cost includes computational resources, energy, and hardware costs \cite{Shuvo9985008}. Deploying GNNs involves infrastructure costs, such as purchasing and maintaining edge devices (e.g., NVIDIA Jetson, Raspberry Pi) or cloud-based resources (e.g., AWS, Google Cloud). Operational costs, such as energy consumption and maintenance, also add to the overall expense \cite{Duc3341145}. Meanwhile, scaling GNN-based solutions to large IoT networks requires additional investments in hardware, software, and network infrastructure \cite{Murshed3469029}.
        \item \textit{Sustainability:} Training and deploying GNNs on IoT devices, especially edge devices, can be energy-intensive. For example, training even lightweight GNN models on edge devices can strain battery life and increase operational costs \cite{strubell-etal-2019-energy}. Besides, the energy consumption of GNNs contributes to carbon emissions, particularly when non-renewable energy sources are used \cite{strubell-etal-2019-energy}. Accordingly, this poses sustainable challenges limiting the adoption of GNNs in NG-IoT networks, especially in regions with strict environmental regulations or limited access to renewable energy. Green AI initiatives, such as energy-efficient algorithms and renewable energy-powered data centers, are critical to addressing these challenges
    \end{itemize}}
    \subsubsection{State-of-the-arts}
    {\color{black}\textbf{Model Optimization for Edge and Cloud Deployment:} To streamline the deployment of GNNs in NG-IoT networks, various studies have explored different key aspects. In attempt to embed GNN, Zhang \textit{et al.} \cite{Zhang3686539} have explored how (FPGAs) can accelerate message-passing operations in GNNs, leading to faster inference times while maintaining model accuracy. Besides, the authors also summarizes the challenges that FPGA-based GNNs accelerators need to address. Meanwhile, Wang \textit{et al.} \cite{Wang3614955} proposes a quantization-based approach to address scalability challenges in GNNs. By compressing models with low-bit quantization and controlling message propagation, it reduces computational and memory costs while maintaining accuracy. The method achieves significant speedups ($5.11\times$ for INT2, $4.70\times$ for INT4), offering a cost-effective solution for deploying GNNs in NG-IoT networks. Xie \textit{et al.} \cite{xie2024lightweight} proposed Graph sparsification and network pruning (GASSIP), a novel framework for lightweight graph neural architecture search (GNAS) tailored for resource-constrained scenarios. By combining operation-pruned architecture search and curriculum graph sparsification, GASSIP efficiently identifies optimal lightweight GNNs with reduced model parameters and sparser graphs. Extensive experiments show that GASSIP achieves comparable or superior performance with significantly fewer resources, making it highly relevant for cost-effective GNN deployment in NG-IoT applications. In \cite{wu2024edgefreestructureawareprototypeguidedknowledge}, Wu \textit{et al.}  introduced Prototype-Guided Knowledge Distillation (PGKD), a novel method for distilling high-accuracy GNNs into low-latency MLPs without relying on graph edges. By leveraging class prototypes and structure-aware distillation losses, PGKD effectively transfers graph structural knowledge from GNNs to MLPs, enabling edge-free learning. 
    
    \textbf{Multi-modal data processing in NG-IoT networks:} 
    Li \textit{et al.} \cite{LI2024119815} proposed GNNMR, a multi-modal recommendation framework that integrates GNNs with deep mutual learning to address modality bias. By training separate GNNs on uni-modal user-item graphs and using mutual knowledge distillation, GNNMR synchronizes latent semantic relationships across modalities, improving multi-modal embeddings. It outperforms existing methods in Top-K recommendation tasks, making it suitable for NG-IoT applications requiring efficient user-item interaction modeling. Similarly, Zhang \textit{et al.} \cite{Zhang10860124} introduced a GNN framework for multimodal data integration, leveraging feature, decision, and deep fusion techniques to achieve high accuracy ($98$\%), recall ($86.9$\%), and F1 score ($0.964$). This approach demonstrates robust performance in complex scenarios, highlighting the potential of multimodal GNNs for NG-IoT applications. 

    \textbf{Cost consideration:} Recent studies have explored various strategies to address the socio-economic challenges of deploying GNNs in NG-IoT systems, particularly focusing on reducing deployment and operational costs, enhancing scalability, and improving energy efficiency. Shuvo \textit{et al.} \cite{Shuvo9985008} emphasized the cost-efficiency benefits of deploying deep learning models on edge devices in NG-IoT systems. While cloud-based HPC clusters provide high computational capacity, they incur high transmission costs, latency, and privacy concerns. In contrast, edge computing with devices like IoT sensors and wearables can offer real-time processing at reduced operational costs.  Duc \textit{et al.} \cite{Duc3341145} addressed operational costs in edge-cloud deployments by proposing ML-based resource provisioning frameworks. Their study examined workload characterization, elastic resource management, and component placement to minimize energy consumption and infrastructure usage, which are key challenges in large-scale NG-IoT applications. In \cite{Murshed3469029}, Murshed \textit{et al.} provided a comprehensive overview of deploying ML models at the edge to reduce latency and offloading costs. They reviewed compression techniques, frameworks, and hardware accelerators suited for real-time ML applications in constrained environments, supporting scalable and low-latency NG-IoT deployment. In parallel, Strubell \textit{et al.} \cite{strubell-etal-2019-energy} quantified the environmental costs of training large AI models and advocated for sustainable AI practices. Their findings support the shift toward low-power, energy-efficient GNN deployment strategies in NG-IoT, aligning with green AI initiatives to reduce carbon footprint and promote equitable access to AI technologies.}

    \subsubsection{Challenges and future direction}
    {\color{black}Despite recent advances, several challenges remain in deploying GNNs efficiently within real-world NG-IoT systems. One of the primary issues is the computational and energy constraints of edge devices, such as IoT sensors and mobile nodes. These resource limitations demand advanced model optimization techniques, including pruning, quantization, and knowledge distillation, alongside hardware-software co-design strategies that tailor GNN architectures to run efficiently on FPGAs, TPUs, and neuromorphic chips. Furthermore, as NG-IoT systems scale in size and complexity, ensuring real-time inference and learning becomes increasingly difficult. This necessitates scalable training frameworks, decentralized and federated GNN models, and intelligent graph partitioning and sampling techniques.

    Another significant challenge is the integration and processing of multimodal data (e.g., sensor readings, logs, and user interactions), which requires effective fusion strategies and synchronized graph representations. In addition, from the socio-economic perspective, cost-efficiency and sustainability remain key concerns, which hinder wide-scale adoption. Hence, promoting energy-efficient algorithms, carbon-aware deployment strategies, and global standardization efforts will be essential for the practical, scalable, and equitable integration of GNNs in NG-IoT applications.}

    \section{Applications of GNN for NG Technologies in IoT Networks}\label{GNN4_6G}
    In this section, we address Open Questions 4, 5, and 6, focusing on how GNNs empower core NG technologies Open Question 4 covers resource management and communication efficiency in massive MIMO, RIS, Satellite, MEC, THz, and URLLC systems, with GNNs optimizing user association, beamforming, and signal propagation. By contrast, Open Question 5 investigates how GNNs enhance integrity and security in blockchain. Lastly, Open Question 6 discusses the integration of GNNs with distributed systems, examining their collaboration to enable privacy-preserving, scalable, and efficient learning across distributed NG-IoT networks.
     
    \begin{table*} \scriptsize%
    \def\arraystretch{1.25}
    \caption{Related studies in the use of GNN on different NG technologies for IoT networks}
    \begin{tabular}{|p{0.08\textwidth}|p{0.08\textwidth}|p{0.27\textwidth}|p{0.24\textwidth}|p{0.08\textwidth}|p{0.09\textwidth}|}
    \hline
    NG key technologies & Reference & IoT applications & Problems & Task Level & GNN model \\
    \hline
    \multirow{8}{*}{\makecell[{{p{0.08\textwidth}}}]{Massive MIMO}} & \cite{RA_Wireless_Communications_Theory_to_Practice_Yifei_May_2023} (2023)& Uplink cell-free massive MIMO & Power control for max-min rate & Node & MPGNN\\
    \cline{2-6}
                     & \cite{cfmMIMO_IoT_Li_Jun_2023} (2023)& Multicarrier-division duplex cell-free mMIMO  & Power control for maximizing average rate & Node & GAT\\
    \cline{2-6}
                     & \cite{cf_mMIMO_Salaun_Dec_2022} (2022)& Downlink cell-free massive MIMO  & Power control for max-min rate & Node & HetGNN\\
    \cline{2-6}
                     & \cite{RA_Vertex_vs_Edge_Peng_Jan_2024} (2024)& Downlink MIMO & Power control for max sum rate & Edge, node & HetGNN \\
    \cline{2-6}
                     & \cite{cf_mMIMO_Ranasinghe_Dec_2021} (2023)& Cell-free massive MIMO & AP selection & Node & MPGNN\\
    \cline{2-6}
                     & \cite{mMIMO_IoT_Chien_Jan_2024} (2024)& Integrated satellite-terrestrial cell-free massive mimo IoT systems  & Power control for max-min rate & Node & HetGNN\\
    \cline{2-6}
                     & \cite{Coordinated_Sum_Rate_Deep_Unrolling_Schynol_Apr_2023} (2023) & Massive MIMO  & Power control  & Node & GCN-WMMSE\\
    \hline
    \multirow{7}{*}{RIS}              &      \cite{RIS_Singh_Dec_2023} (2023) & RIS aided communication  & Channel estimation  & Sub-graph & GCN + Transformer model\\
    \cline{2-6}
                     &      \cite{RIS_Zhang_May_2022} (2022) & RIS support multiuser downlink  & RIS phase shift and power allocation for long-term error minimization  & Node & HetGNN\\
    \cline{2-6}
                     &      \cite{RIS_Satellite_IoT_Cao_Feb_2022} (2022) & RIS support Fed procedure  & RIS phase shift and satellite's beamforming optimization  & Node & GAT\\
    \hline
    \multirow{8}{*}{Satellite} & \cite{SAGS_GNN_IoT_Huang_Nov_2023} (2023)& LEO satellite networks & Satellite routing and network traffic optimization & Node & RL + MPGNN \\
    
    \cline{2-6}
                     & \cite{SAGS_GNN_IoT_Chen_Mar_2023} (2023)& Mega-constellations Satellite Networks Model & Satellite routing & Node & GCN + GRU \\
    \cline{2-6}
                     & \cite{SAGS_GNN_IoT_He_Jan_2024} (2024)& LEO satellite networks & The SFC orchestration & Node & RL + GAT \\
    
    \cline{2-6}
                     & \cite{SAGS_GNN_Wang_Apr_2023HotICN} (2023)& LEO satellite constellation & Topology optimization & Node & RL + MPGNN \\
    
    \cline{2-6}
                     & \cite{SAGS_GNN_IoT_Asheralieva_Jul_2023} (2023)& space-air-ground integrated network with MEC & The network slicing allocation & Node & DL + MPGNN \\
    
    \cline{2-6}
                     & \cite{SAGS_GNN_IoT_Tekbyk_Jul_2021} (2021)& UAV-assisted hybrid satellite-terrestrial network & Trajectory design and link selection & Node & GAT \\
    
    \cline{2-6}
                     & \cite{SAGS_GNN_IoT_Chen_Mar_2023} (2023)& RIS-assisted satellite IoT communications & Channel estimation & Node & GAT \\
    \hline
    \multirow{5}{*}{THz}              & \cite{THz_GNN_IoT_Zhang_Sep_2023} (2023)& Digital Twin (DT) network with the THz band & Weighted mean rate maximization problem & Node & MPGNN \\
    \cline{2-6}
                     & \cite{THz_GNN_IoT_Mar_2024} (2024)& RIS-aided multiuser mimo THz system & Sub-band allocation, the phase shift, and the precoding to maximize the system sum rate & Node & HetGNN with self-attention \\
    \cline{2-6}
                     & \cite{THz_GNN_Li_Oct_2023} (2023)& Integrated communication and sensing for vehicle communication & Operation mode selection & Node & HetGNN \\
    \hline
    \multirow{13}{*}{MEC} & \cite{Edge_general_Sun_Oct_2021} (2021)& MEC support IoT networks & Task off-loading & Edge & RL + MPGNN \\
    \cline{2-6}
                     & \cite{Edge_general_Wang_Oct_2023} (2023)& MEC support D2D communication & Task off-loading & Graph & RL + GAT\\
    \cline{2-6}
                     & \cite{Edge(UAV)_general_Li_Jun_2022} (2022)& MEC based UAV & Task off-loading & Node & RL + MPGNN\\
    \cline{2-6}
       & \cite{Edge_Transpo_Xu_Nov_2023} (2023)& MEC support IoV & Task off-loading & Node & Graph weighted convolution network \\
    
    \cline{2-6}
       & \cite{Edge_Transpo_Zhou_Jan_2023} (2023)& MEC support IoV & Task off-loading & Node & STGNN + GRU + Transformer \\
    \cline{2-6}
       & \cite{Edge_Transpo_Liu_Apr_2022} (2022)& MEC support marine-based IoT & Trajectory prediction at Edge computing & Node & STMGCN + Self-attention \\
    \cline{2-6}
                     & \cite{Edge_health_Fei_Jan_2024} (2024)& MEC support healthcare applications  & Classification at Edge computing &  Node & GCN\\
    \cline{2-6}
                     & \cite{Edge_smarthome_Sun_Apr_2023} (2023)& MEC support smart home & Intrusion detection & Edge  & GraphSAGE\\
    \cline{2-6}
                     & \cite{Edge_industrial_Tang_Oct_2023} (2023)& MEC support industrial IoT  & Anomaly detection & Node & SPGNN\\
    \hline
    \multirow{5}{*}{URLLC}              & \cite{URLLC_IoT_GNN_Liu_Oct_2023} (2023)& Cellular network & QoS violation probability minimization & Node & REGNN \\
    \cline{2-6}
                     & \cite{URLLC_IoT_GNN_Jiaqi_Feb_2024} (2024)& OFDMA wireless network for URLLC services & Maximizing number of successful transmissions & Node & GraphSAGE \\
    \cline{2-6}
                     & \cite{URLLC_IoT_GNN_Liu_Dec_2021} (2021)& A factory automation scenario & Packet loss probability minimization & Node & Edge-wise gate update + FNN \\
    \cline{2-6}
                     & \cite{III_B_Graph_Gu_Y_2024} (2024)& Distributed URLLC system & Decoding error probability minimization & Node & GraphSAGE \\
    \hline
    \multirow{4}{*}{Blockchain}              & \cite{Blockchain_IoT_GNN_Cai_Jul_2023} (2023) & A secure smart blockchain IoT network & User privacy and data processing time & Node & MPGNN \\
    \cline{2-6}
                     & \cite{Blockchain_IoT_GNN_Kim_Nov_2020} (2020)& IoT smart blockchain & Application distribution among IoT networks & Node & GraphSAGE \\
    \cline{2-6}
                     & \cite{Blockchain_IoT_GNN_Ziyu_Aug_2020} (2020)& IoT smart blockchain for healthcare applications & Malicious node detection & Node & GCN \\
    \hline
    \multirow{14}{*}{\makecell[{{p{0.08\textwidth}}}]{Distributes systems}}  & \cite{jianping2024federated} (2024)& Federated learning for network attack detection & Attack detection & Node & GAT\\
    \cline{2-6}
                     & \cite{10026810} (2023)& Federated learning  & Anomaly detection & Node & MPGNN\\
    \cline{2-6}
                     & \cite{10433561} 2024 & Federated learning for UAV assisted MEC systems  & Classification inference & Node & Simplified graph convolutional network \\
    \cline{2-6}
                     & \cite{10734080} (2025)& Federated learning for classification tasks & Node classification tasks & Node & GCN \\
    \cline{2-6}
                     & \cite{khanna2025grl} (2025)& Federated learning for edge computing & Node classification tasks & Node & RL + MPGNN\\
    \cline{2-6}
                     & \cite{10855737} 2025 & Decentralized tasks in wireless networks  & Channel impairments in GNNs  & Node & AirGNN\\
    \cline{2-6}
                     & \cite{9919905} 2022 & Decentralized federated learning in V2V system  & Power control  & Node & GCN\\
    \cline{2-6}
                     & \cite{Edge_general_Zeng_Jul_2023} 2024 & Distributed fog servers for smart {IoT} services  & Communication efficiency  & Node & GCN, GAT, GraphSAGE\\
    \cline{2-6}
                     & \cite{10364739} 2024 & Federated learning over interference-limited wireless networks  & Power control  & Node & GCNs\\
    \cline{2-6}
                     & \cite{III_B_Node_Wang_Z_2023} (2023) & RIS support federated learning & Resource allocation & Node & HetGNN\\
    \cline{2-6}
                     & \cite{10437172} (2023) & Federated learning over wireless network  & Power control  & Node & GCN\\
    \cline{2-6}
                     & \cite{III_B_Graph_Gu_Y_2024} (2024) & Distributed URLLC system  & Beamforming and power control & Node & GraphSAGE\\
    \hline
    \end{tabular}
    \end{table*}

    \subsection*{Open Question 4: How Do GNNs Enhance Communication and Computation Efficiency in Massive MIMO, RIS, Satellite, THz, MEC, and URLLC Systems?}
    \setcounter{subsubsection}{0}
    \begin{figure*}[t]
        \includegraphics[width=\textwidth]
        {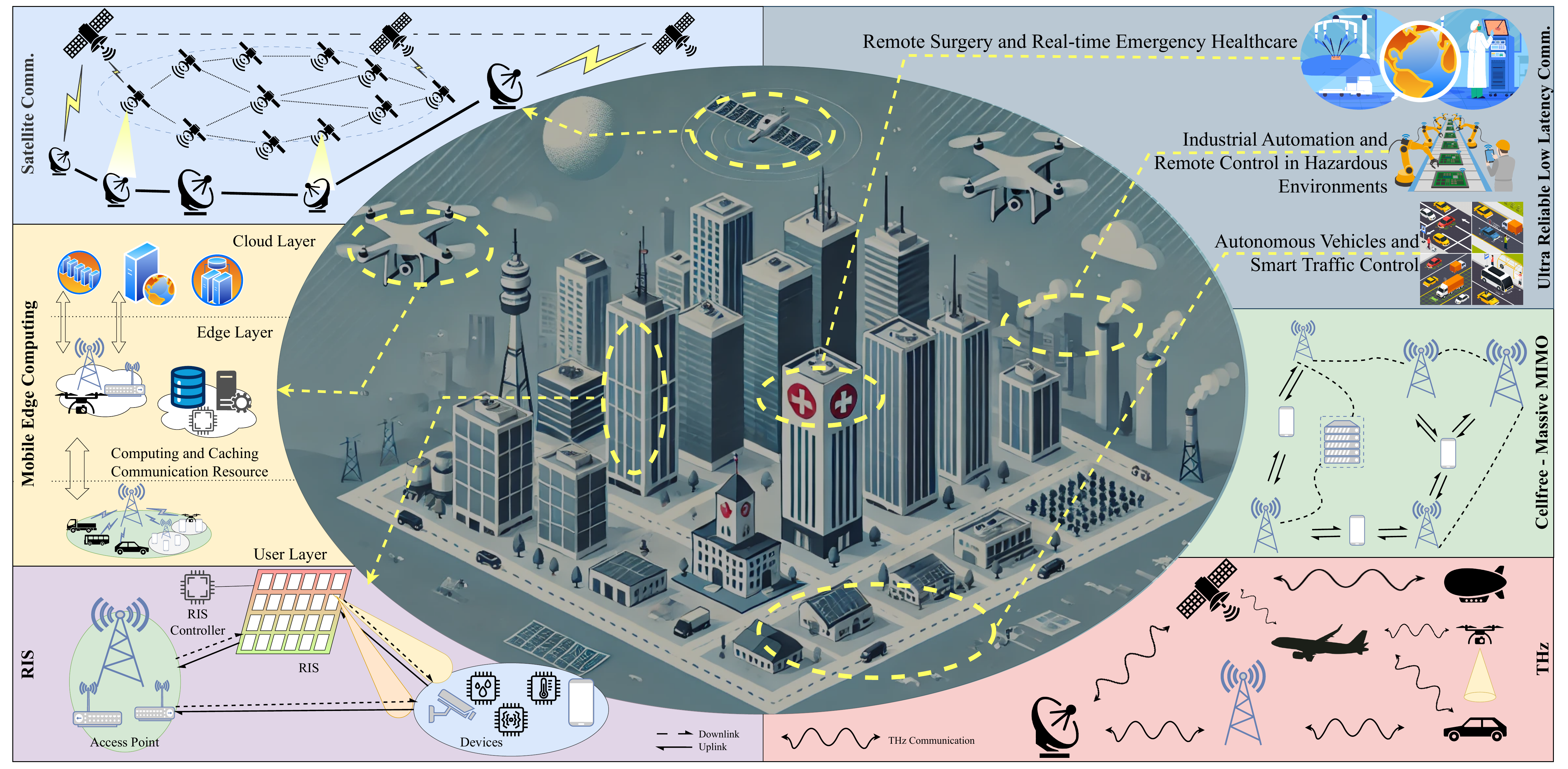}
        \caption{Next generation IoT networks.}
        \label{fig: GNN_Survey-Q3_new}
        \vspace{-15pt}
    \end{figure*}
    \subsubsection{Background:}
    Massive MIMO, RIS, satellite, THz, mobile edge computing, and URLLC systems are key NG technologies aimed at enhancing system performance, including energy, spectrum, and computational efficiency, as well as coverage. These technologies and their applications are comprehensively illustrated in Fig.~\ref{fig: GNN_Survey-Q3_new}, which highlights their roles in NG IoT networks. For instance, massive MIMO systems improve spectral efficiency by employing a large number of antennas, which allows for improved spatial multiplexing and user throughput \cite{cf_mMIMO4IoT_Yan_Apr_2021, cf_mMIMO4IoT_Sep_2021, cf_mMIMO4IoT_Ke_Aug_2020, cf_mMIMO_Ngo_Jan_2017}. RIS creates programmable propagation environments capable of dynamically adjusting the phase, amplitude, and polarization of radio waves. This enhances coverage by reducing interference without relying on power amplifiers or complex signal processing \cite{RIS_Liu_May_2021, RIS_Chien_Dec_2022, RIS_IoT_Nguyen_May_2022, RIS_IoT_Sagir_Jan_2023, RIS_IoT_Sep_2022}. Satellite networks have become crucial in supporting non-terrestrial networks by extending coverage to remote areas, hence providing global connectivity \cite{SAGS_IoT_Cao_Nov_2023, SAGS_IoT_Sana_Jun_2023, SAG_IoT_Hong_Jan_2020}. THz communication leverages the unlicensed spectrum ranging from $0.1 \text{ to } 10$ THz for opening up new frequency bands. The abundance of free spectrum in this band circumvents spectrum scarcity, hence offering the high throughput needed for advanced applications like virtual reality and ultra-high-definition video \cite{THz_Survey_May_2022, THz_IoT_Nabil_Jun_2017, THz_IoT_Yilmaz_Dec_2015, THz_IoT_Liu_Jan_2020}, albeit at a high path-loss and limited coverage. Mobile Edge Computing (MEC) brings data processing capabilities closer to end users by carrying out computation at the network edge instead of centrally, thus reducing latency and improving response times for time-sensitive applications. Harnessing MEC is essential in scenarios where quick decision-making and real-time analytics are needed, such as smart cities, autonomous vehicles, and industrial IoT environments \cite{Edge_general_Sun_Oct_2021, Edge_general_Wang_Oct_2023, Edge_general_Zeng_Jul_2023, Edge(UAV)_general_Li_Jun_2022}. It helps offload computational tasks from centralized cloud servers, optimizing resource utilization and enhancing the overall network efficiency. Meanwhile, URLLC (Ultra-Reliable Low-Latency Communication) ensures low-latency, high-reliability communication for mission-critical applications like autonomous driving, robotics, and healthcare. It is crucial in near real-time scenarios, such as industrial automation and remote medical procedures, using bespoke resource allocation for meeting stringent performance requirements \cite{URLLC_IoT_Schulz_Feb_2020, URLLC_IoT_Ren_Jun_2020}. Despite these advantages, the above sophisticated technologies face common challenges in managing high-dimensional spatial data and optimizing spectrum usage. GNNs are becoming popular in addressing these challenges by learning spatial dependencies, optimizing resource management, and maintaining efficient, low-latency communication in dynamic environments.
    \subsubsection{State-of-the-art:}
    \begin{itemize}
        \item \textbf{Massive MIMO:} GNNs have been increasingly harnessed for addressing challenges in wireless networks using massive MIMO schemes, particularly in resource allocation problems \cite{RA_Wireless_Communications_Theory_to_Practice_Yifei_May_2023, cfmMIMO_IoT_Li_Jun_2023, Tung10750215}. Shen \textit{et al.} \cite{RA_Wireless_Communications_Theory_to_Practice_Yifei_May_2023} formulated tangible guidelines for designing GNNs to solve power allocation problems by maximizing the sum rate of the uplink in a cell-free massive MIMO network. They treated this heterogeneous network associated with two types of nodes corresponding to APs and users. Similarly, Li \textit{et al.} \cite{cfmMIMO_IoT_Li_Jun_2023} explored the use of heterogeneous GNNs in a multicarrier-division duplex cell-free mMIMO system, proposing a sophisticated technique for differentiating between communication and interference links. An attention mechanism was applied for extracting critical information from both the APs and users. In another study, Salaun \textit{et al.} \cite{cf_mMIMO_Salaun_Dec_2022} tackled the max-min power control problem of a cell-free massive MIMO network by constructing a heterogeneous graph. In contrast to the approach in \cite{RA_Wireless_Communications_Theory_to_Practice_Yifei_May_2023, cfmMIMO_IoT_Li_Jun_2023}, the represented graph constructed by Salaun \textit{et al.} composed of $(M \times K)$ nodes, where $M$ and $K$ represent the number of APs and users, respectively. An edge is formed between two nodes if these nodes share the same AP or user, and the graph has two types of edges. Note that for the downlink power control problem, the node embedding approach of Shen \textit{et al.} \cite{RA_Wireless_Communications_Theory_to_Practice_Yifei_May_2023} will fix the output dimension based on the number of APs and users during the training phase. Hence, scalability cannot be guaranteed, while the solution conceived by Salaun \textit{et al.} \cite{cf_mMIMO_Salaun_Dec_2022} succeeded in guaranteeing the scalability of the network. Another method that was put forward by Peng \textit{et al.} in \cite{RA_Vertex_vs_Edge_Peng_Jan_2024} also guaranteed the scalability of the downlink in a MIMO system. Briefly, the MIMO system is represented by a heterogeneous graph associated with two types of nodes: APs and users. The edge-GNN concept was proposed for learning the graph capable of generating edge embedding, which is used to infer the power allocation of an AP for a user. In \cite{cf_mMIMO_Ranasinghe_Dec_2021}, Ranasinghe \textit{et al.} addressed the AP selection problem by utilizing a pair of different graph representations: a homogeneous graph consisting only of APs and a heterogeneous graph that includes both APs and users. They employed graph sample and aggregation (GraphSAGE) \cite{hamilton2018inductiverepresentationlearninglarge} for learning the wireless graph, capable of generating node embeddings that were then used for calculating confidence scores for classifying the links between APs and users. In a related study, Chien \textit{et al.} \cite{mMIMO_IoT_Chien_Jan_2024} explored a system, where multiple IoT users are simultaneously served by both a satellite and access points. The focus was on optimizing the sum of the ergodic uplink throughput with the aid of the most appropriate power allocation across all IoT users. The authors harnessed a heterogeneous GNN for tackling the optimization problem formulated. Notably, Schynol \textit{et al.} \cite{Coordinated_Sum_Rate_Deep_Unrolling_Schynol_Apr_2023} introduced an innovative technique for optimizing the total data rate of a massive MIMO system. They constructed a GNN architecture inspired by the algorithmic unfolding of the weighted minimum mean squared error (WMMSE) method, providing a powerful tool for enhancing the performance of a massive MIMO system.
        
    \item \textbf{Reconfigurable Intelligent Surfaces:} GNN-based RIS-assisted IoT systems have been characterized in \cite{RIS_Singh_Dec_2023, RIS_Zhang_May_2022, III_B_Node_Wang_Z_2023}. Singh \textit{et al.} \cite{RIS_Singh_Dec_2023} combined GNNs with a transformer model, which uses self-attention mechanisms for capturing long-range dependencies in data, in support of channel estimation. Specifically, the GNN layers within the transformer are used for generating embedded vectors representing the groups of RIS elements, which are assumed to have the same channel. These are then processed by the transformer's attention mechanism for accurately predicting the unknown channels. This method significantly reduces the pilot overhead while maintaining high estimation accuracy. As a further advance, Zhang \textit{et al.} \cite{RIS_Zhang_May_2022} proposed a joint optimization procedure for user scheduling, RIS configuration, and base station beamforming to maximize the weighted sum rate in the downlink enhanced by RISs. They utilized a pair of GNNs for user scheduling and RIS configuration, with the final beamforming harnessing the WMMSE algorithm. Wang \textit{et al.} \cite{III_B_Node_Wang_Z_2023} studied the benefits of federated learning (FL) for distributed IoT networks. The RIS was employed for enhancing the FL process by minimizing long-term errors, hence improving accuracy. A heterogeneous GNN was proposed for learning the network's graph structure, enabling the optimization of RIS phases, client power allocation, and denoising factors at the server. Cao \textit{et al.} \cite{RIS_Satellite_IoT_Cao_Feb_2022} investigated the RIS-aided downlink of Low Earth orbit (LEO) satellite IoT networks. They proposed an attention graph neural network for learning the network topology and received pilots, optimizing the RIS phase shifts and satellite beamforming, thereby improving the overall network performance.

    \item \textbf{Satellite Communication:} GNNs are also capable of addressing challenges in satellite-based IoT networks, efficiently handling their dynamic topology and resource constraints \cite{SAGS_GNN_IoT_Huang_Nov_2023, SAGS_GNN_IoT_Chen_Mar_2023, SAGS_GNN_IoT_He_Jan_2024}. Huang \textit{et al.} \cite{SAGS_GNN_IoT_Huang_Nov_2023} studied a suite of multipath routing optimization problems under both bandwidth and flow constraints. The topology of the LEO satellite and ground stations is represented by a spatio-temporal graph. The study proposed a GNN-based multiPath traffic engineering algorithm relying on edge embeddings for distributing traffic across the candidate paths identified by a custom algorithm. Chen \textit{et al.} \cite{SAGS_GNN_IoT_Chen_Mar_2023} also considered the routing issues of the LEO system supporting IoT users. As the number of elements in the network increases, the memory requirement becomes a challenge. Therefore, the authors proposed combining a GCN and a gated recursive unit (GRU) \cite{cho2014learningphraserepresentationsusing} for reducing the memory requirement while still predicting the topology of the LEO system. He \textit{et al.} \cite{SAGS_GNN_IoT_He_Jan_2024} investigated the service function chain orchestration problem of LEO networks with the objective of maximizing the user service acceptance rate. They proposed a GAT-based hierarchical RL technique for solving the problem, in which the GAT model served as a feature extraction module. Similarly, Wang \textit{et al.} \cite{SAGS_GNN_Wang_Apr_2023HotICN} utilized a GNN as an extraction module within a reinforcement learning model to optimize routing in LEO satellite networks. 

    Integrated networks combining satellites with aerial or ground systems have also garnered significant attention. Asheralieva \textit{et al.} \cite{SAGS_GNN_IoT_Asheralieva_Jul_2023} investigated the space-air-ground integrated networking (SAGIN) concept concerning network slicing aided MEC systems designed for IoT and mobile applications. The SAGIN system, which includes aerial, LEO satellite, and terrestrial networks, aims for providing seamless service for IoT devices. However, the rapidly fluctuating dynamic topology can lead to instability and unreliable nodes. To address this, Asheralieva \textit{et al.} proposed a deep learning model based on MPGNN for acquiring node embeddings that the DL model will use for solving the associated slicing problem. 
    
    In \cite{SAGS_GNN_IoT_Chen_Apr_2022}, the hybrid satellite-terrestrial networks have been studied by Chen \textit{et al.} with the support of unmanned aerial vehicles (UAVs) as relay stations. The objective was to maximize the number of IoT devices served along the UAV's trajectory and activating user scheduling. The GAT model was employed for extracting node embeddings, which were used for predicting user scheduling for IoT devices. This scheduling information was then fed into a Q-learning model to determine the UAV's optimal trajectory. To further enhance the satellite transmission performance, RISs may also be considered as a potential solution \cite{RIS_Satellite_IoT_Cao_Feb_2022, SAGS_GNN_IoT_Tekbyk_Jul_2021}. For instance, Tekbyk \textit{et al.} \cite{SAGS_GNN_IoT_Tekbyk_Jul_2021} utilized a GAT network to learn the relationship between the pilot signal and the phase shift of RIS, which was then exploited for channel estimation. Leveraging the GAT network allows the system to estimate all channel coefficients simultaneously and this procedure can be generalized to diverse network configurations.

    \item \textbf{THz Communication:} Recent studies have applied GNNs for optimizing wireless systems operating in the THz band, focusing on addressing unique challenges, such as their high path loss and dynamic channel conditions. For instance, GNNs have been utilized for enhancing resource allocation to optimize the performance of wireless systems utilizing the THz band \cite{THz_GNN_IoT_Zhang_Sep_2023, THz_GNN_IoT_Mar_2024,  THz_GNN_Li_Oct_2023}. Zhang \textit{et al.} \cite{THz_GNN_IoT_Zhang_Sep_2023} proposed integrating Digital Twin (DT) technology with the THz band. The associated weighted mean rate maximization problem subject to power allocation and user association is formulated as a graph optimization problem that is then solved using a distributed message propagation algorithm. Briefly, $K$ message passing layers are utilized to infer the node embedding. The power allocation, $P_i$, and the user association, $\mu_i$, of the $i$-th user are inferred from the node embedding, $x_i^K$, at the last layer $K$, where we have ${P_{i}} = {\mathrm{MLP}}\left ({{x_i^K} }\right)$ and $\mu_i = \sigma\Big(\frac{x_i^T}{\gamma}\Big)$, respectively. Here, $\sigma$ is the sigmoid function, and $\gamma$ is a hyperparameter. 
    
    For mitigating the path loss and improving the propagation distance, Mehrabian \textit{et al.} \cite{THz_GNN_IoT_Mar_2024} suggested using a RIS system. They simultaneously optimized the THz sub-band allocation, the phase shift, and the transmit precoder for maximizing the system's sum rate. The authors proposed a heterogeneous graph-transformer network based on the self-attention mechanism for learning the input features of the RIS, the BS, and all users, resulting in embedding vectors before applying deep neural network (DNN) to predict the specific output for each node. Moreover, to guarantee the minimum required data rate, the authors applied a penalty term when the achievable rate fell below the minimum required rate $r_{i}^{\min}$. 
    
    Li \textit{et al.} \cite{THz_GNN_Li_Oct_2023} harnessed the THz band for vehicular networks. In particular, a set of provider vehicles offer services to several communication and sensing vehicles using the THz band. An integrated sensing and communication problem was considered and the data rates of all communicating vehicles were optimized, while meeting the specifications of the associated sensing task. The system was represented by a heterogeneous network having three types of nodes, including provider, communication, and sensing vehicles. A GNN model was proposed for learning the represented graph to yield embedding vectors for all provider nodes and then to use them for calculating the probability of operating in the sensing and the communication mode or being dormant.    

    \item \textbf{MEC:} GNNs have been widely adopted for solving various problems in MEC-aided IoT networks, \cite{Edge_general_Sun_Oct_2021, Edge_general_Wang_Oct_2023, Edge_general_Zeng_Jul_2023, Edge(UAV)_general_Li_Jun_2022}. To elaborate, Sun \textit{et al.} \cite{Edge_general_Sun_Oct_2021} proposed a graph reinforcement learning-based offloading (GRLO) framework to solve the task offloading problem in a collaborative Artificial Intelligence of Things (AIoT) system consisting of wireless devices equipped with intelligent sensors and MEC servers. They constructed a GNN as an actor network, which learns the policy relying on the relationship between nodes. Wang \textit{et al.} \cite{Edge_general_Wang_Oct_2023} invoked fog computing for task offloading in MEC systems supporting device-to-device (D2D) communication. A real-time GNN inference framework, termed as Foggraph, was proposed for maximizing the servers' performance. In particular, the authors designed an attention mechanism for GNNs to calculate a reward for the proposed inverse reinforcement learning relying on GNNs. Similarly, Li \textit{et al.} \cite{Edge(UAV)_general_Li_Jun_2022} employed a UAV as a mobile-edge server. Their study investigated the joint optimization of UAV trajectory and task allocation using a GNN within an actor-critic structure in order to train real-time actions. In contrast to previous treatises, where a GNN was used as an actor-network in reinforcement learning, Li \textit{et al.} \cite{Edge(UAV)_general_Li_Jun_2022} utilized the GNN as a pre-trained network to harness the associated network feature correlations. 
    
        The application of GNNs in MEC-based IoT scenarios extends to specific use cases such as transportation. In \cite{Edge_Transpo_Xu_Nov_2023, Edge_Transpo_Zhou_Jan_2023}, the authors studied the task offloading problem in the Internet of Vehicles (IoV) utilizing edge servers. Both \cite{Edge_Transpo_Xu_Nov_2023} and \cite{Edge_Transpo_Zhou_Jan_2023} utilized a graph-weighted convolution network (GWCN) for predicting the traffic flow based on the connectivity and distance relations between road segments. This information was then used for optimizing the edge resources within each region using a deep deterministic policy gradient (DDPG) approach. Similarly, Zhou \textit{et al.} \cite{Edge_Transpo_Zhou_Jan_2023} introduced a computation offloading method incorporating demand prediction and reinforcement learning, using an STGNN for accurate predictions. In maritime IoT applications, Liu \textit{et al.} \cite{Edge_Transpo_Liu_Apr_2022} proposed a so-called Spatio-Temporal Multigraph Convolutional Network (STMGCN) for vessel trajectory prediction. This approach uses three distinct graphs based on social force, time to closest approach, and the size of surrounding vessels, demonstrating robust performance in predicting future vessel positions.
        
        Apart from transportation, GNNs have also been applied in other areas, such as healthcare \cite{Edge_health_Fei_Jan_2024} and smart home systems \cite{Edge_smarthome_Sun_Apr_2023}. Fei \textit{et al.} \cite{Edge_health_Fei_Jan_2024} introduced the so-called MedGCN system, which utilizes IoT edge computing for real-time analysis of patient data. A novel graph convolutional network is harnessed by the MedGCN system for predicting and diagnosing occlusive vascular diseases. The authors also considered patient privacy; therefore, this framework has a high potential. Sun \textit{et al.} \cite{Edge_smarthome_Sun_Apr_2023} proposed an edge gateway, which is an important intermediary between edge computing and IoT devices for intrusion detection in smart home applications. Briefly, the graph-type network traffic is fed into the proposed RF-GraphSAGE model in order to predict attack types along the edge between devices. Tang \textit{et al.} \cite{Edge_industrial_Tang_Oct_2023} considered various Cloud-Edge Industrial Internet of Things (IIoT) scenarios, where anomalies happen more often when users generate service function chains (SFCs). They proposed a distributed knowledge distillation framework for time-series anomaly detection. In their teacher model, the authors proposed to utilize spatial graph convolution for capturing spatial topology information in their SFC anomaly detection schemes.

        \item \textbf{URLLC:} By involving the computational efficiency of GNNs, researchers have explored their potential in accelerating the computational process to meet the stringent latency requirements of URLLC applications. For example, in \cite{URLLC_IoT_GNN_Liu_Oct_2023}, Liu \textit{et al.} minimized the packet loss probability by harnessing a mechanism that transmits multiple copies of a packet without waiting for acknowledgment from the receiver, thereby enhancing reliability. To determine the optimal number of slots reserved for each packet and the number of repetitions, a pair of cascaded random-edge GNN networks (REGNN) was constructed. The first REGNN will learn the traffic state, the network state, and the channel state information in support of determining the number of slots for each packet. The second REGNN utilizes the results from the first REGNN and the network state to predict the most appropriate number of repetitions. 
        
        In \cite{URLLC_IoT_GNN_Jiaqi_Feb_2024}, Jiaqi \textit{et al.} aimed for maximizing the success probability of URLLC data transmissions by formulating the resource allocation problem as a Markov decision process. They used a reinforcement learning framework along with a GraphSAGE encoder to extract networking information and feed it into the actor-critic network for decision-making. In \cite{URLLC_IoT_GNN_Liu_Dec_2021}, Liu \textit{et al.} proposed a user association solution employing edge-wise gated GNNs (EG-GNN) for modeling the network as a bipartite graph of the BS and IoT devices. The BS nodes included estimated collision and delay violation probabilities, while the UE nodes had packet loss probabilities. The EG-GNN predicted the most appropriate gate values for the device-BS connections, with the devices selecting the particular gate promising the most beneficial association. As a further advance, Gu \textit{et al.} \cite{III_B_Graph_Gu_Y_2024} studied a massive URLLC (mURLLC) relying on multiple transmit antennas, and aiming for minimizing the decoding error probability of the worst link in their beamforming design. The mURLLC network considered was represented by a fully connected graph, where each communication link corresponds to a node in the graph, and the interference link is represented by an edge. The authors proposed a distributed GNN for a mURLLC system, allowing each node to determine its policy based solely on the channel state information gathered from the previous frame. Accordingly, the system can reduce both the signaling overhead and the computational delay by updating graph embeddings based on the correlation of CSI between two consecutive frames, rather than harnessing multiple updates within the same frame, as in previous GNN models.
    \end{itemize}

    \subsubsection{Challenges and future directions}
    One of the key challenges is the heterogeneity of both the IoT networks and of the additional infrastructure, such as the associated MEC and satellites. This further complicates prediction and optimization, because each entity requires specific resource allocation strategies. Secondly, NG-IoT networks tend to rely on dynamic resource allocation. The complexity increases in the face of multiple constraints, such as power, bandwidth, and interference, which makes the problem harder to represent as a standard graph problem. Each technology requires careful consideration of the unique characteristics of the network, such as the spatial correlation experienced by massive MIMO, the dynamic channel conditions of THz, or the strict latency and reliability requirements of URLLC communication. These constraints must be incorporated into the GNN models for ensuring accurate and efficient resource allocation across different entities, which requires novel methods for graph representation and message passing.

    Additionally, the scalability of GNNs becomes critical, since all these technologies involve a large number of network entities, such as numerous antennas in massive MIMO schemes, reflecting elements in RISs, or other devices in satellite and THz communication systems. Managing these vast networks in the presence of complex interdependencies challenges GNN models, which must balance the computational efficiency vs. the need to capture detailed spatial and temporal correlations. Future research should focus on developing scalable GNN architectures capable of handling larger graphs and integrating techniques like graph partitioning, hierarchical GNNs, and distributed learning for ensuring that GNN-based models succeed in handling the complex constraints of these massive networks effectively.

    \subsection*{Open Question 5: How Can GNNs Enhance Integrity, Security, and Scalability Along with Blockchain for IoT Systems?}
    \begin{figure}[t]
        \includegraphics[trim=0.5cm 0cm 0cm 0cm, clip=true, width=3.5in]{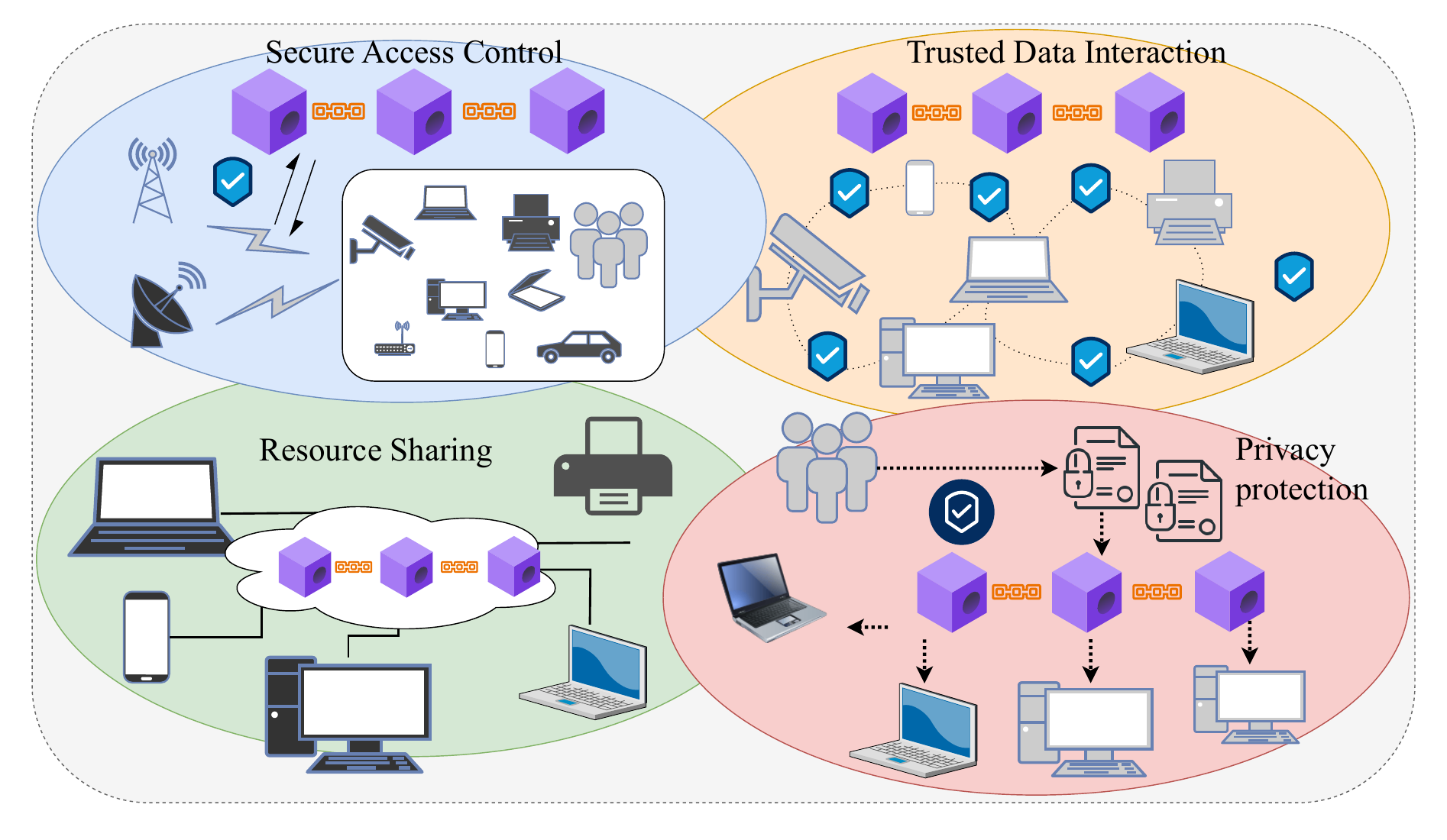}
        \caption{ Blockchain for NG-IoT networks.}
        \label{fig: GNN_Survey-Q4}
        \vspace{-15pt}
    \end{figure}
    \setcounter{subsubsection}{0}
    \subsubsection{Background:}
    The blockchain concept has emerged as a powerful solution for addressing security and privacy challenges in IoT systems, thanks to its decentralization, traceability, trustworthiness, and immutability \cite{Blockchain_IoT_Dai_Jun_2019}. These features make blockchain an ideal candidate for enhancing security in IoT networks, which are increasingly vulnerable to attacks and privacy breaches. {\color{black}As shown in Fig. \ref{fig: GNN_Survey-Q4}, the blockchain provides a flexible framework for secure IoT applications, allowing users to control connection permissions for IoT devices, base stations, and satellites, enabling trusted data exchange.} However, leveraging blockchain in IoT environments also presents several challenges, including their scalability and limited computational resources.
    \subsubsection{State-of-the-art:}
    GNNs have been explored as a solution to address these challenges in blockchain-based IoT systems \cite{Blockchain_IoT_GNN_Cai_Jul_2023, Blockchain_IoT_GNN_Kim_Nov_2020, Blockchain_IoT_GNN_Ziyu_Aug_2020}. In \cite{Blockchain_IoT_GNN_Cai_Jul_2023}, Cai \textit{et al.} introduced a technique termed as GTxChain, which is a secure IoT smart blockchain framework based on GNNs. To elaborate a little further, after collecting data, the blockchain data structure is exploited for eliminating unnecessary data. At the blockchain nodes, a GNN model is utilized to learn as well as maintain the information on the blockchain and the information stored in the so-called InterPlanetary File System. For the graph represented, each block is treated as an object, and the connection between blocks represents the edge. Both the computational resources and the data harnessed for training the GNN are allocated based on the nodes' workload. Kim \textit{et al.} \cite{Blockchain_IoT_GNN_Kim_Nov_2020} focused their attenuation on transaction exchanges between blockchain network nodes in IoT environments. A GNN was designed for node classification, determining whether nodes should spread, skip, or specific activate transactions, thus facilitating efficient distributed applications across blockchain networks. 
    
    As a further development, Ziyu \textit{et al.} \cite{Blockchain_IoT_GNN_Ziyu_Aug_2020} designed a decentralized blockchain-aided system for maintaining data privacy in health applications. Their solution, termed as GuardHealth, combines blockchain and smart contracts for achieving secure data storage and sharing. {\color{black}As shown in the lower-right quadrant of Fig.~\ref{fig: GNN_Survey-Q4}, blockchain allows users to encrypt their data and grant access only to authorized institutions with their permission.} The authors represented the network by an undirected graph having $N$ nodes, including patient nodes, institute nodes, and cloud service providers, where the nodes and edges connected them. The nodes have different features trust assessment mechanism. The GNN model was proposed for malicious node detection in order to reduce transactions with nodes, which are eventually removed from the network. These studies highlight the potential of GNNs in addressing the unique challenges of integrating blockchains into IoT systems for improving their security, scalability, and efficiency across a range of applications.
    
    \subsubsection{Challenges and future directions:}
    Despite these advances, numerous challenges remain in applying GNNs to the blockchain-aided IoT. The large scale and extreme heterogeneity of IoT networks require more scalable and computationally efficient GNN architectures. The integration of hybrid storage mechanisms, as highlighted in \cite{Blockchain_IoT_GNN_Cai_Jul_2023}, points toward reducing on-chain storage load, but significant computational and communication burdens still persist. Again, improving the privacy and security of GNN-based blockchain solutions is a key concern, particularly in environments having constrained resources, where computational capabilities are limited. Future research should focus on developing lightweight GNN models capable of handling large-scale blockchain data, while incorporating robust privacy-preserving techniques, such as zero-knowledge proofs \cite{Sun9520375}, to ensure trust and data integrity in decentralized IoT networks.

    \subsection*{Open Question 6: How Can GNNs Be Integrated with Distributed Systems in NG-IoT Networks?}
    \setcounter{subsubsection}{0}
    \subsubsection{Backgrounds:}
\begin{figure}[t]
        \centering
        \includegraphics[width=\columnwidth]{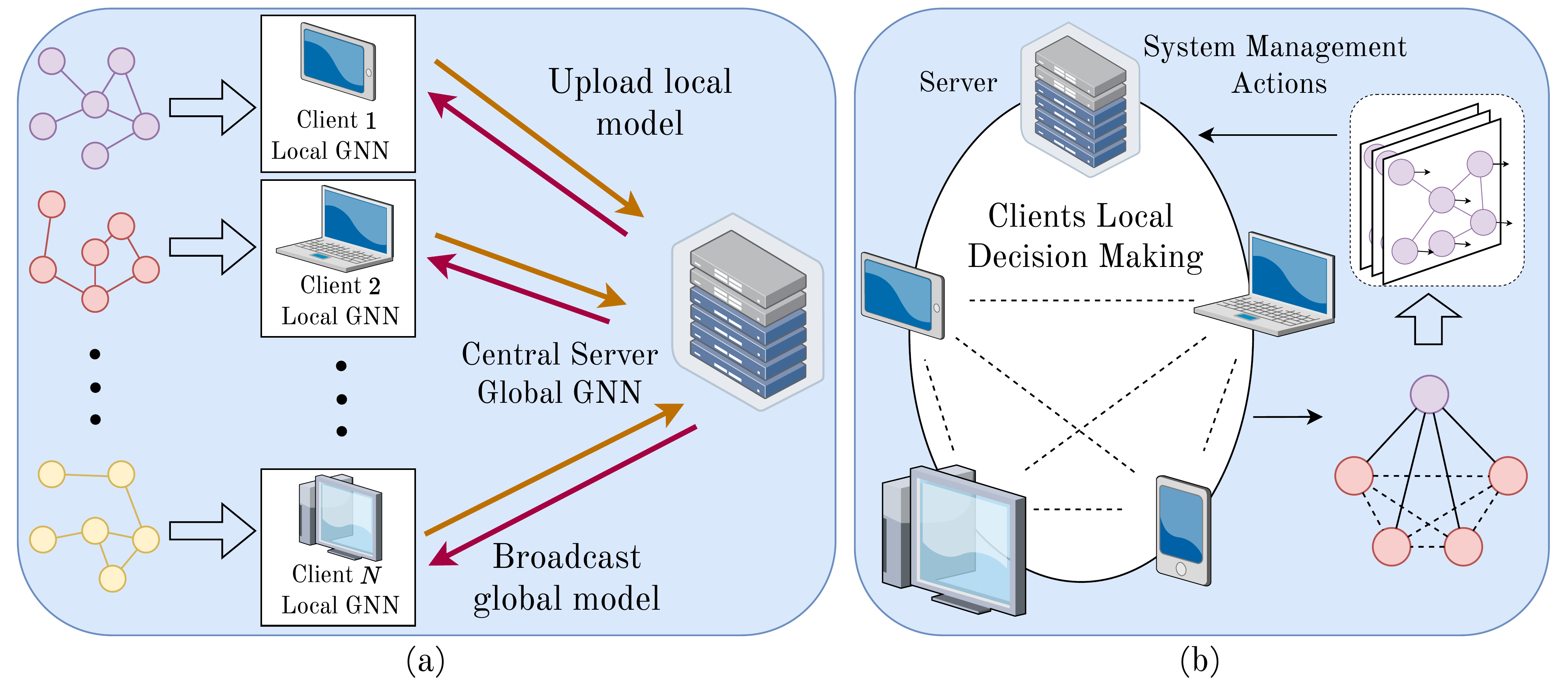}
        \caption{\textcolor{black}{The incorporation of GNN and FL: (a) FL-assisted GNN, and (b) GNN-assisted a decentralized system} }
        \label{fig:GNN_FL}
    \end{figure}

    {\color{black}As NG-IoT environments consist of diverse, distributed, and often resource-constrained edge devices, centralized data collection and processing pose critical challenges in terms of data privacy, communication costs, and model scalability. To address these issues, distributed paradigms, such as federated learning (FL) and decentralized structures, have gained significant attention. FL allows multiple edge devices to collaboratively train a shared model without exchanging raw data, preserving privacy and reducing communication overhead \cite{mammen2021federated, Li9084352, Nguyen9829327, Hieu3628826, nguyen2025federated, nguyen2025federateddomaingeneralizationdatafree, Junewoo10946888}. Meanwhile, decentralized architectures eliminate the need for a central server by enabling peer-to-peer collaboration among devices \cite{Burger8643507, Tung10750215}. The integration of GNNs with distributed systems has emerged as a promising paradigm to meet the privacy, scalability, and efficiency requirements of next-generation IoT (NG-IoT) networks. Combining these two paradigms empowers edge nodes in NG-IoT systems to collaboratively learn graph-based models with minimal communication overhead and stronger privacy guarantees. This synergy opens two prominent research directions: (1) leveraging distributed architectures to optimize GNN(a) and (2) using GNNs to enhance distributed system in NG-IoT systems. Fig.~\ref{fig:GNN_FL} illustrates the integration of GNN with distributed frameworks in NG-IoT: (a) Federated learning-assisted GNN, where a central server aggregates local GNN models trained on distributed clients; (b) GNN-enabled decentralized system, where local agents make decisions collaboratively based on peer-to-peer graph reasoning without a central server.}
    \subsubsection{State-of-the-art:}
    {\color{black}\textbf{Distributed architecture for GNN in NG-IoT networks:} Distributed systems, especially FL, provide robust frameworks for collaboratively training GNNs across distributed nodes while preserving privacy. For instance, Wu \textit{et al.} \cite{jianping2024federated} developed FedGAT, a federated graph attention network for privacy-preserving network attack detection, leveraging attention to improve internal traffic interaction modeling. Similarly, Zhang \textit{et al.} \cite{10026810} employed FL to train a GNN-based anomaly detection system for controller area networks (CAN bus), preserving privacy without compromising detection accuracy. Addressing resource constraints in mobile edge computing (MEC), Zhong et al. \cite{10433561} proposed a lightweight federated GNN framework for UAV-assisted environments, combining adaptive information bottlenecks and compact GNNs for accelerated classification. Meanwhile, Huang \textit{et al.} \cite{10734080} optimized both client and cross-client edge selection to improve convergence performance and cost-efficiency in training GCNs over distributed data. Khanna \textit{et al.} \cite{khanna2025grl} combine FL with graph reinforcement learning to optimize edge caching in MEC, demonstrating latency and cache hit improvements. 
    
    In \cite{10533257}, Li \textit{et al.} presented DHTGL, a distributed hierarchical temporal graph learning framework, reducing communication in IIoT systems. For decentralized training environments, Gao \textit{et al.} \cite{10855737} introduced AirGNNs, a communication-aware GNN architecture that incorporates wireless channel effects and enables over-the-air decentralized GNN training. The framework is robust to fading and noise and supports decentralized tasks such as multi-robot flocking and source localization. Li \textit{et al.} \cite{9919905} proposed a GCN-based dynamic topology selection strategy in vehicle-to-vehicle (V2V) networks to ensure reliable model sharing in decentralized federated learning (DFL), thereby improving training stability in mobile environments. In \cite{Edge_general_Zeng_Jul_2023}, Zeng \textit{et al.} proposed FoGraph, a distributed GNN inference platform deployed over fog nodes. By leveraging compression-aware execution planning, FoGraph reduces communication bottlenecks compared to traditional cloud-based model serving and accelerates GNN inference by over $5\times$.

    \textbf{GNN for the distributed systems in NG-IoT networks:} Several studies utilize GNNs to improve the design and performance of decentralized systems and FL. Li \textit{et al.} \cite{10364739} proposed a GCN-based power allocation policy designed for wireless FL systems operating under channel variability and non-IID data distributions. Their method parameterizes power control using a GCN and solves the associated optimization via a primal-dual approach, ensuring communication efficiency while maintaining learning accuracy. Wang \textit{et al.} \cite{III_B_Node_Wang_Z_2023} incorporated GNNs into over-the-air FL by learning efficient mappings from wireless channel states to RIS-assisted transceiver configurations, thus mitigating aggregation bottlenecks and improving spectral efficiency. In \cite{10437172}, Yang \textit{et al.} introduced a two-stage GNN-based collaborative FL framework aimed at minimizing energy consumption. The model enables each device to learn an optimal set of communication links for local parameter exchange and corresponding transmit power levels, significantly outperforming traditional iterative optimization techniques in both speed and energy efficiency. In another work, Zhang \textit{et al.} \cite{III_B_Graph_Gu_Y_2024} leveraged GNN-based policies to solve the joint distributed beamforming and power control problem in massive URLLC networks, targeting a reduction in decoding error probabilities under tight latency constraints.}
    

    \subsubsection{Challenges and future directions:}
    {\color{black}Despite these advances, several challenges remain in integrating GNNs with the distributed systems in IoT networks. First, heterogeneity in data and system conditions, such as device capabilities, communication channels, and data distributions, restricts effective collaboration. Besides, devices in NG-IoT systems possess different capabilities in computational power, memory, and energy constraints. This results in non-uniform convergence behavior and degraded model performance in federated GNN training, especially in highly dynamic environments like UAVs, mobile edge networks, or vehicular systems.

    Another significant problem is about communication efficiency, although approaches such as over-the-air computation (e.g., AirGNNs \cite{10855737}, Air-MPNN \cite{10068338}) and graph compression techniques (e.g., LW-FGL \cite{10433561}, DHTGL \cite{10533257}) aim to mitigate communication overhead, the scalability of these methods in extremely large or densely connected networks is still underexplored. Besides, distributed nodes requires synchronization and consistency, which remains difficult to achieve when topology evolves rapidly (e.g., in V2V networks), requiring more robust protocols for model aggregation and graph updates. Finally, dynamic graph learning under limited feedback and partial observability remains a key challenge, as NG-IoT graph structures often evolve due to mobility and device failures. Future GNN architectures must support online updates and continual learning without full graph re-computation to meet real-time demands.}
    
    \section{Adversarial Attacks and Defense Mechanisms Conceived for GNN-Based NG-IoT Networks}
    \label{Section: Adversarial Graph Neutral Network}
    In this section, we address a pair of critical questions: Open Question $9$ focuses on the nature of adversarial attacks targeting GNN-based systems, while Open Question $10$ explores defense techniques designed for safeguarding these systems. We commence by examining how adversarial attacks exploit vulnerabilities in GNN models within NG-IoT networks, targeting both homogeneous GNNs (HoGNNs) and heterogeneous GNNs (HeGNNs). These attacks can severely impact the performance and security of systems in applications such as smart cities, autonomous transportation, and healthcare. Next, Open Question 6 explores defense mechanisms that have been developed for countering these attacks, enhancing both the robustness and reliability of GNN-based systems. Finally, we review the latest strategies of adversarial defense and provide insights into future research directions for securing GNNs in the complex dynamic environments of NG-IoT networks.
    
    \subsection*{Open Question 7: How Do Adversarial Attacks Exploit Vulnerabilities in GNN-Based NG-IoT Networks?}
    \subsubsection{Background:} 
    \begin{figure}[t]
        \includegraphics[trim=0.5cm 0.2cm 0cm 0.2cm, clip=true, width=3.6in]{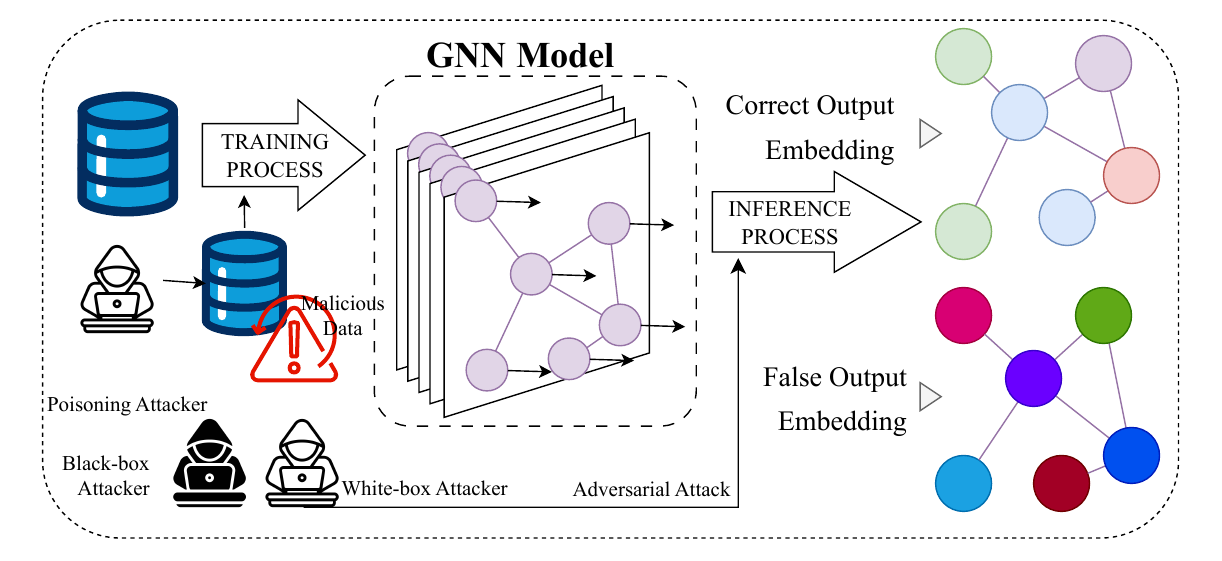}
        \caption{Adversarial attack applied into GNNs.}
        \label{fig: GNN_Survey-Q5}
        \vspace{-15pt}
    \end{figure}
    Graph neural networks have become pivotal in enhancing the performance of NG-IoT networks across various domains, including smart cities, healthcare, autonomous transportation, and communication systems relying on massive MIMO schemes, RIS, and URLLC. Despite having numerous benefits, GNNs are vulnerable to adversarial attacks, when malicious actors introduce subtle perturbations into the data to manipulate the model's output. Fig.~\ref{fig: GNN_Survey-Q5} outlines the process of adversarial attacks on GNNs, showcasing three main types: poisoning attacks, where malicious data is injected during training, and white-box and black-box attacks, which target the inference phase. It also highlights how these attacks can lead to incorrect embeddings and outputs, compromising the system's integrity. These attacks may erode the system's performance, compromise data integrity, and pose significant security risks, especially in critical applications like autonomous driving and industrial automation. The ability to defend GNN-based systems from such attacks is essential for ensuring secure and robust operations in complex and dynamic NG-IoT environments.
    \subsubsection{State-of-the-art:}
    \begin{itemize}
        \item \textbf{Adversarial homogeneous graph neural network:} Ma \textit{et al.} \cite{ma2020towards} proposed a novel setup for black-box attacks on GNNs representing one of the most practical approaches available at that time. By exploring the structural inductive biases of GNNs, which can be leveraged for adversarial black-box attacks, they introduced a practical greedy method for adversarial attacks targeting node classification tasks. In another study, Sharma \textit{et al.} \cite{sharma2023task} proposed the so-called TANDIS algorithm in the context of the evasion attack-based targeted black box scenario. {\color{black}As shown in Fig.~\ref{fig: GNN_Survey-Q5}, these attacks target the training or inference process, leading to incorrect outputs and degraded performance. The figure highlights how attackers, including poisoning and inference-based (white-box and black-box), exploit vulnerabilities in the GNN pipeline.} In the realm of HoGNNs, some authors have mitigated adversarial attacks \cite{dai2018adversarial}, \cite{chang2020restricted}, \cite{wang2020evasion}, but their solutions often suffered from limitations in three key areas:
        
            \textit{1. They focused on specific tasks.}
            
            \textit{2. The adversaries had knowledge about the GNN models.}
            
            \textit{3. Their methods rely heavily on node or edge labels.}

        The experiments showed that the TANDIS algorithm of \cite{sharma2023task} outperformed other evasion attack-based black-box algorithms, despite running approximately 1000 times faster and achieving up to 50\% higher effectiveness in terms of Drop-in-Accuracy (DA\%), which quantifies the percentage reduction in model accuracy before and after an attack. Despite being model-agnostic and task-independent, this algorithm highlights the vulnerability of HoGNNs \cite{sharma2023task} when faced with adversarial attacks, proving their susceptibility.

        \item \textbf{Adversarial heterogeneous graph neural network:} Sun \textit{et al.} \cite{sun2020adversarial} introduced a hierarchical-learning-based method that enables adversaries to execute data poisoning attacks without relying on reinforcement learning techniques. They examined a novel graph node injection attack, which adversely impacts the accuracy of heterogeneous GNNs, even though it does not alter the link structure of the original graph. Additionally, their framework was tested on several real-world graph datasets, including Cora \cite{mccallum2000automating}, Citeseer \cite{Sen_Citeseer}, and Pubmed \cite{mccallum2000automating}, and it was shown to gravely degrade the model accuracy. 
        
        Additionally, H. Zhao \textit{et al.} introduced the so-called HGAtack concept of \cite{zhao2024hgattack}, which operated under the gray-box scenario, where attackers have limited knowledge of the targeted models. Suffice to note that Sun \textit{et al.} \cite{sun2020adversarial} conducted experiments under white-box scenarios, when attackers have full knowledge of the targeted models. The results acquired by HGAttack \cite{zhao2024hgattack} showed that the proposed attack method was effective in gray-box evasion attacks in the context of the ACM, IMDB, and DBLP datasets. Therefore, H. Zhao \textit{et al.} \cite{zhao2024hgattack} showed the potential opportunities for applying adversarial attacks to mislead the heterogeneous GNN models in the real world black-box scenarios, when attackers only know the input and output of the targeted models.
    \end{itemize}

    \subsubsection{Challenges and future directions:} 
    Let us now discuss the family of adversarial attacks designed for damaging key GNN technologies. Firstly, we classify adversarial attacks into three levels: white, grey, and black. Briefly, in a white box scenario, the adversary has complete knowledge of the victim model, hence facilitating grave destruction even upon using straightforward algorithms like fast gradient descent or projected gradient descent. However, most real-world adversarial attacks occur in grey- or black-box scenarios, where the adversary's knowledge is limited or nonexistent. Nevertheless, even in these cases, there are potent techniques of attacking the model; a possible approach is to generate various white boxes to find the most suitable model similar to the targeted model. Notably, the adversary may be able to infer the model input and output, subject to the practical trade-offs between destruction performance and the computational complexity. For example, to generate adversarial perturbations for attacking model-based federated learning, the adversary may harness centralized or distributed attacks. Each family of attack has its advantages and disadvantages as regard to the potential resources required. Additionally, multiple attack algorithms may be combined. Overall, we must carefully guard against adversarial attacks. 

    \subsection*{Open Question 8: What Are the Most Effective Defense Techniques Against Adversarial Attacks in GNN-Based NG-IoT Networks?}
    \setcounter{subsubsection}{0}
    \subsubsection{Background:}
    \begin{figure}[t]
        \includegraphics[trim=0.5cm 0.2cm 0cm 0.2cm, clip=true, width=3.6in]{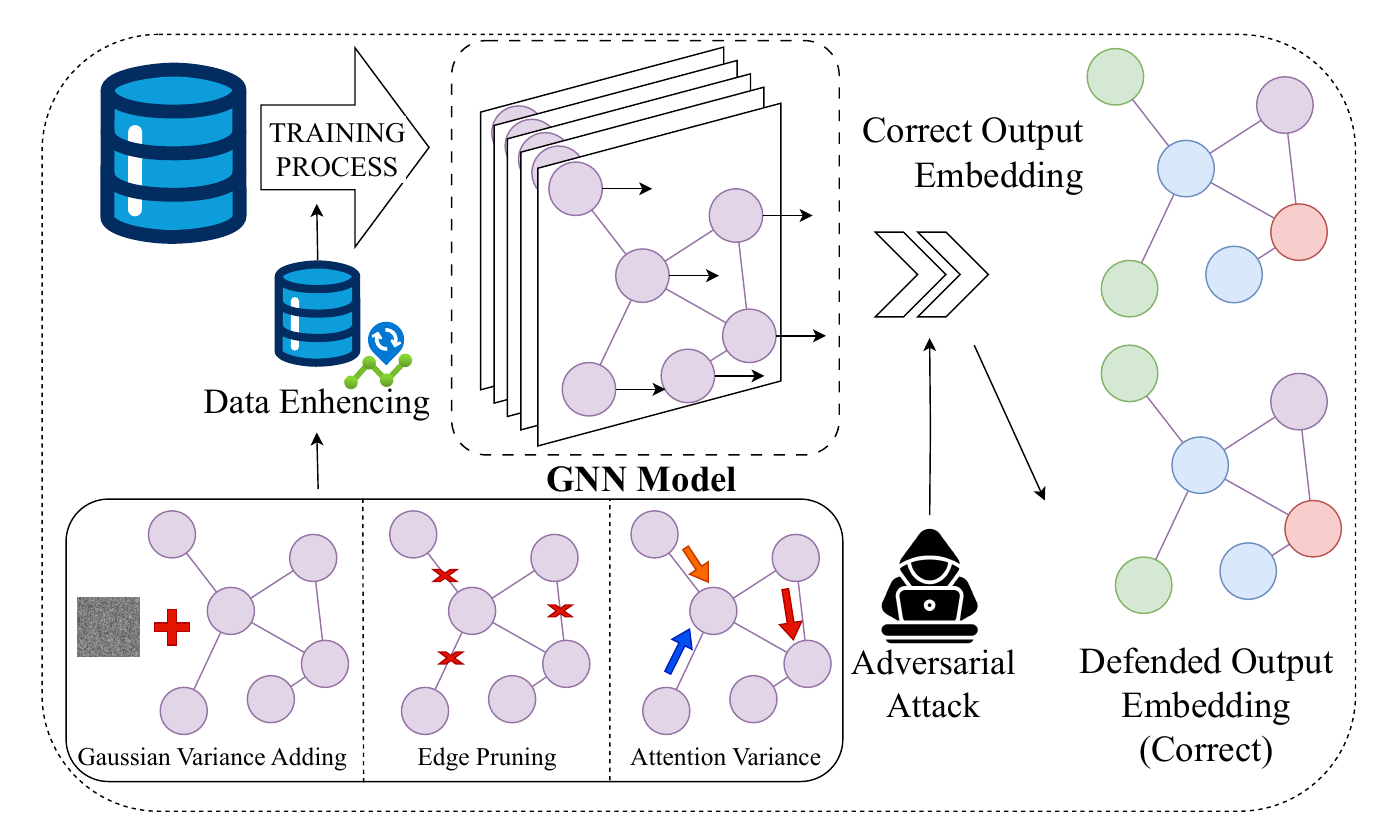}
        \caption{Adversarial defense for GNN.}
        \label{fig: GNN_Survey-Q6}
        \vspace{-15pt}
    \end{figure}
    As GNNs become increasingly integral to NG-IoT networks, their vulnerability to adversarial attacks poses a significant threat to system reliability and security. Defending against such attacks is vital for ensuring the integrity and robustness of GNN-based systems, particularly in dynamic and distributed NG environments. Both homogeneous GNNs and heterogeneous GNNs are susceptible to these threats, necessitating a range of defense techniques tailored to their specific vulnerabilities. {\color{black}Fig.~\ref{fig: GNN_Survey-Q6} demonstrates how these defenses protect the inference process, maintaining the accuracy and reliability of the system. Adversarial defenses focus on enhancing data during the training process through pre-training techniques. These approaches aim to mitigate the impact of adversarial attacks, ensuring the GNN model produces correct and secure embeddings despite malicious attempts.}
    \subsubsection{State-of-the-art:}
    \begin{itemize}
        \item \textbf{Adversarial HoGNN defense}: To enhance the robustness of GCNs against adversarial attacks, Zhu \textit{et al.} \cite{zhu2019robust} proposed a novel model, where the hidden representations of nodes are modeled by Gaussian distributions. Their approach allows the model to mitigate the impact of adversarial structure alterations by incorporating these changes into the variances of the Gaussian distributions. Additionally, they introduced a variance-based attention mechanism for mitigating the impact of adversarial attacks within GCNs. This involves assigning particular weights to specific node neighborhoods based on their variances during the associated convolution operations. Their experimental results demonstrated that the proposed method was capable of significantly improving the robustness of GCNs, leading to enhanced node classification accuracy, as demonstrated with the aid of three benchmark datasets: Cora \cite{mccallum2000automating}, Citeseer \cite{Sen_Citeseer}, and Pubmed \cite{mccallum2000automating}. 
        
        As a further development, Zhang \textit{et al.} \cite{zhang2020gnnguard} developed GNNGuard, a general defense algorithm for securing discrete graph structures. GNNGuard can be readily integrated into any GNN model. The primary objective of GNNGuard was to minimize the adverse effects of adversarial attacks by inferring the relationship between the node features and the graph structure. GNNGuard facilitates the robust propagation of the neural message by using revised edges, which was achieved by learning the most appropriate weights for linking the node, whereas pruning edges between irrelevant nodes. Experimental results derived for five types of GNNs showed that GNNGuard outperforms other existing defense strategies, including GNN-Jaccard \cite{wu2019adversarialexamplesgraphdata}, RobustGCN \cite{zhu2019robust}, and GNN-SVD \cite{Entezari3371789}, with an average improvement of 15.3\% in the accuracy over the Cora \cite{mccallum2000automating}, Citeseer \cite{Sen_Citeseer}, ogbn-arxiv \cite{hu2021opengraphbenchmarkdatasets}, and DP datasets \cite{agrawal2017largescaleanalysisdiseasepathways}.

        \item \textbf{Adversarial HeGNN defense}: Zhang \textit{et al.} in \cite{zhang2022robust} identified a pair of key issues contributing to the vulnerabilities of heterogeneous GNNs (HeGNNs): perturbation enlargement effect and soft attention mechanism. Their experiments, conducted across three specific types of HeGNNs, revealed that perturbation enlargement is less significant in meta-path aggregated graph neural networks and graph transformer networks than in heterogeneous graph attention networks. To improve the robustness of HeGNNs, Zhang \textit{et al.} \cite{zhang2022robust} proposed the concept of Robust Heterogeneous GNNs (RoHe), which may be used for purifying the node-level aggregation framework by harnessing an attention purifier against topology adversarial attacks. Another notable contribution is by Sang \textit{et al.} \cite{10124876}, who introduced a model called AHGNNRec designed for robust recommendation systems based on HeGNNs. The authors applied adversarial training for optimizing hierarchical HeGNN layers by generating perturbed nodes from clean nodes in order to explore the weaknesses of their system. The experimental results based on  YouTube and Yelp datasets illustrated the power of AHGNNRec.
    \end{itemize}
    \subsubsection{Challenges and future directions:}
    While existing defense techniques offer promising solutions, there is still substantial room for improving their efficiency and robustness in NG-IoT networks. Adversarial training is a widely used method, which strengthens the models by exposing them to adversarial perturbations. Additionally, defensive distillation constitutes another potent method where knowledge is distilled from a complex model to a simpler one with the objective of enhancing the performance. Briefly, this approach seeks for creating a more robust classifier that is better prepared to guard against adversarial attacks by relying on precise gradient information.
    Nonetheless, this method's effectiveness may be compromised by attacks that do not depend on gradients or use gradient approximation techniques. A combination of defense methods can also be used for providing stronger protection against adversarial attacks, but this can make the model more complex and increase its carbon footprint. 
    
    \section{The Role of GNNs in Future Integrated Networks and Quantum Computing}\label{Challenges}
    In this section, we address a pair of critical questions related to the future of GNN applications in NG-IoT networks. Open Question 7 focuses on how GNNs enhance the performance and scalability of future integrated networks, such as SAGSINs and ISAC. Figure \ref{fig: GNN_Survey-Q7} illustrates an ISAC network and a SAGSIN, showcasing their integration within next-generation IoT networks supported by GNNs. These emerging technologies present unique opportunities for IoT networks, but also pose significant challenges for GNN-based solutions. Open Question 8 explores how GNNs can be combined with future computational technologies, including quantum computing, in order to support various NG applications. Quantum computing is maybe expected to revolutionize the computational landscape for NG systems by accelerating tasks such as encryption, signal processing, and resource optimization. By integrating GNNs with quantum technologies, NG networks may realize more efficient and scalable solutions, further enhancing capabilities across different network layers and applications.
    
    \subsection*{Open Question 9: How Can GNNs Enhance the Performance and Scalability of Future Integrated Sensing and Communication and Space-Air-Ground-Sea Integrated Networks?}
    \subsubsection{Background:}
    \begin{figure*}[t]
        \includegraphics[trim=0.0cm 0cm 0.0cm 0cm, clip=true, width=7.0in]
        {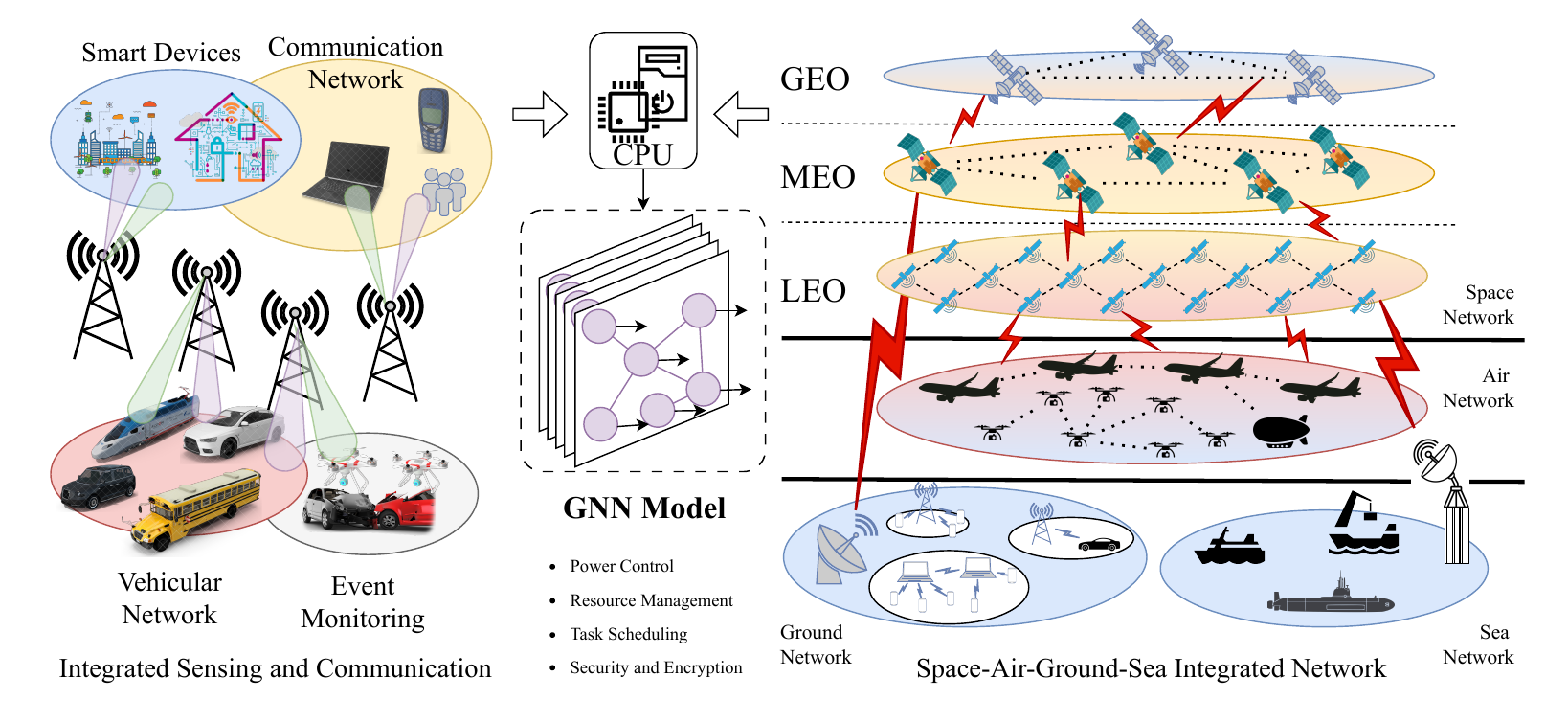}
        \caption{Integrated Sensing and Communication (ISAC) and Space-Air-Ground-Sea Integrated Network (SAGSIN) in NG frameworks.}
        \label{fig: GNN_Survey-Q7}
        \vspace{-15pt}
    \end{figure*}
    \begin{itemize}
        \item \textbf{Integrated sensing and communications:} In the realm of NG networks, the concept of ISAC network emerges as a transformative technology, seamlessly integrating sensing capabilities with communication functionalities \cite{OpenIssue_JSAC_Lu_Feb_2024}. Leveraging the extensive coverage of the operational network infrastructures, ISAC becomes capable of providing sensing capabilities across the entire communications network at a modest additional cost, effectively using the network as a sensor array. {\color{black}As illustrated in the left sub-figure of Fig.~\ref{fig: GNN_Survey-Q7}, ISAC integrates vehicular networks, event monitoring, and smart device communication into a unified framework, ensuring efficient data collection and processing. Sensing is also capable of significantly enhancing communications by providing improved accuracy in localization, imaging, and environment reconstruction, leading to accurate beamforming and CSI tracking, thus improving the overall communication performance.} ISAC exhibits several key characteristics, including dynamic resource allocation, heterogeneous data processing, and real-time adaptability, which are crucial for NG IoT applications. However, these demanding requirements also impose grave challenges and necessitate careful optimization and network design. 
        
        \item \textbf{Space-air-ground-sea integrated networks:} Space-Air-Ground-Sea Integrated Networks have emerged as an intriguing solution in NG-IoT research and development, extending connectivity to all corners of the Earth. This includes challenging environments such as mountainous regions, oceans, underwater areas, and space. SAGSINs achieve this by integrating satellite, aerial, terrestrial, and marine communication networks \cite{Guo9628162, Jahanbakht9328873}. {\color{black}The need for a comprehensive global coverage and the growing demands of NG-IoT applications drive the transition from SAGINs to SAGSINs. While a SAGIN covers the vast majority of terrestrial and aerial needs, incorporating sea-based communication networks into SAGSINs addresses the unique requirements of maritime and underwater environments. The right sub-figure of Fig.~\ref{fig: GNN_Survey-Q7} illustrates how SAGSIN integrates satellite, aerial, terrestrial, and marine networks into a layered architecture. Each layer plays a specific role: satellites provide global connectivity, aerial networks enhance communication coverage and flexibility, terrestrial networks handle dense urban demands, and sea networks enable maritime and underwater communication. This extension ensures robust and reliable connectivity for maritime operations, deep-sea exploration, and remote IoT applications, thus creating a truly global communication infrastructure.}

    \end{itemize}

    \subsubsection{State-of-the-art:}
    \begin{itemize}
        \item \textbf{Integrated communications and sensing:} In the context of ISAC, GNN is a strong candidate as a benefit of its capabilities in terms of modeling complex relationships and dependencies. GNNs can offer insights into network dynamics and resource allocation. In \cite{OpenIssue_JSAC_Lee_Oct_2022}, Lee \textit{et al.} proposed a framework for joint radar and communication in an intelligent vehicle based system, enabling high-performance radar detection, while balancing performance without extensive knowledge or specialized hardware. To solve the joint radar and communications problem, the authors designed a multi-agent deep reinforcement learning (DRL) algorithm combined with GNNs for learning inter-agent coordination. They demonstrated superior results compared to traditional algorithms dispensing with learning. Similarly, Li \textit{et al.} \cite{THz_GNN_Li_Oct_2023} utilized GNNs to extract graph information from integrated sensing and communication in vehicle networks for deciding between the sensing and communication modes of the vehicles. 
        
        \item \textbf{Space-air-ground-sea integrated networks:} To fully harness the benefits of SAGSINs, leveraging GNNs has become an attractive proposition for researchers. A notable example is presented in \cite{Edge_Transpo_Liu_Apr_2022}, where Liu \textit{et al.} considered a maritime IoT system that collects vessel trajectories using a satellite-terrestrial based automatic identification system. These trajectories were processed using the proposed framework for traffic management. The authors represented vessel trajectories through three distinct graphs: the social force graph, the time-to-closest-point-of-approach graph, and the vessel-size graph. Their solution learned these graphs for the sake of extracting vessel features, which were then fed into a self-attentive temporal convolutional layer for vessel trajectory prediction. Beyond solving network management problems, GNNs can facilitate cross-domain integration of SAGSINs by modeling their interactions and optimizing their cooperation, resulting in seamless data transmission and interoperability across the entire network. 
    \end{itemize}

    \subsubsection{Challenges and future directions:}
    \begin{itemize}
        \item \textbf{Integrated sensing and communications:} The employment of the use of GNNs for ISAC in IoT networks is still in its infancy. ISAC presents a myriad of challenges, including scalability concerns, heterogeneous data processing, privacy and security issues associated with sensitive IoT data, and the need for robustness against adversarial attacks in dynamic network environments. Overcoming these challenges is paramount for unlocking the full potential of GNNs in ISAC-aided NG-IoT networks, enabling seamless integration of sensing and communication to facilitate transformative advances in IoT applications.
        
        \item \textbf{Space-air-ground-sea integrated networks:} Despite the promising potential of GNNs in SAGSINs, numerous challenges remain. Firstly, guaranteeing the scalability of GNNs in SAGSINs is a grave challenge, given the dynamic nature of the network. Secondly, security vulnerabilities in SAGSINs pose significant challenges, especially in low Earth orbit satellite communication systems. Potential attacks may compromise the integrity and reliability of the network. Harnessing robust security measures for these satellite systems is critical for preventing disruptions and safeguarding data transmission \cite{Yue10209551}. Finally, conceiving a standardized protocol for integrating GNNs into SAGSINs is essential for ensuring that different systems and technologies can work together seamlessly.
    \end{itemize}
    
    \subsection*{Open Question 10: How Can GNNs and Future Computational Technologies, Like Quantum Computing, Work Together to Enhance the Capabilities of NG-IoT Networks?}
    \setcounter{subsubsection}{0}
    \subsubsection{Background:} 
    
    Quantum computing exploits the principles of quantum mechanics to solve complex problems at a potentially lower number of cost-function evaluations than classical computers. In contrast to classical bits, which represent information as either 0 or 1, qubits can exist in a superposition of both states simultaneously. This capability of processing multiple possibilities simultaneously gives quantum computing a distinct edge over classical computers for certain problems. While classical computing requires performing a new calculation each time a variable changes, yielding a single result, quantum computing is capable of exploring the entire solution space in parallel. The advanced computational capability of quantum computing makes it a promising solution for wireless communication applications, where classical optimization methods struggle with scalability.

    Recent research has increasingly focused on combining the strengths of quantum computing with advanced classical machine learning algorithms, giving rise to the interdisciplinary field of quantum machine learning (QML)~\cite{10274707, JEONG2024608, hur2022quantum, Simeone2000000118}. An area of particular interest is the integration of quantum computing with GNNs~\cite{verdon2019quantum}. Although GNNs excel at processing graph-structured data, they face scalability and computational challenges for large-scale graphs, leading to high training and inference costs~\cite{10094894, 10036979}. To address these limitations, researchers have developed quantum graph neural networks (QGNNs), which combine the structural advantages of GNNs with the computational power of quantum computing.

    \subsubsection{State-of-the-art:}

    \begin{figure}[t]
        \includegraphics[trim=0cm 0cm 0cm 0cm, clip=true, width=0.5\textwidth]{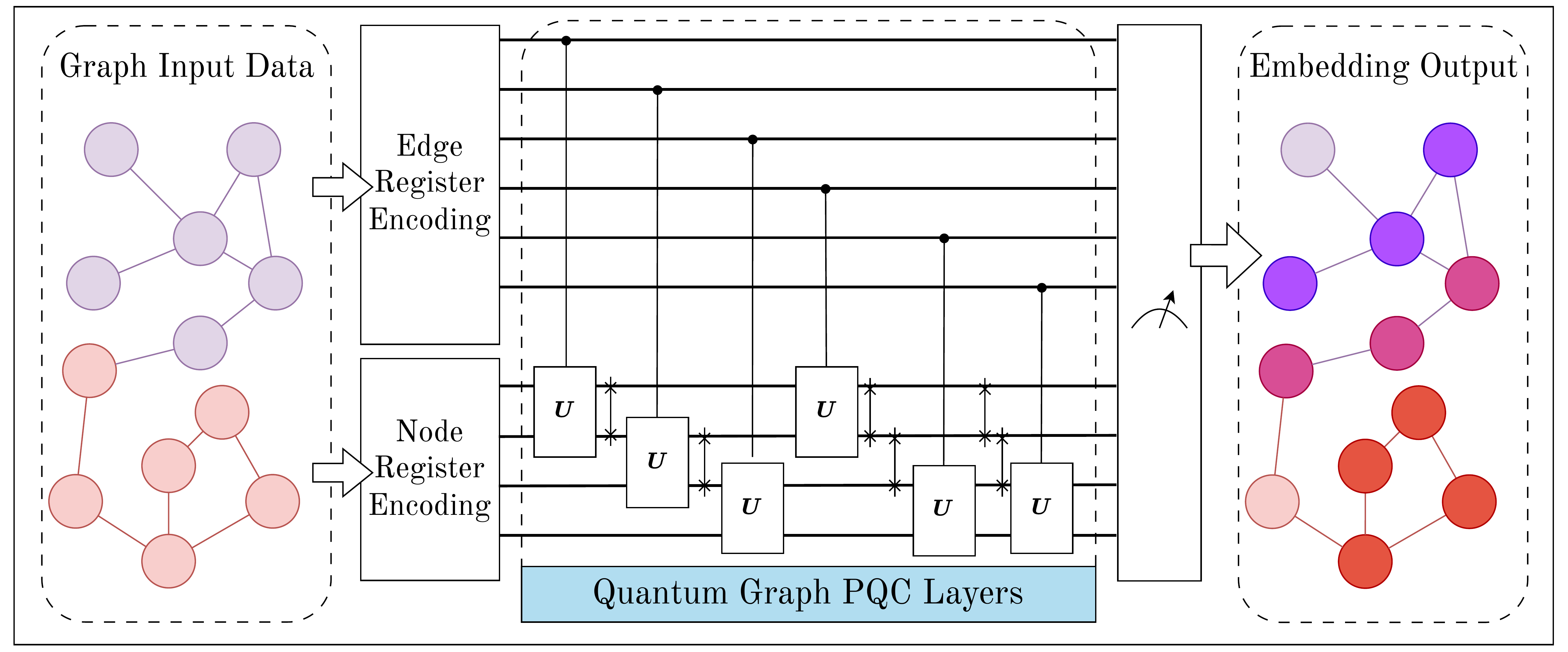}
        \caption{Quantum GNN relying on a quantum graph convolutional layer circuit, where U is a unitary operator.} 
        \label{fig: GNN_Survey-Q8.b}
        \vspace{-15pt}
    \end{figure}
    
    {\color{black}The integration of quantum computing with GNNs involves designing quantum circuits that emulate the layers of a GNN, particularly the message-passing mechanism \cite{10499715Zheng_QuantumGNN_VQC}. Fig.~\ref{fig: GNN_Survey-Q8.b} illustrates the architecture of a QGNNs that relies on a quantum graph convolutional layer circuit. The process begins with the encoding of graph input data, where node and edge features are transformed into quantum states through the node register encoding and edge register encoding blocks. These encoded quantum states are then processed through layers of quantum gates, represented by unitary operators (U), within the quantum graph convolutional layers. By leveraging quantum properties like superposition and entanglement, the QGNN efficiently represents the relationships between nodes in the graph. Moreover, the adjacency matrix of a graph, which defines node connections, can be mirrored by the entanglement patterns between qubits, potentially enabling a parallelized representation of the graph structure. }
    \begin{figure}[t]
        \includegraphics[trim=0cm 0cm 0cm 0cm, clip=true, width=0.5\textwidth]{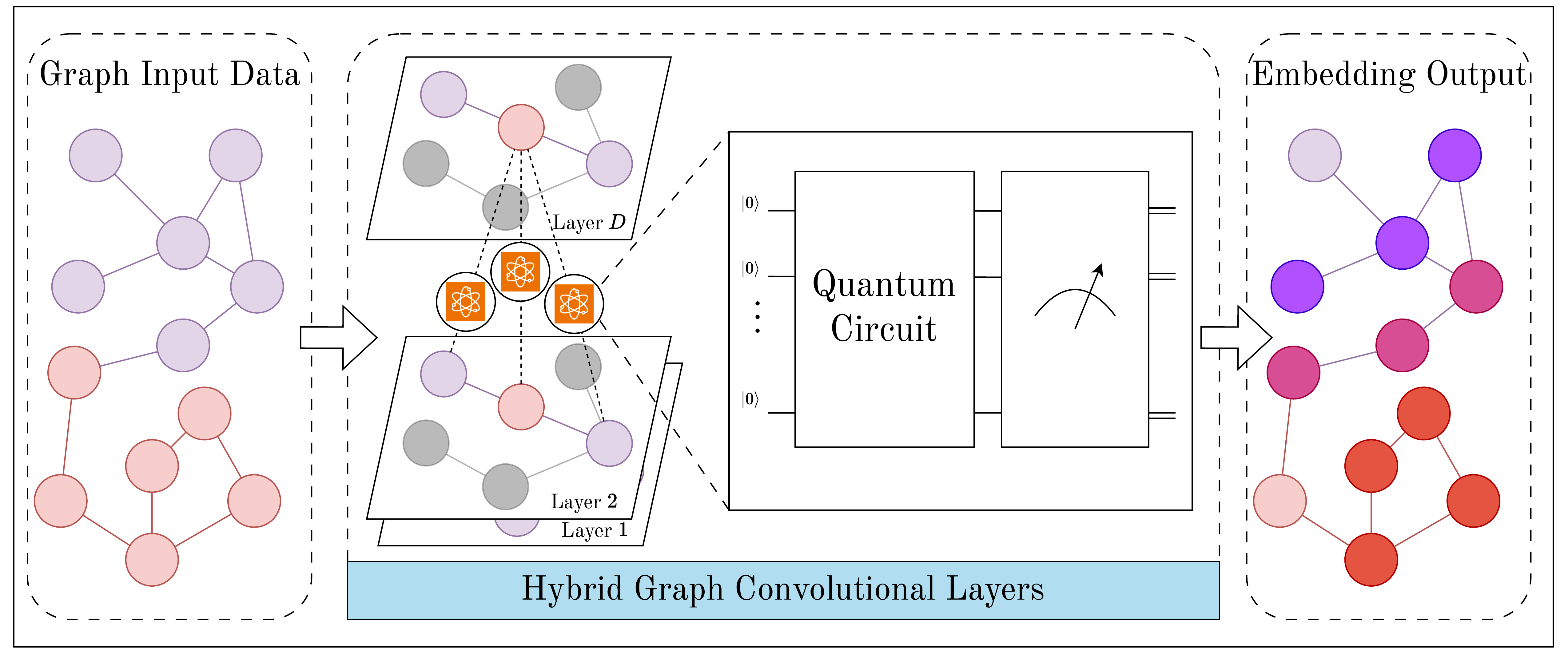}
        \caption{Hybrid quantum GNN.} 
        \label{fig: GNN_Survey-Q8.a}
        \vspace{-15pt}
    \end{figure}
    
    {\color{black}Another promising scheme is constituted by the hybrid quantum graph neural network (HQGNN), which replaces the classical MLP in GNNs by quantum neural networks (QNNs)~\cite{liao2024graph, 2025_ICAIIC_QGNN_Giang}. In details, as illustrated in Fig.~\ref{fig: GNN_Survey-Q8.a}, the HQGNN begins with graph input data, which is processed through hybrid graph convolutional layers that combine classical and quantum components. The quantum circuit performs updates on quantum states during training, effectively encoding node and edge relationships into the embedding space. By employing variational quantum circuits (VQC), QNNs are capable of facilitating quantum state updates during training, allowing for more efficient solution space exploration, while reducing the parameters and computational resources for large-scale graphs.} This hybrid quantum-classical approach allows for more efficient exploration of the solution space than classical DNNs, hence reducing both the number of parameters and the computational resources required for processing large-scale graphs. By incorporating quantum circuits within the DQN, the computationally intensive parts of GNNs can be offloaded to quantum hardware, while the rest of the GNN relies on classical processors. This hybrid quantum-classical approach offloads computationally intensive tasks to quantum hardware, leaving the remaining calculations for classical processors, making it ideal for applications like NG networks. These include optimizing resource allocation, dynamic slicing, and traffic management. The simulated results presented in Fig.~\ref{fig: Max-min-fairness} highlighted the potential of the proposed hybrid quantum GNN. This approach demonstrates the potential of hybrid quantum GNNs and paves the way for future advances.
    
    \subsubsection{Challenges and future directions:} QGNNs hold promising potential in terms of addressing the key challenges of NG networks, but they also face limitations in the era of noisy intermediate-scale quantum (NISQ) devices. At the current state of quantum computing, these NISQ devices have a limited qubit and significant quantum domain impairments. Naturally, the performance of a quantum computer depends on three key factors: scale, fidelity, and speed \cite{wack2021qualityspeedscalekey}. \textbf{Scale} refers to the number of qubits available in the quantum computer, which determines the dimension of the problem it can solve. \textbf{Fidelity} represents the level of quantum impairments in a quantum computer~\cite{PhysRevA.100.032328}. The so-called quantum volume characterizes the dimensions of quantum circuits that can be effectively run on a quantum computer, providing insight into the practical limits of these devices \cite{pelofske2022quantum}. \textbf{Speed}, quantified in terms of the number of circuit layer operations per second (CLOPS), indicates the computational efficiency of quantum circuits~\cite{wack2021qualityspeedscalekey}. Accordingly, implementing QGNNs on NISQ devices faces challenges such as representing high-dimensional problems using quantum circuits limited by the scale and quality of current quantum hardware~\cite{periyasamy2022incremental}. Beyond optimization, QGNNs can significantly enhance the security and privacy of communications in NG-IoT networks by incorporating quantum cryptography. This would enable unbreakable encryption methods, providing robust defenses against sophisticated cyber threats. Quantum key distribution and quantum secure direct communications are already quite mature \cite{Cao9684555, Pan10440135, Hosseinidehaj8439931}, but their standardization requires substantial future efforts.
    
    \section{Generic Design Guidelines} \label{Sec: Generic Design Guidelines}
    \begin{figure}[t]
        \includegraphics[trim=0.2cm 0.3cm 1.1cm 0.7cm, clip=true, width=0.5\textwidth]{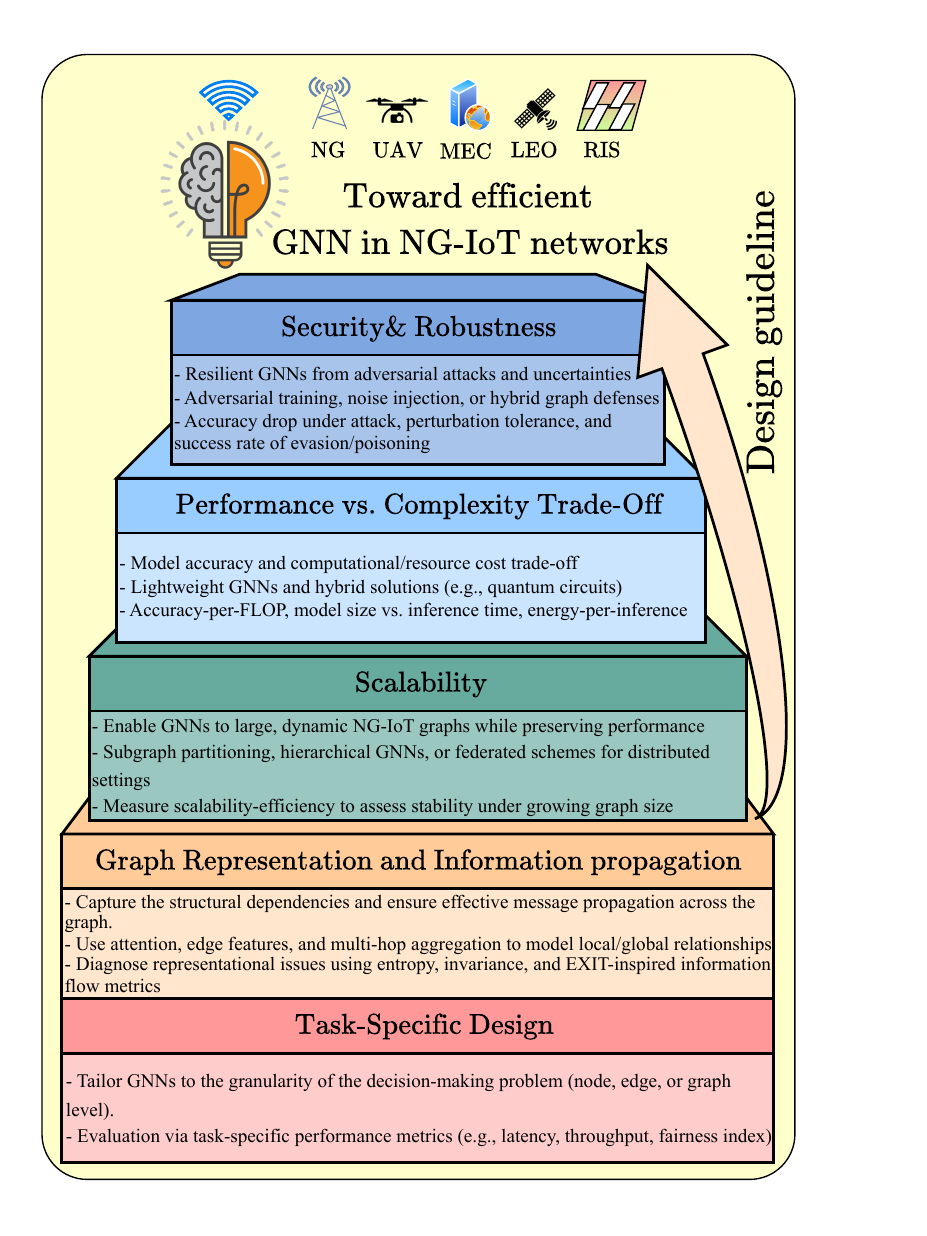}
        \caption{{Design guidelines for efficient GNNs in NG-IoT networks.}}
        \label{fig: GNN_Survey-3-Generic-Design-Guideline}
    \end{figure}
    To guide the development of efficient GNN models tailored for NG IoT networks, we highlight key design principles, as summarized in Fig.~\ref{fig: GNN_Survey-3-Generic-Design-Guideline}, and elaborated below.
    \begin{itemize}
        {\color{black}\item \textbf{Task-specific design:} GNN models are highly effective for node and edge-level tasks, such as user association, resource allocation, and network routing, providing precise predictions and improved network performance. For larger-scale problems, graph-level tasks offer a holistic view of the network, capable of holistic system optimization, for example, throughput and task offloading in complex environments like edge computing.

        \textbf{Evaluation metrics:} The effectiveness of this design can be validated using task-specific performance metrics, e.g., accuracy, precision, or latency for edge-level tasks; and throughput, fairness index, or load balance for graph-level tasks. These metrics indicate whether the GNN architecture is aligned with the intended scope of decision-making, ensuring that the model does not underfit or overfit the level of abstraction.
        
        \item \textbf{Ensuring effective graph representation and information propagation:} The design has to guarantee that the GNN model accurately captures the underlying structure and relationships within a network, reflecting the network's topological and functional characteristics. The model should effectively propagate information across the graph, ensuring that all important connections and dependencies are learned. Techniques such as attention mechanisms, multi-hop message passing, and edge feature enhancement can help ensure that the GNN accurately captures both local and global relationships.

        \textbf{Evaluation metrics:} Metrics such as graph entropy, equivariance, invariance, and attention weight distribution provide valuable insights into how effectively a GNN model captures and preserves the underlying structural information. Additionally, inference accuracy under partial graph observation evaluates the model’s ability to propagate critical information across the network, while EXIT chart-inspired metrics can monitor mutual information flow between layers, helping diagnose issues such as over-smoothing or representation collapse. Overall, these metrics demonstrate the critical role of representational power in achieving robust performance in complex NG-IoT environments.
        
        \item \textbf{Scalability of GNN models:} When designing GNNs for NG IoT networks, ensuring scalability is critical for efficiently handling large and dynamically evolving graph structures. GNNs must be capable of processing vast, evolving networks having numerous nodes and edges, while minimizing both the computational and memory costs. Approaches such as partitioning large graphs into smaller subgraphs or leveraging hierarchical GNN models can help scale solutions, while maintaining accuracy. Moreover, ensuring that the GNN models can generalize across various network sizes without performance degradation is essential for real-time network applications. 

        \textbf{Evaluation metrics:} The key metric is scalability-efficiency, measuring how performance (e.g., accuracy or reward) changes with varying graph sizes. This indicator is essential for validating the model’s practicality in real-time applications and resource-constrained NG-IoT deployments, ensuring it remains effective as the network scales.
        
        \item \textbf{Balancing performance and computational complexity:} The objective is to design lightweight GNNs imposing a low computational overhead, while maintaining high performance. Incorporating quantum computing has the potential of enhancing efficiency by offloading computationally intensive tasks to quantum circuits, in large-scale graphs. Besides, the depth of the GNN model must be carefully designed to strike a balance between achieving node feature uniqueness and avoiding over-smoothing.

        \textbf{Evaluation metrics:} Trade-off metrics like accuracy-per-FLOP, model size vs. latency, and energy-per-inference provide quantitative insights into this balance. These metrics will evaluate the deployment feasibility on constrained hardware (e.g., FPGAs or Edge TPUs), hence facilitating compact and effective GNN designs in NG-IoT networks.

        \item \textbf{Security and robustness:} It is vital to incorporate adversarial defense mechanisms, such as adversarial training or hybrid defense techniques for ensuring robustness in critical applications like healthcare or autonomous vehicles.

        \textbf{Evaluation metrics:} The main evaluation metrics will be accuracy drop (DA\%), perturbation tolerance thresholds, and resistance to evasion or poisoning attacks. Evaluating model reliability under simulated threats provides empirical justification for integrating defense mechanisms, proving that the model can maintain trustworthiness in volatile or adversarial environments.}
    \end{itemize}
    \section{Lessons Learned, Future Research Direction, and  Conclusions}\label{Conclusion}
    \subsection{Lessons Learned and Future Research Directions}
    {\color{black}Based on the comprehensive discussion of each open question, we summarize the key lessons learned and highlight promising future research directions to further strengthen the integration of GNNs with NG-IoT networks.
    \begin{itemize}
        \item GNNs can provide a flexible and expressive modeling tool to capture spatial, temporal, and structural dependencies in NG-IoT systems, enabling intelligent decision-making across edge, cloud, and hybrid platforms. However, the integration of GNNs into these systems is challenged by scalability limitations, heterogeneous data, and the constrained resources of IoT devices. Therefore, future researches should focus on developing lightweight, adaptive GNN architectures, using techniques such as model pruning, quantization, and dynamic graph sparsification, enabling real-time and energy-efficient deployment in practical NG-IoT scenarios.
        \item {\color{black}Privacy preservation in GNN-based NG-IoT systems presents unique challenges distinct from adversarial robustness. The message-passing mechanisms in GNNs may unintentionally expose critical structural patterns, such as influential node connections, bridge links, or tightly-knit communities, making generic privacy preservation techniques like simple anonymization or random link concealment ineffective. Such indiscriminate defenses often degrade model utility while failing to protect high-impact, context-sensitive relationships that adversaries are most likely to target. Future research must further advance context-aware privacy frameworks that selectively protect sensitive elements of NG-IoT networks while maintaining GNN utility. A key future direction is developing topology-aware DP mechanisms for graphs, since traditional DP methods inject excessive noise due to structural correlations. Recent advances demonstrate decentralized DP for subgraph statistics, personalized DP-GNNs enabling heterogeneous privacy budgets, and privacy-preserving embeddings \cite{DecentralizedDPSubgraph2023, PersonalizedDPGNN2023, PrivacyPreservingGraphEmbedding2023}. These approaches collectively pave the way for context-aware privacy solutions that balance strong privacy guarantees with high utility in NG-IoT networks. Another direction is topology-aware graph perturbation, where measures like betweenness centrality or edge criticality guide selective link obfuscation to protect high-impact connections, while avoiding unnecessary anonymization \cite{ijcai2021p310}. Complementary to DP, homomorphic encryption \cite{ran2022cryptogcn, ran2023penguin} enables privacy-preserving GCN inference by processing encrypted graph data on cloud servers. These methods protect both graph features and topology without revealing raw data, albeit at the cost of additional computational complexity.}

        
        \item The involvement of multimodal and non-Euclidean data with complex interdependencies causes traditional ML models ineffective in capturing topological structures and dynamic interactions. Meanwhile, GNNs are naturally suited to learn from these structures, but their application to real-time and distributed settings still requires innovation in data fusion strategies, synchronization mechanisms, and dynamic graph representation learning. This requires a robust multimodal GNN pipelines with attention-based fusion, continual learning capabilities, and temporal adaptation for evolving graphs.
        \item From a system-level perspective, the deployment of GNNs at the edge introduces a trade-off between performance, energy, and cost. We learn that embedding GNNs into edge devices (e.g., FPGAs, TPUs) and optimizing them through compression and hardware-software co-design can dramatically reduce latency and communication overhead. This necessitates research into green AI strategies, edge-aware GNN design, and scalable training techniques for large-scale IoT environments. Moreover, real-world adoption also depends on developing standardized frameworks for benchmarking GNN models in terms of accuracy, efficiency, and sustainability across heterogeneous NG-IoT settings.
        \item Distributed systems offer a natural fit for GNNs in NG-IoT environments where data privacy, communication cost, and scalability are critical. Yet, federated GNN training faces challenges from device heterogeneity, partial observability, and synchronization issues in dynamic graphs. This highlights the need for new protocols and architectures supporting structure-aware aggregation, over-the-air computation, and robust federated updates in non-IID, time-varying settings. Besides, an unified frameworks and public benchmarks will be essential to accelerate reproducibility and progress in this domain.
        \item Security and robustness remain pressing concerns. GNNs are vulnerable to adversarial attacks that manipulate graph structures or features, threatening reliability in mission-critical NG-IoT applications. Existing defense techniques provide partial solutions, but many are tailored to specific attack models or fail under dynamic graph conditions. Future directions should include designing universally robust GNNs using adversarial training, explainable AI, and quantum-safe encryption techniques, especially for sensitive applications in smart grids, autonomous transport, and blockchain-integrated systems.
        \item In emerging applications like ISAC and SAGSINs, GNNs show promise in managing massive interconnectivity and optimizing cross-domain coordination. However, challenges in handling highly dynamic topologies, spatial-temporal dependencies, and extreme heterogeneity remain. Future work should investigate hierarchical and modular GNN models that can seamlessly operate across different communication layers and adapt to time-varying resources.
        \item {\color{black}Semantic communications represent an emerging frontier for GNN integration, where learning-based models extract, process, and transmit meaning rather than raw data. Future research should explore how GNNs can support knowledge graph reasoning, semantic-aware encoding, and context-driven signal processing in NG-IoT applications. Papers by Zhang \textit{et al.} \cite{Zhang10614087}, Guo \textit{et al.} \cite{Guo9814464}, and Hello \textit{et al.} \cite{Hello10694291} offer early evidence of this synergy, but more robust GNN architectures are needed to maintain semantic fidelity under dynamic and resource-constrained conditions.}
        \item Finally, quantum computing opens new frontiers in accelerating GNN training, enhancing privacy via quantum cryptography, and expanding computational capacity for large-scale NG-IoT graphs. Despite limitations of current noisy quantum hardware, hybrid quantum-classical GNN models and quantum-enhanced optimization can offer long-term benefits. Additionally, designing scalable QGNN frameworks is critical to fully harness the potential of quantum computing for large-scale NG-IoT applications.
    \end{itemize}}

    \subsection{Conclusions}
    {\color{black}A critical appraisal of the application of Graph Neural Networks (GNNs) in NG-IoT networks was presented. We commenced by introducing the underlying fundamental graph concepts and GNN paradigms, establishing the foundational principles necessary for understanding their relevance. Building on this, we highlighted the pivotal role of GNNs in NG-IoT networks and modeled them as dynamic constrainted graphs. To further justify the GNN's applicability, we related them to LDPC decoding, framing GNNs as learnable generalizations of belief propagation, hence supporting convergence and information flow analysis via density evolution and EXIT charts. We further discussed how selecting appropriate task levels, among node-, edge-, or graph-level, can significantly enhance problem-solving efficiency across diverse NG-IoT scenarios. We also examined the application of GNNs to key NG technologies, including massive MIMO schemes, RISs, satellites, THz, MEC, and URLLC solutions, highlighting their capability to circumvent the limitations of traditional methods. We also analyzed their integration with distributed systems, highlighting their role in supporting distributed intelligence with privacy preservation and reduced communication overhead. Furthermore, we explored the integration of GNNs with emerging technologies, including integrated sensing and communication, space-air-ground-sea integrated networks, and quantum computing, showcasing their potential to enhance NG-IoT networks. Additionally, we addressed a range of critical security concerns by discussing adversarial attacks on GNN-based systems and by appraising the family of effective defense strategies. By systematically addressing a suite of ten open questions, this survey fills a crucial gap in understanding how GNNs can be efficiently leveraged in NG-IoT networks. In conclusion, our findings have emphasized the transformative potential of GNNs in optimizing, scaling, and securing NG-IoT networks, laying the groundwork for future research in this rapidly evolving field.}



 
\scriptsize
\bibliographystyle{IEEEtran}
\bibliography{Bib1}
\end{document}